\documentclass[10pt,onecolumn,draft]{IEEEtran}

\usepackage{amsmath}
\usepackage{amsthm}
\usepackage{amssymb}
\usepackage{graphicx}
\usepackage{rotating}
\usepackage{flushend}
\usepackage{verbatim}
\allowdisplaybreaks
\usepackage{mathtools}
\usepackage{tikz}

\newsavebox{\tempbox}

\usepackage{xcolor}

\newcommand{\logt}{\log_{2}}

\newcommand{\abs}[1]{\left| #1 \right| }

\newcommand{\set}[1]{\left\{ #1 \right\}}

\newcommand{\defn}{\triangleq}

\newcommand{\mbf}[1]{\mathbf{#1}}

\newcommand{\mcf}[1]{\mathcal{#1}}
\newcommand{\idc}[1]{1\{#1\} }

\makeatletter
\newcommand{\subalign}[1]{%
  \vcenter{%
    \Let@ \restore@math@cr \default@tag
    \baselineskip\fontdimen10 \scriptfont\tw@
    \advance\baselineskip\fontdimen12 \scriptfont\tw@
    \lineskip\thr@@\fontdimen8 \scriptfont\thr@@
    \lineskiplimit\lineskip
    \ialign{\hfil$\m@th\scriptstyle##$&$\m@th\scriptstyle{}##$\crcr
      #1\crcr
    }%
  }
}
\makeatother

\DeclareMathOperator{\markov}{\setlength{\unitlength}{.5cm} \begin{picture}(1,1)  \put(0,.22){\line(1,0){1}}  \put(.5,.22){\circle{.3}}   \end{picture}}
\DeclareMathOperator*{\argmin}{argmin} 
\DeclareMathOperator*{\argmax}{argmax} 

\DeclareMathAlphabet\mbcf{OMS}{cmsy}{b}{n}

\newtheorem{theorem}{Theorem}
\newtheorem{define}[theorem]{Definition}
\newtheorem{lemma}[theorem]{Lemma}
\newtheorem{cor}[theorem]{Corollary}
\newenvironment{remark}[1][Remark]{\begin{trivlist}
\item[\hskip \labelsep {\bfseries #1}]}{\end{trivlist}}

\IEEEoverridecommandlockouts

\setkeys{Gin}{draft=false}

\begin{document}

\title{Secret key authentication capacity region, Part \textrm{I}: average authentication rate}
\author{ Jake Perazzone, Eric Graves, Paul Yu, and Rick Blum
\thanks{This research was presented in part at the International Symposium on Information Theory (ISIT) 2018.
This material is based upon work partially supported by the U. S. Army Research Laboratory and the U. S. Army Research Office under grant number W911NF-17-1-0331 and by the National Science Foundation under grants ECCS-1744129 and CNS-1702555.}
\thanks{Eric Graves and Paul Yu are with the Army Research Lab,  Adelphi, MD
  20783, U.S.A. \texttt{\{eric.s.graves9.civ, paul.l.yu.civ\}@mail.mil } }%
\thanks{Jake Perazzone and Rick Blum are with the Department of Electrical and Computer Engineering,
 Lehigh University,
 Bethlehem, PA 18015, U.S.A. 
\texttt{\{jbp215,rb0f\}@lehigh.edu}}%
}
\maketitle

\begin{abstract}
This paper investigates the secret key authentication capacity region. 
Specifically, the focus is on a model where a source must transmit information over an adversary controlled channel where the adversary, prior to the source's transmission, decides whether or not to replace the destination's observation with an arbitrary one of their choosing (done in hopes of having the destination accept a false message). 
To combat the adversary, the source and destination share a secret key which they may use to guarantee authenticated communications.
The secret key authentication capacity region here is then defined as the region of jointly achievable message rate, authentication rate, and key consumption rate (i.e., how many bits of secret key are needed).

This is the first of a two part study, with the parts differing in how the authentication rate is measured. 
In this first study the authenticated rate is measured by the traditional metric of the maximum expected probability of false authentication. 
For this metric, we provide an inner bound which improves on those existing in the literature. 
This is achieved by adopting and merging different classical techniques in novel ways. 
Within these classical techniques, one technique derives authentication capability directly from the noisy communications channel, and the other technique derives its' authentication capability directly from obscuring the source.

\end{abstract}

\section{Introduction}

Authentication is inherently a physical layer problem; any protocol that labels data as valid or invalid naturally creates a bifurcation of the physical layer observations. 
What is clear is that this labeling should degrade the performance of the communication system in comparison to a system which does not require authentication since any possible observation which is labeled as inauthentic can no longer contribute to the probability of reliably decoding.
Our goal with this series of papers is to explore this trade-off.
In particular, we focus on a model previously considered by Lai et al.~\cite{lai2009authentication} and as a sub-case by Gungor and Koksal~\cite{gungor2016basic}.
In this model, the source sends information in the presence of an adversary that listens to the communication and can change what is received by the true destination. 
On the other end, the communicating parties are allowed to share a secret key prior to communications.
For this model, our goal is to derive a classical information theoretic ``rate region'' that describes the trade-off between information rate, authentication rate (to be defined), and the amount of secret key required (termed the key consumption rate).

This work has been split into two papers since, in the course of our efforts to obtain the desired rate region characterization, we discovered that the traditional metric for authentication rate (the maximum probability of false authentication) does not necessarily represent how strong the authentication capability of the system is for a typical use.
In fact, the traditional metric is beholden to extremely unlikely events occurring in the communication channel; for example, a noisy binary symmetric channel acting as a noiseless channel.
Upon this discovery, we formulated a new metric which only considers ``typical'' behaviour of the communication channel, with all other behavior being written off as loss. 
Nevertheless, we still consider the traditional metric here, and favorably compare our results to those existing in the literature.  
The splitting of the papers based upon choice of metric is done to allow flexibility in how the results are presented, with the traditional metric's dependence on unlikely empirical channel distributions dictating a notation where the information theoretic terms are functions of probability distributions, while the new metric allows for a (in our opinion) simpler presentation where the information theoretic terms are functions of random variables.

Authentication is an important topic considered outside of the information theoretic literature. 
Some examples include: Yu et al.~\cite{Yu08} who used spread spectrum techniques in addition to a covert channel to ensure authentication, 
Xiao et al.~\cite{Xiao08} who used the unique scattering of individual users in indoor environments to authenticate packets, and Korzhik et al.~\cite{Korzhik07} who make use of a (possibly noisy) initialization setup to create unique correlations which then allow for detection. 
These methods, while perhaps more suitable for application, use tools that are insufficient in determining the various information theoretic measures considered here.
Instead, what they highlight is a concern for authentication that should not be ignored. 
With this work, we hope to provide insight into the general problem, and provide baselines to what is possible.

On the other hand, authentication has only somewhat been considered from the information theoretic viewpoint.
In particular, it can be argued that Blackwell et al.~\cite{blackwell1960capacities} and their study of the  \emph{arbitrarily varying channel} (AVC) was the first true study of authentication. 
In particular, in the AVC, an adversary can at will choose the state of the communication channel between the two communicating parties.
This classic work and those that followed, such as \cite{wolfowitz2012coding,dobrushin1975coding,csiszar1988capacity}, all considered the maximum communication rate that can be obtained subject to an arbitrarily small probability of error (over any choice of communication states by the adversary). 
Note, this indeed implies that a decoded message would be authentic because the probability of error must take into account the adversaries actions.
In this vein, Ahlswede~\cite{ahlswede1978elimination} considered the communication rate over an AVC when the source and destination share a secret key. 
More specifically, Ahlswede gave the two communicating parties access to shared randomness, which must be kept private from the adversary prior to transmission.
For Ahlswede, allowing this secret key dramatically improved the communication rate, essentially transforming AVCs into a compound channel.

While these papers do examine an aspect of authentication, one can also argue that they are much too strict in their operational requirement. 
Today, the detection of the adversary's involvement is a strong enough result for many fields of security; for example, in quantum key distribution, a system is considered operational even though the adversary can reduce the key rate to zero by measuring the data.
In our case, it makes even more practical sense to forgo such a harsh operational requirement. 
That is, if an adversary wanted to reduce the communication rate to zero between two parties in practice, they would simply need a strong enough jammer. 
Of course, simply jamming a signal is different than trying to have a node accept a fabricated message as authentic. 
This is the stance we adopt here: when the adversary is attacking, a system is operational if it can decode the correct message \emph{or} detect the attack; when the adversary is not attacking, we want the system to communicate as much data as possible.

Adopting this viewpoint, works by Jiang~\cite{jiang2014keyless,jiang2015optimality}, Graves et al.~\cite{graves2016keyless}, and Kosut and Kliewer~\cite{kosut2018authentication} all consider authentication over an AVC without a secret key.
In particular, Jiang considered the sub-case of AVC where the output of the AVC was independent of the legitimate parties input for all but a single channel state. 
Graves et al. considered a general AVC where the adversary is given the side information of which message is being transmitted while Kosut and Kliewer considered the general AVC case\footnote{We shall also be adopting the terminology of Authentication Capacity, first coined by Kosut and Kliewer. Although it should be noted we are looking at a region, we adopt this terminology since the root requirement for a system to be operational is equivalent.}.
Each of these works avoids looking at the strength of the authentication capability, and instead only considers the data rate given the maximum probability of false authentication goes to zero.

Works considering a secret key and strength of the authentication capability have a genesis in that of Simmons~\cite{Auth}, who considered a special case of the model presented here with all channels noiseless. 
Since all channels were noiseless, all authentication capability had to be derived from the secret key.
This distinction fundamentally separates the problems of keyless and secret key authentication; the former relies on exploiting the nature of the communication channels, while the latter relies on exploiting a finite resource.
Later came the works of Lai et al.~\cite{lai2009authentication} and that of  Gungor and Koksal~\cite{gungor2016basic}, who both consider generalizations of Simmons' model with noisy channels. 
Each of these works has aspects which could be strengthened.
Lai et al. require the amount of secret key bits to be asymptotically negligible when compared with the blocklength of the transmission.
In doing so, though, they can make no distinction in the importance of verifying $10$ versus $10,000$ bits of data. 
Meanwhile, Gungor and Koksal's coding scheme is inefficient and mismanages the key by unnecessarily using it in a way that favors the adversary.
Furthermore, their work does not attempt\footnote{Although we did endeavour to extract such a rate region from their works, we were unable to do so and instead had to settle for an outer bound. Nevertheless, our results improve on this outer bound.} to explicitly derive such a region, instead opting for a presentation of error exponents. 

Our works look to characterize this trade-off between information, authentication, and key consumption rate.
Towards the efforts (under the traditional metric), we are only able to derive an inner bound on the rate region when considered over an arbitrary number of rounds of communication.
To the credit of this inner bound, it improves on all inner bounds previously appearing in literature (even that of our earlier work~\cite{perazzone2018inner}). 
In comparison to Lai et al., we shall measure the authentication relative to the block length, hence allowing the level of protection to scale with the amount of data while our coding scheme will strictly improve on the rate region obtained by Gungor and Koksal, and be presented as a rate region that is, in principal, computable. 
Our results are achieved by combining a broadcast channel with confidential communications (see Csisz{\'a}r and K{\"o}rner~\cite{csiszar1978broadcast}) code adapted using a strategy similar to that in~\cite{lai2009authentication}, with the coding scheme of Simmons~\cite{Auth}. 
While it would be easy to construct a new ``novel'' code, and present it as such, we feel that it is more important to emphasize that the coding scheme can be derived through previously established concepts (with appropriate modifications).

We begin this discussion by thoroughly presenting the problem formulation (notation, model, operational definitions) in Section~\ref{sec:for}.
It should be noted that these sections will differ between the two papers.
With these in place, we circle back to discuss in finer detail~\cite{lai2009authentication},~\cite{Auth},~\cite{gungor2016basic}, and~\cite{csiszar1978broadcast} before presenting our results in Section~\ref{sec:main_cont}, and giving examples in Section~\ref{sec:examples}.
Proofs can be found in appendices.


\section{Formulation}\label{sec:for}

\subsection{Notation}\label{sec:notation}

Uppercase letters will be used to denote \emph{random variables} (RVs) and lowercase letters will be used to denote constants. 
The probability of event $\mcf{A}$ is denoted $\Pr (\mcf{A} )$. 
Function $p$ with subscript RV will be used to denote the probability distribution over the RV (i.e., $p_{X}(x) = \Pr (X=x)$). 
To simplify presentation, the subscript may be suppressed when clear. 
Calligraphic font or curly brackets will be used to denote sets, for instance $\mcf{Y} = \set{1,\dots, 10}$. 
The only exceptions to this are the set of positive real numbers, denoted $\mathbb{R}^+$, and the set of positive integers, denoted $\mathbb{Z}^+$. 
Subscripts will generally be used for bookkeeping purposes, 
while $|$ denotes the word ``given,'' and $:$ ``subject to.''

The function $\times$ will be used to denote the Cartesian product.
We will frequently need to use the Cartesian product of $n$ (where $n$ will denote the block length of a given code) correlated RVs, constants, and sets.
This need arises so frequently that we denote these Cartesian products by bold face.
For instance, $\mbf{X} = \times_{i=1}^n X_i = (X_1,\dots,X_n)$ and $\mbcf{X} = \times_{i=1}^n \mcf{X}.$
When using this notation with a probability distribution, the terms in the product are uncorrelated. 
For example, given a probability distribution $t$ over $\mcf{X}$
\[
\mbf{t}(\mbf{x}) = \prod_{i=1}^n t(x_i)
\]
for each $\mbf{x} \in \mbcf{X}$.

The indicator function of an event $\mcf{A}$ is denoted $\idc{\mcf{A}}$, that is $\idc{\mcf{A}} = 1$ if $\mcf{A}$ occurs, otherwise $\idc{\mcf{A}} = 0$.

The set of all probability distributions on a certain set, say $\mcf{X}$, is denoted by $\mcf{P}(\mcf{X})$, likewise $\mcf{P}(\mcf{Y}|\mcf{X})$ denotes the probability distributions of $\mcf{Y}$ conditioned on elements of $\mcf{X}$. 
The set $\mcf{P}(\mcf{Y}\gg \mcf{X})$ represents a special subset of $\mcf{P}(\mcf{Y}|\mcf{X})$, where for each $v \in \mcf{P}(\mcf{Y} \gg \mcf{X}) $ and $y \in \mcf{Y}$ there exists at most one $x \in \mcf{X}$ such that $v(y|x) > 0$. 
Note, for random variables $X,Y,Z$, if $p_{Y|X} \in \mcf{P}(\mcf{Y}\gg \mcf{X})$, then $X,~Y,~Z$ form a Markov chain, $X \markov Y \markov Z$.

Another special subset of the distributions is the possible ``empirical distributions'' (or type classes) for a given $n$-length sequence, denoted $\mcf{P}_{n}(\cdot)$. 
The empirical distribution of sequence $\mbf{x}$, denoted $p_{\mbf{x}}$, is the distribution defined by the proportion of occurrences of $x$ in sequence $\mbf{x}$.
In other words 
\[ p_{\mbf{x}}(b) \defn  \frac{\sum_{i=1}^n \idc{x_i = b} }{n}, \quad\quad \forall b \in \mcf{X}.
\]
This follows similarly for empirical conditional distributions, but we further list the empirical distribution of the conditioning value, such as $\mcf{P}_n(\mcf{Y}|\mcf{X};\rho)$ for $\rho\in \mcf{P}_n(\mcf{X})$. 
Here the empirical conditional distribution of $\mbf{y}$ given $\mbf{x}$ is defined by 
\[ p_{\mbf{y}|\mbf{x}}(b|a) \defn  \frac{\sum_{i=1}^n \idc{y_i =b} \idc{x_i = a} }{\sum_{i=1}^n \idc{x_i =a }}, \quad\quad \forall a\times b \in \mcf{X} \times \mcf{Y} .
\]
For each $\mu \in \mcf{P}_{n}(\mcf{Y}|\mcf{X};\rho)$ and $\rho \in \mcf{P}_{n}(\mcf{X})$ the type class of $\mu$ given a $\mbf{x}$ such that $p_{\mbf{x}} = \rho$ is denoted
\[
\mbcf{T}_{\mu}(\mbf{x}) \defn \set{\mbf{y} : p_{\mbf{y}|\mbf{x}} = \mu }.
\]

Next, we need to define particular functions of probability distributions.
For $\rho\in \mcf{P}(\mcf{X}|\mcf{U})$ and $v \in \mcf{P}(\mcf{Y}|\mcf{X},\mcf{U})$, the distributions $v\rho \in \mcf{P}(\mcf{Y}|\mcf{U})$ and $v \times \rho \in \mcf{P}(\mcf{Y},\mcf{X}|\mcf{U})$ are defined by
\[
v\rho(y|u) \defn \sum_{x\in \mcf{X}} v(y|x,u) \rho(x|u), \quad\quad \forall (y,u)\in \mcf{Y}\times \mcf{U}
\]
and
\[
(v\times \rho) (y,x|u) \defn v(y|x,u) p(x|u), \quad \quad \forall (x,y,u) \in \mcf{X}\times \mcf{Y} \times \mcf{U}.
\]
When there is a dimensional mismatch in the above notation, it is to be treated as if the distribution are independent of the missing dimension. 
For example, when $ v \in \mcf{P}(\mcf{Y}|\mcf{X})$ and $\rho \in \mcf{P}(\mcf{X}|\mcf{U})$, then 
\[
v\rho(y|u) = \sum_{x \in \mcf{X}} v(y|x) \rho(x|u).
\]
Black board bold (other than the two exceptions discussed earlier) is used to denote functions which are averaged over RVs or their distribution.
Of particular importance is $\mathbb{E}$ which denotes the expectation operator.
Other important functions are entropy, mutual information, and Kullback-Leibler divergence, denoted (respectively) by
\begin{align*}
\mathbb{H}(\rho|\sigma) &= -\sum_{\subalign{u &\in \mcf{U}\\ x & \in \mcf{X}}}  \rho(x|u)\sigma(u) \logt \rho(x|u), \\
\mathbb{I}(q,\rho|\sigma) &= \sum_{\subalign{y&\in \mcf{Y},\\x &\in \mcf{X},\\ u&\in \mcf{U}}} q(y|x,u)\rho(x|u) \sigma(u) \logt \frac{q(y|x,u)}{q\rho(y|u)}\\
&= \mathbb{H}(q\rho|\sigma) - \mathbb{H}(q|\rho\sigma) ,\\
\mathbb{D}(\rho||  \omega |\sigma ) &= \sum_{\subalign{x&\in \mcf{X},\\ u &\in \mcf{U}}}  \rho(x|u)\sigma(x) \logt \frac{ \rho(x|u)}{ \omega (x|u)}.
\end{align*}
for $q \in \mcf{P}(\mcf{Y}|\mcf{X},\mcf{U})$, $\rho \in \mcf{P}(\mcf{X}|\mcf{U}),$ $ \omega \in \mcf{P}(\mcf{X}|\mcf{U}),$  and $\sigma \in \mcf{P}(\mcf{U}).$
Additionally, four other functions will be extremely useful
\begin{align} 
\mathbb{F}( \mu|| t,\rho|\sigma) &\defn  \min_{\substack{\zeta \in \mcf{P}(\mcf{Y},\mcf{X}|\mcf{U}) : \\ \subalign{ \sum_{y\in \mcf{Y}}\zeta(y,x|u) &= \rho(x|u) , \\ \sum_{x\in \mcf{X}}\zeta(y,x|u) &= \mu(y|u) }}} \mathbb{D}(\zeta || t \times \rho | \sigma)  \notag\\ 
&= \min_{\substack{\zeta \in \mcf{P}(\mcf{Y},\mcf{X}|\mcf{U}) : \\ \subalign{ \sum_{y\in \mcf{Y}}\zeta(y,x|u) &= \rho(x|u) , \\ \sum_{x\in \mcf{X}}\zeta(y,x|u) &= \mu(y|u) }}}   \sum_{\subalign{y&\in \mcf{Y},\\ x &\in \mcf{X}, \\u &\in \mcf{U}}} \zeta(y,x|u) \sigma (u) \logt \frac{ \zeta(y,x|u)}{t(y|x)\rho(x|u)},  \notag \\
\mathbb{F}_n( \mu|| t,\rho|\sigma) &\defn  \min_{\substack{\zeta \in \mcf{P}_n(\mcf{Y},\mcf{X}|\mcf{U};\sigma) : \\ \subalign{ \sum_{y\in \mcf{Y}}\zeta(y,x|u) &= \rho(x|u) , \\ \sum_{x\in \mcf{X}}\zeta(y,x|u) &= \mu(y|u) }}} \mathbb{D}(\zeta || t \times \rho | \sigma) , \notag\\
\mathbb{S}_{a,b}(\mu,\nu|\rho,\sigma) &\defn  \mathbb{I}(\mu,\rho|\sigma) + a - \mathbb{I}(\nu,\rho|\sigma) + \left| \mathbb{I}(\mu\rho,\sigma) + b - \mathbb{I}(\nu\rho,\sigma) \right|^+ , \notag\\
\mathbb{S}_{a,b}(\nu|\rho,\sigma) &\defn a - \mathbb{I}(\nu,\rho|\sigma) + \left|  b - \mathbb{I}(\nu\rho,\sigma) \right|^+ \notag
\end{align} 
where, in addition to the previously defined distributions, $\mu \in \mcf{P}(\mcf{Y}|\mcf{U})$ and $t \in \mcf{P}(\mcf{Y}|\mcf{X})$.
The need for function $\mathbb{F}$ and $\mathbb{F}_n$ arises through Lemma~\ref{lem:btx}, while $\mathbb{S}$ allows us to measure the secrecy of a channel.

For any $\mbf{a} \in \mathbb{R}^n$, or set $\mcf{B}\subseteq \mcf{X}$, define the following absolute values:
\begin{align*}
\abs{\mbf{a}} &= \sqrt{\sum_{i=1}^n |a_i|} \\
\abs{a_i}^+ &= a_i \idc{a_i > 0} \\
\abs{a_i}^- &= a_i \idc{a_i < 0} \\
\abs{\mcf{B}} &= \sum_{x \in \mcf{X}} \idc{x \in \mcf{B}}.
\end{align*}

Finally the $O$ function from the Bachmann-Landau notation will be employed here. 
That is by writing $g(x,n) = f(x,O(h(n)))$, we are saying that there exists a constant $\zeta$, independent of $n$, such that
$$| g(x)|  \leq \max_{ t \in [-\zeta h(n),\zeta h(n)]} f(x,t).$$

\subsection{Model}

\tikzstyle{block} = [draw, fill=white, rectangle, 
    minimum height=20pt, minimum width=20pt, text centered]

\begin{figure}
 \begin{center}
\begin{tikzpicture}
\node[block] (enc) at (0,0) {$\begin{array}{c} \text{Alice} \\ f_i \end{array}$};
\node[block] (chan) at (3.2,0) {$\begin{array}{c} \text{Channel} \\ t \end{array}$};
\node[block] (chanz) at (1,-3) {$\begin{array}{c} \text{Channel} \\ q \end{array}$};
\node[block] (dec) at (8,-1.5){$\begin{array}{c} \text{Bob} \\ \varphi_i  \end{array}$};
\node[block] (adver) at (3.2,-3){$\begin{array}{c} \text{Gr{\'i}ma} \\ \psi \end{array}$};
\node[circle] (switch) at (6,-1.5)  {$\begin{array}{c}~  \end{array}$ };
\draw[->,thick] (-1.5,0) -- node[above]{$M$} (-1.5,0)  -- (enc.west) ;
\draw[->,thick] (enc.east)  node[above right]{$\mbf{X}$}  -- (chan.west) ;
\draw[->,thick] (1,0) -- (chanz.north) ;
\draw[->,thick] (chan.east) -- (6,0) -- (switch.north) ;
\draw[->,thick] (adver.east) -- (6,-3) -- (switch.south) ;
\draw[-,thick] (switch.south) -- (switch.east);

\draw[->,thick] (dec.east) node[above right]{$\hat M$} -- (9.5,-1.5) ;
\draw[->,thick] (chanz.east) node[above right]{$\mbf{Z}$}  --(adver.west);
\draw[<->,thick,dashed] (enc.north) -- (0,1.2) -- (4,1.2) node[above]{$K$} -- (8,1.2)-- (dec.north) ;
\draw[->,thick,dashed] (adver.north) -- (3.2,-1.5) 
-- 
(switch.east) ;
\draw[->,thick] (switch.east) node[above right] {$\mbf{Y}$}  -- (dec.west);
\end{tikzpicture}
 \caption{Channel model in the $i$-th round of communication where Gr{\'i}ma has chosen to interlope.}
\label{fig:II}
\end{center}
\end{figure}
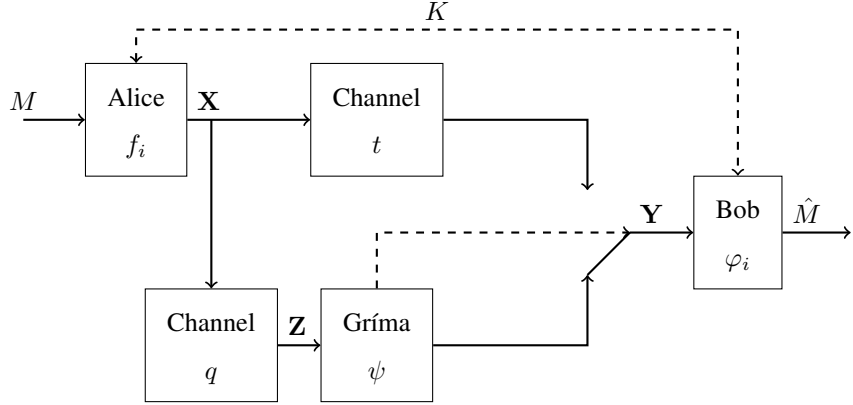

In this studies' model, for $j \in \mathbb{R}^+$ rounds of communication, Alice (representative of the source) is connected with Bob (representative of the destination) via a \emph{discrete memoryless-adversarial interlope channel$(t,q)$} (DM-AIC$(t,q)$), where $t \in \mcf{P}(\mcf{Y}|\mcf{X})$ and $q \in \mcf{P}(\mcf{Z}|\mcf{X})$.
The DM-AIC$(t,q)$, pictured in the $i$-th round in Figure~\ref{fig:II}, is controlled by Gr{\'i}ma\footnote{Chosen for Gr{\'i}ma Wormtongue from \emph{Lord of the Rings} by J.R.R. Tolkien. Gr{\'i}ma was an advisor to the King of Rohan, while secretly an agent of Saruman. Thus his role was to listen to information presented to the King and manipulate it towards Saruman's agenda. This seemed more appropriate than ``Eve,'' since the adversary does not only take the role of eavesdropper.} (representative of the adversary) who, prior to the communication round, decides if Bob will receive the transmission from Alice or himself.
When Gr{\'i}ma decides to send a message to Bob, instead of allowing Alice to send the message to Bob, it will be called interloping. 
The $t$ and $q$ in a DM-AIC$(t,q)$ specify the memoryless channels that connect Alice to Bob (if Gr{\'i}ma does not interlope) and Alice to Gr{\'i}ma respectively. 
Hence when Gr{\'i}ma does not interlope, Bob will receive $\mbf{y}$ with probability $\mbf{t}(\mbf{y}|\mbf{x}) \defn \prod_{\ell=1}^n t(y_\ell|x_\ell)$ given Alice transmitted $\mbf{x}$.
On the other hand, when Gr{\'i}ma does interlope, Gr{\'i}ma arbitrarily decides the value of $\mbf{y}$, and does so as a function of his own current and previous rounds' observations.
In other words if Gr{\'i}ma interlopes in the $\ell$-th, Bob will receive $\mbf{y}$ with probability $\psi_{\ell}(\mbf{y} | \times_{i=1}^\ell \mbf{z}_i)$, where $\psi_\ell \in \mcf{P}(\mbcf{Y}|\times_{i=1}^\ell \mbcf{Z})$ minimizes the authentication measure (to be discussed more later), given Gr{\'i}ma received $\mbf{z}_1,\dots,\mbf{z}_{\ell}$ in the first through $\ell$-th rounds respectively. 
Regardless of whether or not Gr{\'i}ma interlopes, he will observe $\mbf{z}$ with probability $\mbf{q}(\mbf{z}|\mbf{x})$ given Alice transmitted $\mbf{x}$.
One final point, the DM-AIC$(t,q)$ is not a true memoryless channel, as Gr{\'i}ma is not required to act in a memoryless fashion; the ``memoryless'' part of the DM-AIC$(t,q)$ only refers to the component channels $t$ and $q$.

Now, over these $j$ rounds of communication Alice wishes to send messages $M_1,\dots,M_j$ to Bob, where each message is independent and uniformly distributed on the set $\mcf{M} = \{1,\dots,2^{nr}\}$, with $n \in \mathbb{Z}^+$  and $r \in\mathbb{R}^+$.
To assist in communication, prior to the first round, Alice and Bob share a secret key, $K$, distributed uniformly over $\mcf{K} \defn \set{1, \dots, 2^{n j \kappa}}$, where $\kappa \in \mathbb{R}^+$.
For simplicity, we assume that $2^{nr}\in \mathbb{Z}^+ $ and $2^{nj\kappa}\in \mathbb{Z}^+$.
To send these messages, Alice uses an \emph{encoder} that selects the channel input sequence $\mbf{X}= \times_{i=1}^n X_{i}$ as a stochastic function of the message $M$ and key $K$.
Throughout this paper, the encoder will be identified by the stochastic relationship between message, key, and channel input sequence. 
More specifically, $f_i \in \mcf{P}(\mbcf{X}|\mcf{M} ,\mcf{K})$ will be used to denote the encoder used in the $i$-th round.

On the other end, Bob uses a \emph{decoder} to estimate the message (or detect that Gr{\'i}ma interloped) as a function of the received sequence from the DM-AIC$(t,q)$ and the shared key. 
In the case that the decoder does declare that Gr{\'i}ma interloped, it does so by producing the symbol ``$\mbf{!}$'' instead of a message.
Similar to the encoder, the decoder in the $i$-th will be identified by a conditional probability distribution $\varphi_i \in \mcf{P}(\mcf{M} \cup \{ \mbf{!} \} |\mbcf{Y},\mcf{K})$, where the $\mbf{!}$ symbol is used by the decoder to signify that the message is false.

\subsection{Operational Definitions}

So far the following operational definitions have been introduced.
\begin{center}
\begin{tabular}{ c|c } 
Quantity & Name 
\\
 \hline \hline
 $j$ & total rounds of communication \\
$f_i$ & $i$-th round encoder \\
$\varphi_i$ & $i$-th round decoder \\
$\times_{i=1}^j (f_i,\varphi_i)$ & code \\
$n$ &block size \\
$r$ & message rate \\
$\kappa $ & key consumption rate 
\end{tabular}
\end{center}
Before moving on to values not previously introduced, we would like to draw attention to a key difference in how the message set and possible keys are specified. 
In particular, the size of the message set is determined by the block size and message rate $(\mcf{M}= \{1,\dots,2^{nr}\})$, while the number of possible secret keys is determined by the block size, key consumption rate, and the total number of rounds $(\mcf{K}=\{1,\dots,2^{nj\kappa}\})$. 
This is done so that both the key consumption rate and message rate are measures of resource per transmitted symbol. 
Indeed, a new message is drawn from the message set for each round, while the same secret key must be used for all $j$-rounds of communication.

Another important operational parameter of note is the \emph{authentication rate} $$\alpha\defn \min_{i\in \{1,\dots,j\}} \min_{\psi \in \mcf{P}(\mbcf{Y}|\mbcf{Z}^i)} -n^{-1} \logt  \mathbb{E}  \left[ \omega_{f^j,\varphi^j,i}(\mbf{Z}^i, M_i,K) \right], $$
where 
$$
\omega_{f^j,\varphi^j,i}(\mbf{z}^i,m,k) \defn \sum_{\mbf{y}} \psi(\mbf{y}|\mbf{z}^i) \varphi_{i}(\mcf{M} - \set{m} |\mbf{y},k ),
$$
and for each $\mbf{z}^i,m,k$ the probability $(\mbf{Z}^i,M_i,K) = \mbf{z}^i,m,k$  is
$$ 2^{-n(ir +j\kappa)} \left( \sum_{\mbf{x} \in \mbcf{X}} \mbf{q}(\mbf{z}_i | \mbf{x}) f_i (\mbf{x}|m,k) \right)  \prod_{\ell=1}^{i-1} \sum_{\substack{\mbf{x} \in \mbcf{X} \\ m \in \mcf{M}}} \mbf{q}(\mbf{z}_\ell | \mbf{x}) f_\ell (\mbf{x}|m,k)  .$$
Note the quantity $\omega_{f^j,\varphi^j,i}(\mbf{z}^i,m,k)$ is the probability of Gr{\'i}ma being successful in his attack given that he has received $\mbf{z}^i$, the message in the $i$-th round is $m$, and the secret key shared between Alice and Bob is $k$.
As for the authentication rate itself, it is once again a resource per transmitted symbol. 
That is, it is a measure of the exponent of the worst case average probability of false authentication that is normalized by block length.

The final fundamental operational parameter is the \emph{probability of message error}, 
\begin{equation*}
\varepsilon_{f^j,\varphi^j} \defn \min_{i\in\{1,\dots,j\}} \varepsilon_{f_i,\varphi_i}  
\end{equation*}
where
\begin{align*}
\varepsilon_{f,\varphi}  &\defn  \mathbb{E}\left[ \varepsilon_{f,\varphi}(M,K) \right]\\
\varepsilon_{f,\varphi}(m,k) &\defn 1- \sum_{\mbf{x},\mbf{y}} \varphi(m|\mbf{y},\mbf{k}) \mbf{t}(\mbf{y}|\mbf{x}) f(\mbf{x}|m,k)
\end{align*}
Unlike previous values, the probability of message error is not measured on a per transmitted symbol basis.
Instead the probability of message error is measured as the probability of error over a single round.

We can now put all of these operational parameters together, to define the operational measure of the code.
\begin{define}\label{def:average code}
Code $(f^j,\varphi^j)$ is a $(r,\alpha,\kappa,\epsilon,j,n)$-\emph{average authentication} (AA) code for DM-AIC$(t,q)$ if it has block-length $n$, message rate at least $r$, authentication rate at least $\alpha,$ key consumption rate at most $\kappa$, and probability of message error less than $\epsilon$.
\end{define}
With the measures of a code defined, for a fixed number of rounds $j$, we wish to know the maximum values of the message rate ($r$), authentication rate $(\alpha)$, with the minimum key consumption rate $(\kappa)$ for error probabilities going to zero. 
Of course, probability of error being zero exactly restricts coding schemes too severely, hence, as is tradition, the authentication capacity region is defined as a limit point of the operational measures as the block length goes to infinity. 
\begin{define}
A triple $(a,b,c)\in \mcf{C}_{\mathrm{AA}(j)}(t,q)$ if there exist a sequence of $(r_i,\alpha_i,\kappa_i,\epsilon_i,j,i)$-AA codes, for DM-AIC$(t,q)$ such that 
\[
\lim_{i \rightarrow \infty} \abs{ (r_i,\alpha_i,\kappa_i,\epsilon_i) - (a,b,c,0)} = 0.
\] 
The set $\mcf{C}_{\mathrm{AA}(j)}(t,q)$ is called the \emph{average authentication capacity region}.
\end{define}
\begin{remark}
This region is closed by definition. 
\end{remark}


\section{Prior methods}\label{sec:prev_work}

For the reader's convenience, we shall briefly describe the coding schemes of Lai et al.~\cite{lai2009authentication}, Simmons~\cite{Auth}, and Gungor and Koksal~\cite{gungor2016basic}. 
While our coding scheme is novel in the sense that it has not previously appeared, it does share a design philosophy with Lai et al. and with Simmons. 
These schemes separate in an intuitive way, with Lai et al.'s scheme exploiting the channel for authenticity, and Simmon's scheme exploiting only the secret key. 

Our results recover those of Lai et al.~\cite{lai2009authentication}, who only considered the problem with values of $\kappa$ which asymptotically vanish (i.e., the key consumption rate must go to zero). 
Our results also strictly improve on those of Gungor and Koksal~\cite{gungor2016basic}. 
Although to demonstrate this, first we need to fix an error in their paper as discussed in Appendix~\ref{app:gungordonescrewedup}.

\subsection{Lai's Strategy} \label{sec:back:lai}

Lai et al.~\cite{lai2009authentication} used the strategy of Alice sending Bob the information about the secret key over the channel.
The idea behind this strategy is that when Bob decodes the transmission he will receive both the secret key and the message, if the secret key matches his own then the message is authentic since no one else could know the key. 
But, in order for this to be a viable strategy, the coding scheme has to ensure two properties.
First, Gr{\'i}ma can not be able to recover the secret key from his observations, since otherwise he could replace the message with one of his own choosing. 
Second, the message and the secret key cannot be separately decoded (i.e. the set of transmitted symbols can not be broken down into those corresponding to the secret key and those corresponding to the transmission) since otherwise Gr{\'i}ma would only need to replace the part of the transmission related to sending the message.

To accomplish both of these objectives, Lai et al.~\cite{lai2009authentication} used a modified wiretap coding scheme where, in particular, they first chose an integer $n$ and distribution $\rho \in \mcf{P}_n(\mcf{X})$ such that $$\begin{array}{lrl}&\mathbb{I}(t,\rho) - \mathbb{I}(q,\rho) &> 0, \\ &|\mcf{M}||\mcf{K}| &< 2^{n \mathbb{I}(t,\rho) }, \\ \text{ and } &|\mcf{K}| &< 2^{n \left[ \mathbb{I}(t,\rho) - \mathbb{I}(q,\rho)\right] }. \end{array}$$ 
Next, they randomly and independently selected approximately $2^{n \mathbb{I}(t,\rho) }$ codewords from the type set of $\mbcf{T}_{\rho}$.
These codewords were then placed into one of $2^{n \left[ \mathbb{I}(t,\rho) - \mathbb{I}(q,\rho)\right] }$ bins at random, giving approximately $2^{n\mathbb{I}(q,\rho)}$ codewords per bin.
Each of these bins were then associated with a particular key, and each codeword in the bin was assigned a message.
Because the capacity of the channel from Alice to Gr{\'i}ma was entirely exhausted sending the information about the message given the secret key, the secret key remained obscured from Gr{\'i}ma and yet still correlated with the message.

While we will use the same strategy as Lai (sending the secret key as a secure message) in the construction of our coding scheme, we will not use their coding scheme. 
In situations other than when $\kappa \rightarrow 0$, their coding scheme produces a poor trade off between the message rate, authentication rate, and key consumption rate.
Instead we shall use a general code for the \emph{discrete broadcast channel with confidential communications,} (DM-BCCC) which we describe in greater detail in Section~\ref{sec:back:bcc}. 
Immediately this code can be seen as at least as good a coding scheme, since it can recover theirs as a special case.
Our scheme improves upon their scheme as well, since the detection bounds remain unchanged except some channels now allow for message rates near $\max_{\rho \in \mcf{P}(\mcf{X})} \mathbb{I}(t,\rho)$ even when $\mathbb{I}(t,\rho) - \mathbb{I}(q,\rho) < 0$ for the $\rho$ which maximizes $\mathbb{I}(t,\rho).$

\subsection{Simmons' strategy}\label{sec:back:simmons}

Simmons~\cite{Auth} considered this problem where all links were noiseless.
As such, you would expect Simmons' strategy to not rely on a noisy channel. 
Simmons' strategy, specifically, was to associate each key $k \in \set{1,\dots,2^{n\kappa}}$ with an independently and randomly chosen subset $\mbcf{X}(k) \subset \mbcf{X}$ where 
\[
\abs{\mbcf{X}(k)} = 2^{- n \kappa/2}|\mbcf{X}| = |\mcf{M}|. 
\]
For each $m \in \mcf{M}$ and $k$, Alice chooses a unique $\mbf{x} \in \mbcf{X}(k)$ to represent the message. 
Hence, the message rate is $$n^{-1} \logt |\mcf{M}| = n^{-1}\logt \abs{\mbcf{\tilde X}} - \kappa/2 .$$ 

On the other hand, consider the scenario where Gr{\'i}ma observes $\mbf{x}$ and replaces it with $\mbf{x}'\neq \mbf{x}$. 
Having observed $\mbf{x}$, Gr{\'i}ma can narrow down the value of the key (since not all $\mbcf{X}(k)$ contain $\mbf{x}$) and use this information in the selection of $\mbf{x}'$. 
On average, there should be $\abs{\mcf{K}} ( \abs{\mbcf{X}(k)} /\abs{\mbcf{X}})^2 =  1$ value of $k$ such that $\mbcf{X}(k)$ contain both $\mbf{x}$ and $\mbf{x}'$. 
At the same time there will be on average $\abs{\mcf{K}} ( \abs{\mbcf{X}(k)} /\abs{\mbcf{X}}) = 2^{n \kappa/2}$ values of $k$ such that $\mbcf{X}(k)$ contains $\mbf{x}$. 
Hence, on average Gr{\'i}ma should only have a $2^{-n\kappa/2}$ chance of selecting a $\mbf{x}'$ which is actually valid for the given secret key.

\subsection{Gungor and Koksal's strategy}

Gungor and Koksal~\cite{gungor2016basic} established inner bounds for $\mcf{C}_{\mathrm{AA}(1)}(t,q)$ by using a two secret key coding scheme, where one of the secret keys is used via Lai's strategy and the second is used to obfuscate the first key.
In essence, one can view their scheme as relating a unique code (whose input is the message and first key) to each value of the second key. 
This helps to obfuscate the first values from Gr{\'i}ma since before decoding the message and first key value, he needs to decode the second to determine which code is being used.
On the other hand, Bob already has the second key and therefore does not need to decode it.

There is, though, one major problem with this scheme.
If the original key cannot already be made secret, it implies that the channel to Gr{\'i}ma is superior.
Using more bits of secret key to push the message to a point where Gr{\'i}ma cannot decode is playing into his strength, and as a result, unnecessarily leaks bits of the keys.
Instead, a composition of Lai's strategy and Simmons' strategy will improve on this region. 
For the remainder of this section, we derive the inner bound for $\mcf{C}_{\mathrm{AA}(1)}(t,q)$ that can be obtained from~\cite{gungor2016basic} (with a correction to their paper found here in Appendix~\ref{app:gungordonescrewedup}).

In our paper's language,\footnote{Indeed this is because our declaration of intrusion, $\mbf{!}$, is equivalent to an erasure in~\cite{gungor2016basic}. Thus, $\varepsilon_{f,\varphi}$ is less than the superposition of the undetected and erasure bounds (with our correction) from~\cite{gungor2016basic}. Next, our measures only require consideration of substitution attacks, since the probability of a successful impostor attack is always less than the probability of a successful substitution attack.}~\cite[Theorem~1]{gungor2016basic} shows that to every $\rho \in \mcf{P}(\mcf{Y}|\mcf{U})$, $\tau \in \mcf{P}(\mcf{U})$, positive real numbers $\kappa_1, \kappa_2$, and large enough $n$, there exists an $(r,\alpha,\kappa_1 + \kappa_2, \epsilon,1,n)$-AA code 
where
\begin{align*}
\epsilon &\leq 2^{-n\left| \mathbb{I}(t\rho,\tau) - r - \kappa_1 \right|^+ + O(\logt n) } \\
\alpha & = \inf_{\nu \in \mcf{P}(\mcf{Y}|\mcf{U}) } \mathbb{D}(\nu||q\rho|\tau) + \min( \kappa_1 , \left| r + \kappa_1 + \kappa_2 - \mathbb{I}(\nu,\tau) \right|^+ ))  + O(n^{-1}\logt n).
\end{align*}
In turn, then, the subset of $\mcf{C}_{\text{AA}(1)}(t,q)$ that can be derived from their scheme is a subset of 
\begin{equation}\label{eq:reg_gungor}
\bigcup_{\tilde \kappa \in \mathbb{R}_+ } \mcf{R}_{G}(\tilde \kappa)  
\end{equation}
where $\mcf{R}_{G}(\tilde \kappa) $ is the set of all non-negative triples $(r,\alpha,\kappa)$ such that 
\begin{equation}
 \begin{matrix*}[l] 
 r+ \kappa &\leq \mathbb{I}(t\rho,\tau) + \tilde \kappa  \\
\alpha - \kappa &\leq - \tilde \kappa \\
\alpha &\leq \inf_{\nu \in \mcf{P}(\mcf{Y}|\mcf{U}) } \mathbb{D}(\nu||q\rho|\tau) + \left|  \tilde \kappa + \mathbb{I}(t\rho,\tau) - \mathbb{I}(\nu,\tau) \right|^+ \,.
\end{matrix*}
\end{equation}
This follows by observing that all $(r,\alpha,\kappa)$ such that 
\begin{align}
r   &\leq \mathbb{I}(t\rho,\tau)  - \kappa_1 \label{eq:gk1}\\
\kappa&=\kappa_1 + \kappa_2 \label{eq:gk2}\\
\alpha&\leq \kappa_1 \label{eq:gk3}\\
\alpha &\leq \inf_{\nu \in \mcf{P}(\mcf{Y}|\mcf{U}) } \mathbb{D}(\nu||q\rho|\tau) + \left|  r + \kappa_1 + \kappa_2  - \mathbb{I}(\nu,\tau) \right|^+ \label{eq:gk4}
\end{align}
cause the upper bound on $\epsilon$ to go to $0$. 
Furthermore, $(r,\alpha,\kappa)\in \mcf{R}_{G}(\tilde \kappa)$ for all $(r,\alpha,\kappa)$ that satisfy Equations~\eqref{eq:gk1}--\eqref{eq:gk4} which comes from replacing Equation~\eqref{eq:gk4} with the looser (due to Equation~\eqref{eq:gk1}) requirement that
\[
\alpha \leq \inf_{\nu \in \mcf{P}(\mcf{Y}|\mcf{U}) } \mathbb{D}(\nu||q\rho|\tau) + \left|  \kappa_2 + \mathbb{I}(t\rho,\tau)  - \mathbb{I}(\nu,\tau) \right|^+   ,
\]
and then applying Fourier-Motzkin elimination to remove $\kappa_1$.

\subsection{Broadcast channel with confidential communications} \label{sec:back:bcc}

 Optimal codes for the \emph{discrete memoryless broadcast channel with confidential communications} (DM-BCC$(t,q)$) channel were first presented and later improved by Csisz{\'a}r and K{\"o}rner~\cite{csiszar1978broadcast},~\cite[Chapter~17]{CK}. 
 We describe what appears in~\cite[Chapter~17]{CK}. 
 In a DM-BCCC$(t,q)$, Bob's observation given Alice's transmission, $\mbf{Y}|\mbf{X}$, is distributed $\mbf{t}(\mbf{y}|\mbf{x})$, while Gr{\'i}ma's observation give Alice's transmission, $\mbf{Z}|\mbf{X}$, is distributed as $\mbf{q}(\mbf{z}|\mbf{x})$, just as they are in the DM-AIC$(t,q)$. 
 In the DM-BCCC, though, Alice is attempting to send three messages.
 First, a message $M_0$, where $\mcf{M}_0 \defn \set{1,\dots,2^{nr_0}}$, which needs to be reliably decoded by both Bob and Gr{\'i}ma. 
 Second, a message $M_s$, where $\mcf{M}_s \defn \set{1, \dots, 2^{nr_s}}$, which needs to be reliably decoded by Bob, but kept secret from Gr{\'i}ma in the sense that
\[
\mathbb{I}( p_{Z^n|M_s},p_{M_s} ) \leq \epsilon.
\]
Finally, a third message $M_1$, where $\mcf{M}_1\defn \set{1,\dots,2^{nr_1}}$, which only needs to be reliably decoded by Bob, and has no secrecy constraint placed on Gr{\'i}ma's observation.

For our coding scheme, we shall employ codes optimal for a DM-BCC$(t,q)$ for use in a DM-AIC$(t,q)$. 
Although these codes are more general than those used by Lai et al., the way in which they will produce a positive measure of authentication is the same. 
Namely, we shall send the secret key using the secure message.
On the other hand, by using $M_0$ and $M_1$ strategically, this coding scheme will allow for higher data rates to Bob by sacrificing the leakage of inconsequential information to Gr{\'i}ma. 

Consider this, by sending the secret key as the secure message, every received sequence at Bob can correspond to at most one secret key. 
Thus, the maximum probability that Gr{\'i}ma is successful is equal to the probability of the most likely key given his own observations.
Now, one might suspect, then, that as long as the code could remain strongly secure it would guarantee that the Gr{\'i}ma would gain negligible information about the key, and thus deliver an authentication rate of $j\kappa$. 
After all, strong security by its definition is a measure of how close to the uniform distribution Gr{\'i}ma's key likelihoods are.
Alas, this is not always the case, as there are rare outlier cases which contribute negligible amounts to the leakage but dominate the authentication rate.
Because of this, optimizing codes around the authentication rate leads to different codes than optimizing around leakage. 
\section{Main Contributions} \label{sec:main_cont}

For our first major contribution, we employ Lai's strategy with DM-BCCC codes to obtain the following inner bound for $\mcf{C}_{\mathrm{AA}(j)}(t,q)$.
\begin{theorem}\label{thm:lai_strong}
If non-negative real numbers $(r,\alpha,\kappa)$ satisfy 
\begin{equation}\label{eq:reg_lai:s}
\begin{matrix*}[l]
r + \alpha &\leq \mathbb{I}(t\rho,\sigma\tau) \\
\alpha & \leq    \mathbb{L}( t,q|\rho,\sigma,\tau) \\
\alpha & \leq  \mathbb{I}(t\rho,\sigma|\tau) \\
\alpha -   j\kappa &\leq 0 
\end{matrix*} ,
\end{equation}
where 
\begin{align*}
\mathbb{L}(t,q|\rho,\sigma,\tau) &\defn  \min_{\nu \in \mcf{P}(\mcf{Z}|\mcf{U})} \mathbb{F}(\nu||q,\rho|\sigma\tau) +  \left| \mathbb{S}(t\rho,\nu|\sigma,\tau) \right|^+, \\
\mathbb{F}( \nu || q,\rho|\sigma) &\defn  \min_{\substack{\zeta \in \mcf{P}(\mcf{Z}|\mcf{X},\mcf{U}) : \\ \sum_{x\in \mcf{X}}\zeta(z|x,u)\rho(x|u) = \nu(z|u)  }} \mathbb{D}(\zeta || q  | \rho\times\sigma),  \\ 
\mathbb{S}(\mu,\nu|\sigma,\tau) &\defn  \mathbb{I}(\mu,\sigma|\tau)  - \mathbb{I}(\nu,\sigma|\tau) + \left| \mathbb{I}(\mu\sigma,\tau)  - \mathbb{I}(\nu\sigma,\tau) \right|^+ ,
\end{align*}
for some $\rho \in \mcf{P}(\mcf{X}|\mcf{U})$, $\sigma \in \mcf{P}(\mcf{U}\gg\mcf{W})$, and $\tau \in \mcf{P}(\mcf{W})$ with $\abs{\mcf{U}}$ and $\abs{\mcf{W}}$ finite, then $(r,\alpha,\kappa) \in \mcf{C}_{\mathrm{AA}(j)}(t,q).$
\end{theorem}
The proof of Theorem~\ref{thm:lai_strong} is contained in Appendix~\ref{sec:p_lai_strong}.
In order to prove the theorem, a unique code must be created for each blocklength, but not for each round.
The limit point of these operational parameters is then taken to establish the theorem.

This result extends those of Lai et al. in~\cite{lai2009authentication} to the case where the key consumption rate is allowed to scale with the blocklength. 
Specifically, the authentication rate can equal the key consumption rate as long as $j \kappa$ is smaller than some threshold. 
When moving beyond that threshold, though, this scheme does not allow for further increases in the authentication rate. 

Another similar aspect to~\cite{lai2009authentication} is that number of communication rounds matters relatively little.
In fact, the only major impact that the number of rounds has on the rate region is allowing the key consumption rate to be smaller and therefore more efficient. 
Intuitively, this makes sense, as the secret key is being kept secret during each round of communication, and thus can be used in later rounds. 
This also implies the decrease in key consumption rate, since the key consumption rate is normalized by the total number of symbols transmitted over each round.

Now, observe from Theorem~\ref{thm:lai_strong} that the authentication rate and information rate share the channel's finite resources. 
Additionally, the authentication rate is dependent on the secrecy of the empirical channel. 
Indeed, observe the second restriction which requires
\begin{align*}
\alpha  \leq \min_{\nu \in \mcf{P}(\mcf{Z}|\mcf{U})} \mathbb{F}(\nu||q,\rho|\sigma\tau) +  \left| \mathbb{S}(t\rho,\nu|\sigma,\tau) \right|^+.
\end{align*}
In this bound, the $\nu$ term can be understood as a communications channel between Alice and Gr{\'i}ma while $t\rho$ is the equivalent channel between Alice and Bob. 
With this in mind, $\mathbb{S}(t\rho,\nu|\sigma,\tau)$ is the amount of secret bits that $t\rho$ provides when the empirical channel to Gr{\'i}ma is $\nu$.
On the other hand, $\mathbb{F}(\nu||q,\rho|\sigma\tau) $ is a penalty representing how unlikely channel $\nu$ is to occur for Gr{\'i}ma given his actual channel is $q$ and all code words are chosen according to a distribution of $\rho$.
Thus, the upper bound on the authentication rate is determined by the channel with the minimum combination of penalty and secrecy.

While Theorem~\ref{thm:lai_strong} makes explicit use of the channel to hide information, it is also possible (as shown by Simmons~\cite{Auth}) to send authenticated information even when there does not exist an advantage in communication channels.
In fact, Simmons' scheme (with a few modifications) is actually universally composable. 
Our next major contribution is a result of this, since being universally composable allows for us to start with any code and apply Simmons' scheme to obtain a new message rate, authentication rate, key consumption rate trade off.

\begin{theorem}\label{thm:-t2}
If $(r,\alpha, \kappa) \in \mcf{C}_{\mathrm{AA}(j)}(t,q)$ then $(r-\beta,\alpha+\beta,  \kappa+ [1+j^{-1}]\beta) \in \mcf{C}_{\mathrm{AA}(j)}(t,q)$, for all non-negative $\beta < r$.
\end{theorem}

The proof of Theorem~\ref{thm:-t2} can be found in Appendix~\ref{app:cc2}.

The major difficulty in establishing Theorem~\ref{thm:-t2} is that every round of communication requires a new encoder. 
Using the same encoder every round would in essence allow for Gr{\'i}ma to effectively send previous rounds' messages. 
It is important to note, here, that the key consumption rate actually decreases with the total number of rounds. 
This is because more and more key can be re-used throughout the transmissions, leading to an increase in efficiency.

Now, from Theorem~\ref{thm:lai_strong} and~\ref{thm:-t2}, it follows that all $(r,\alpha,\kappa)$ that satisfy 
\begin{align*}
\begin{array}{ll}
\hat r + \hat \alpha &\leq \mathbb{I}(t\rho,\sigma\tau)\\
\hat \alpha &\leq \mathbb{L}(t,q|\rho,\sigma,\tau) \\
\hat \alpha &\leq \mathbb{I}(t\rho,\sigma|\tau) \\
\hat \alpha  &\leq j \hat \kappa \\
r &\leq \hat r - \beta \\
\alpha &\leq \hat \alpha + \beta \\
\kappa &\geq \hat \kappa + [1+j^{-1}]\beta \\
0 &\leq \beta 
\end{array}
\end{align*}
are contained in $\mcf{C}_{\mathrm{AA}(j)}(t,q).$
Using Fourier-Motzkin elimination on the above region, specifically eliminating $\hat r,~\hat \alpha, ~\hat \kappa, $ and $\beta$, proves the following theorem. 
\begin{theorem}\label{thm:ib}
If non-negative real numbers $(r,\alpha,\kappa)$ satisfy 
\begin{equation}\label{eq:reg_lai:s2}
\begin{matrix*}[l]
r + \alpha &\leq \mathbb{I}(t\rho,\sigma\tau) \\
[1+j^{-1}]\alpha -   \kappa  & \leq  \mathbb{L}(t,q|\rho,\sigma,\tau)\\
[1+j^{-1}]\alpha -\kappa  & \leq  \mathbb{I}(t\rho,\sigma|\tau) \\
\alpha -j\kappa  & \leq  0 \, ,
\end{matrix*} 
\end{equation}
for some distributions $\rho \in \mcf{P}(\mcf{X}|\mcf{U})$, $\sigma \in \mcf{P}(\mcf{U}\gg\mcf{W})$, and $\tau \in \mcf{P}(\mcf{W})$, with $\abs{\mcf{U}}$ and $\abs{\mcf{W}}$ finite, then $(r,\alpha,\kappa) \in \mcf{C}_{\mathrm{AA}(j)}(t,q)$.
\end{theorem}

The second and third conditions of Theorem~\ref{thm:ib} are more easily understood as 
$$ \alpha - j^{-1} ( j \kappa - \alpha) \leq \min\left[\mathbb{L}(t,q|\rho,\sigma,\tau), \mathbb{I}(t\rho,\sigma|\tau) \right]  .$$
That is, unlike using Lai's strategy alone, the authentication rate can exceed $\min\left[\mathbb{L}(t,q|\rho,\sigma,\tau), \mathbb{I}(t\rho,\sigma|\tau) \right]$, but in order to do so, the key consumption rate $\kappa$ must be increased as well. 
In particular, 
$$\Delta_{\kappa} \defn \kappa - j^{-1} \min\left[\mathbb{L}(t,q|\rho,\sigma,\tau), \mathbb{I}(t\rho,\sigma|\tau) \right]  \geq \frac{j+1}{j} \Delta_{\alpha}, $$
for authentication rates $\alpha  = \Delta_{\alpha} +  \min\left[\mathbb{L}(t,q|\rho,\sigma,\tau), \mathbb{I}(t\rho,\sigma|\tau) \right]$ and $\Delta_{\alpha} > 0$.
Just as with Simmons' scheme, our scheme is more efficient in key usage when the total number of rounds increases\footnote{Although more rounds still requires a larger secret key, $2^{(j+1)n\Delta} 2^{n \min\left[\mathbb{L}(t,q|\rho,\sigma,\tau), \mathbb{I}(t\rho,\sigma|\tau) \right]} $ versus $2^{jn\Delta} 2^{n \min\left[\mathbb{L}(t,q|\rho,\sigma,\tau), \mathbb{I}(t\rho,\sigma|\tau) \right]}$ bits of secret key for $j$ versus $j-1$ rounds.}; the minimum increase in key consumption rate being $\Delta_{\kappa} = \Delta_{\alpha}$.
 
It is easy to wonder if Theorem~\ref{thm:ib} also constitutes an outer bound.
Unfortunately, the answer to this problem is not so easy, since the metric essentially requires bounding information rates of more than two channels.
That is, the authentication rate will always be dominated by the channel which minimizes the combination of empirical channel penalty and unleaked secret key information. 
Hence, when attempting a converse proof, the converse proof must consider all channel simultaneously in absence of a result showing that a particular channel is the worst case regardless of coding scheme.

\section{Examples} \label{sec:examples}
In order to demonstrate the trade-offs between the three parameters that make up the average authentication capacity region and show the improvement over Gungor and Koksal's region outer bound, we provide the following examples for one round of authentication.
For ease of understanding, we consider the case where both $t$ and $q$ are \emph{binary symmetric channels} (BSC). That is, if $t$ is a BSC with parameter $\lambda_t \in [0,1/2]$, then $\mcf{X} = \mcf{Y} = \set{0,1}$ and $t(0|1) = t(1|0) =  \lambda_t$. 
Additionally, we restrict the distributions $\zeta$, $\nu$, $\rho$, $\sigma$, and $\tau$ in Eq. \eqref{eq:reg_lai:s2} to also be binary and symmetric for purposes of computational simplicity.
We hypothesize that the optimal distributions are in fact in this form, but do not prove it here for sake of brevity.
If this does not hold, the plots can simply be interpreted as an additional inner bound to the average authentication capacity region for BSCs.

We provide three different annotated plots to illustrate different factors for the region.
This includes the simple message rate and authentication rate trade-off, the efficiency of consumed key material, and the effects of main channel quality including both the less noisy and more noisy regimes.
A second set of plots contains Gungor and Koksal's region outer bound for comparison with the previous three plots.

\begin{figure}
	\centering
	\includegraphics[scale=.5]{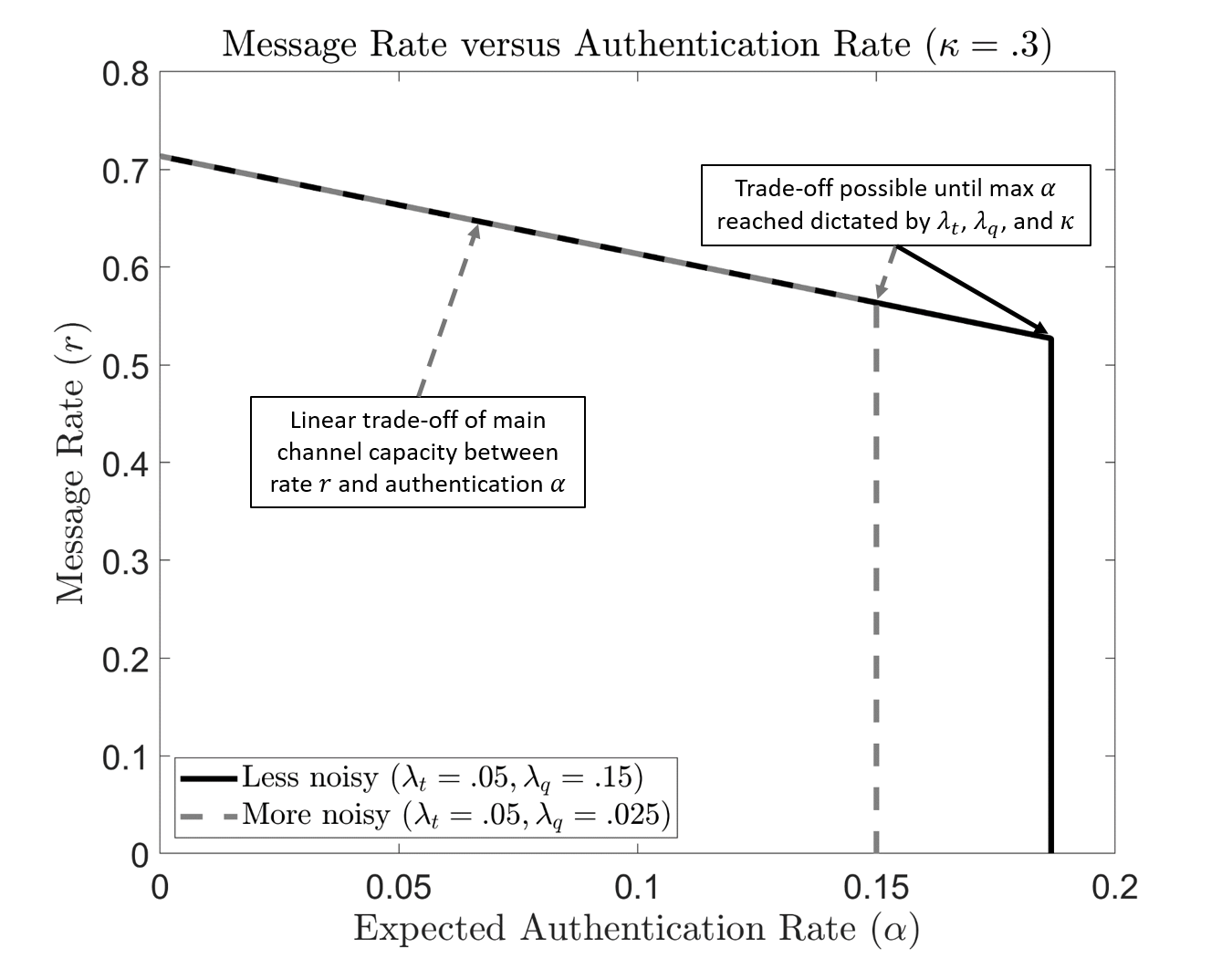}
	\caption{Message rate ($r$) vs. authentication rate ($\alpha$) for both less noisy and more noisy main channel for the inner bound of the average authentication capacity region (Eq. \eqref{eq:reg_lai:s2}).}
	\label{fig:RvsAlpha}
\end{figure}

An important consequence of the inner bound of the capacity region in both \eqref{eq:reg_lai:s} and \eqref{eq:reg_lai:s2} is that communication and authentication must share the main channel capacity.
This trade-off is depicted in Fig. \ref{fig:RvsAlpha} for both the less noisy and more noisy channel cases.
Thanks to the incorporation of Simmons' noiseless strategy in our code, authentication is still possible even when the main channel is more noisy (as opposed to Lai's region \eqref{eq:reg_lai:s}) albeit with a lower maximum possible $\alpha$.
It can be seen that the more noisy case is limited by half the key $\kappa/2$ as dictated by Simmons' strategy while the less noisy case is limited by the second condition in \eqref{eq:reg_lai:s2} which essentially represents the secrecy capacity of the channel pair in combination with Simmons'.
Having nonzero secrecy capacity clearly improves the region.

\begin{figure}
	\centering
	\includegraphics[scale=.5]{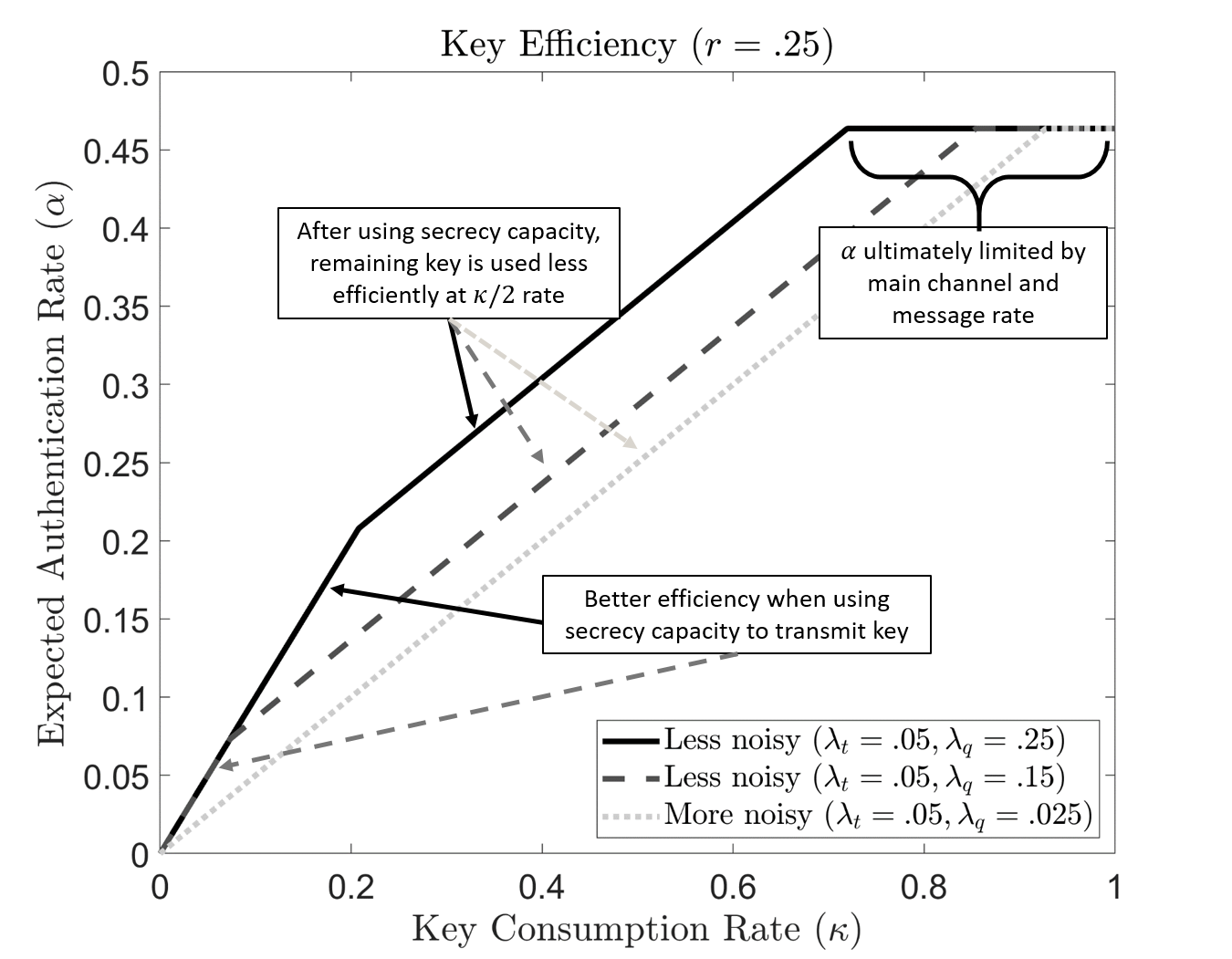}
	\caption{The amount of authentication rate gained per increase in key consumption rate is better when secrecy capacity is nonzero. Curves obtained from inner bound of the average authentication capacity region (Eq. \eqref{eq:reg_lai:s2}).}
	\label{fig:keyEfficiency}
\end{figure}

Next, we examine the efficiency of the key in terms of bits of key consumption to bits of authentication rate in different scenarios.
When secrecy capacity is available, as much key as possible should be sent using the secrecy provided by the channel since it is doubly as efficient as Simmons' strategy.
Fig. \ref{fig:keyEfficiency} demonstrates this effect.
As seen in the two less noisy cases, the initial gain in authentication is directly proportional to the amount of key used.
Once the secrecy capacity has been exhausted, however, the additional gain in authentication per key bit consumed is halved since the less efficient Simmons' scheme must be used.
In the more noisy main channel case, the trade-off is at a constant $1/2$ for all key consumption rates since no secrecy capacity is available.
In all cases, authentication is ultimately limited by the main channel capacity and desired message rate.

\begin{figure}
	\centering
	\includegraphics[scale=.5]{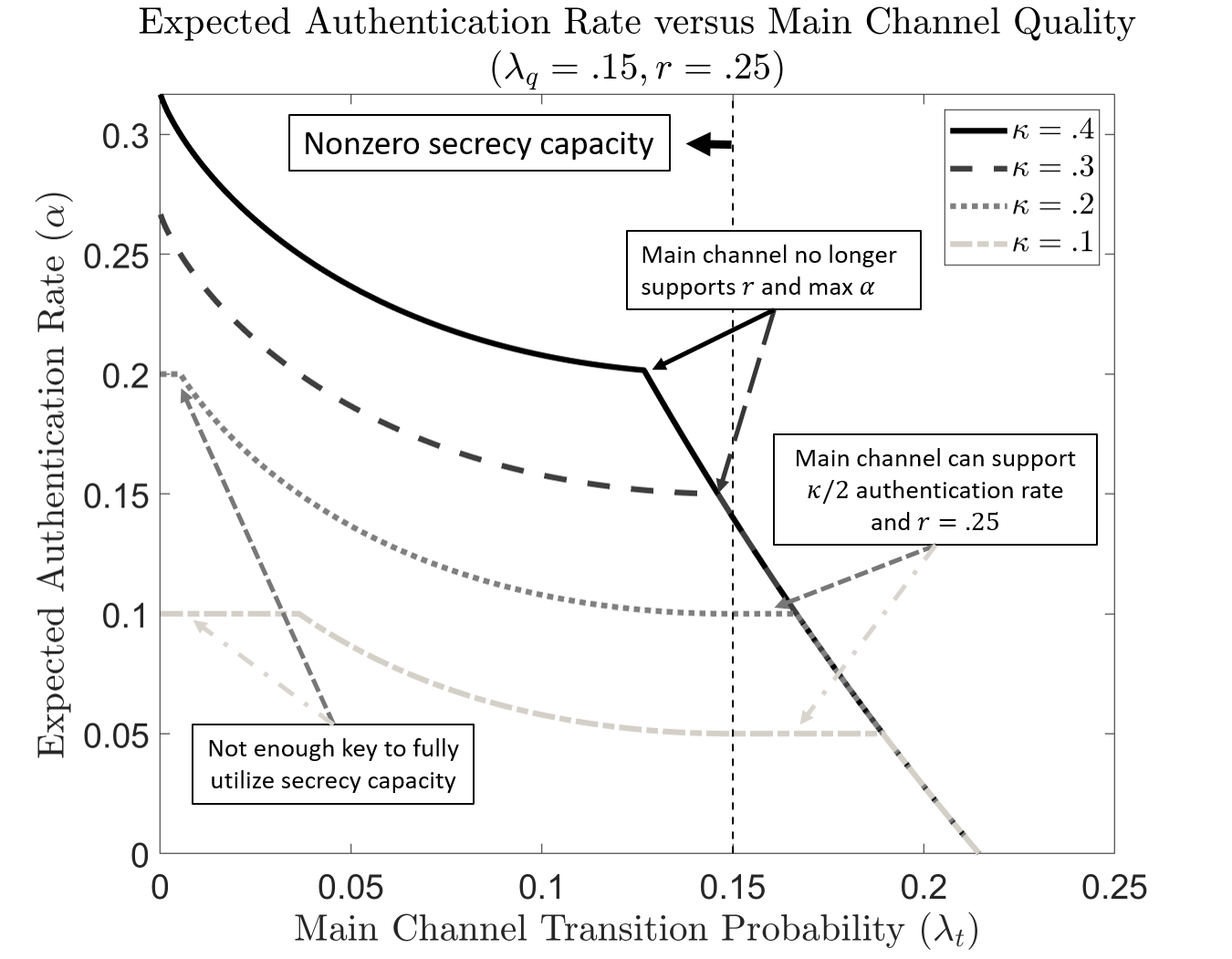}
	\caption{Authentication capabilities decrease with worsening channel conditions. Curves obtained from inner bound of the average authentication capacity region (Eq. \eqref{eq:reg_lai:s2}).}
	\label{fig:varyMainChannel}
\end{figure}

Next, in Fig. \ref{fig:varyMainChannel}, we show how the quality of the main channel directly affects the amount of authentication possible for different key consumption rates.
Naturally, as the main channel decreases in quality, lower authentication rates are obtainable.
For the cases of $\kappa=.2$ and $\kappa=.1$, the amount of secrecy capacity initially available exceeds the amount of key material possessed.
Thus, the channel can accommodate the entire key and $\alpha=\kappa$ until the channel worsens since authentication is always limited by the size of the key.
As we approach the nonzero secrecy capacity point, the authentication rate tends towards half the key consumption rate $\kappa/2$ as previously explained.
However, in the cases of $\kappa=.4$ and $\kappa=.3$, by the time this happens, the main channel capacity cannot support both the desired message rate of $r=.25$ and an authentication rate of $\kappa/2$, so $\alpha$ falls steeply.
Eventually, all four cases converge due to the falling main channel quality and the inability to sustain the desired message rate and authentication rate.

\begin{figure}
	\centering
	\includegraphics[scale=.5]{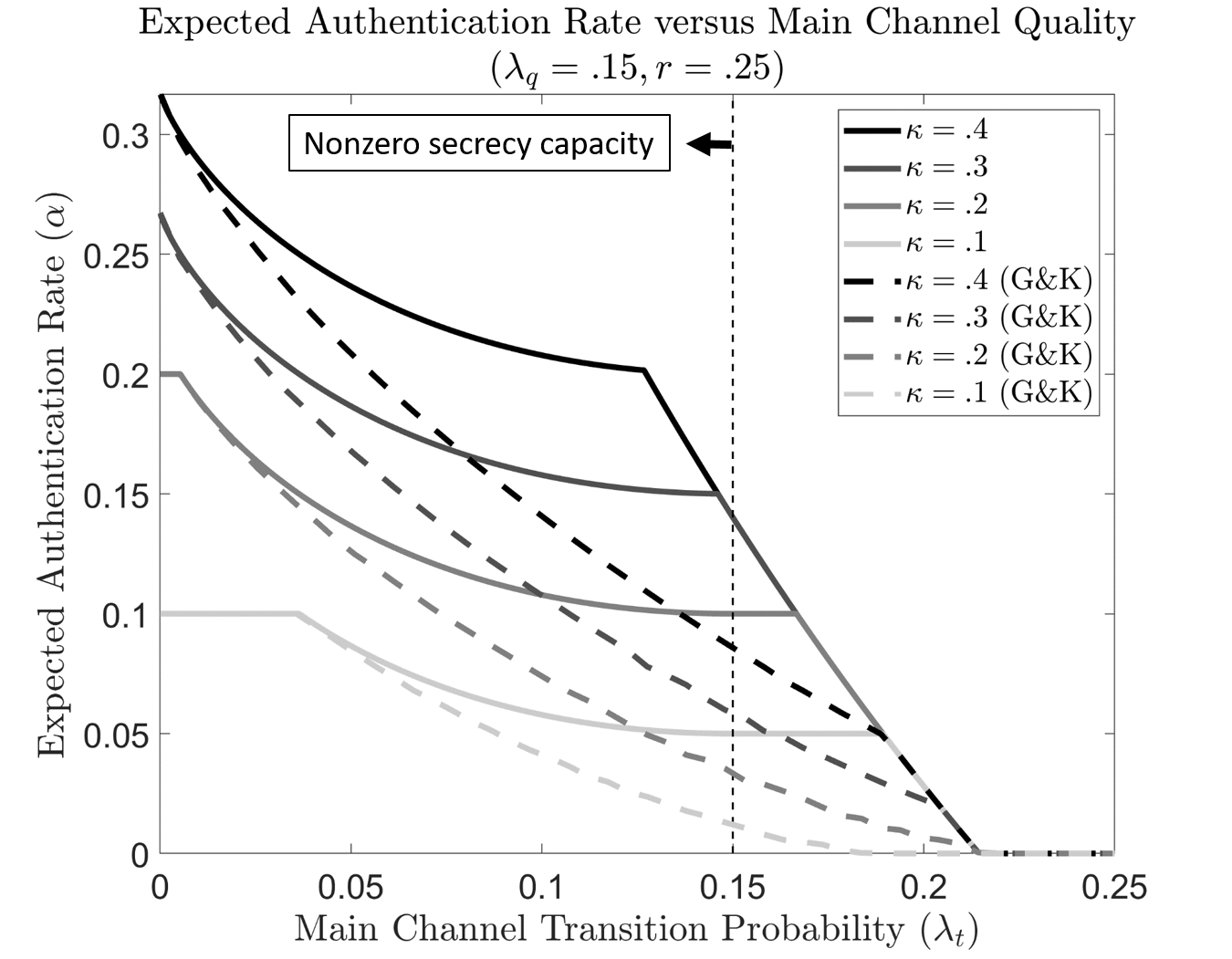}
	\caption{Authentication capabilities decrease with worsening channel conditions. Curves obtained from inner bound of the average authentication capacity region (Eq. \eqref{eq:reg_lai:s2}) and Gungor and Koksal's outer bound (Eq. \eqref{eq:reg_gungor}).}
	\label{fig:varyMainChannel_wGungor}
\end{figure}

Finally, we compare our inner bound of the average authentication capacity region to the outer bound of Gungor and Koksal's scheme in Eq. \eqref{eq:reg_gungor}.
The following figures (\ref{fig:RvsAlpha_wGungor}, \ref{fig:keyEfficiency_wGungor}, and \ref{fig:varyMainChannel_wGungor}) are the same as Figs. \ref{fig:RvsAlpha}, \ref{fig:keyEfficiency}, and \ref{fig:varyMainChannel}, but now with Gungor and Koksal's outer bound included.
The improvement is plainly seen in each case.

\begin{figure}
	\centering
	\includegraphics[scale=.5]{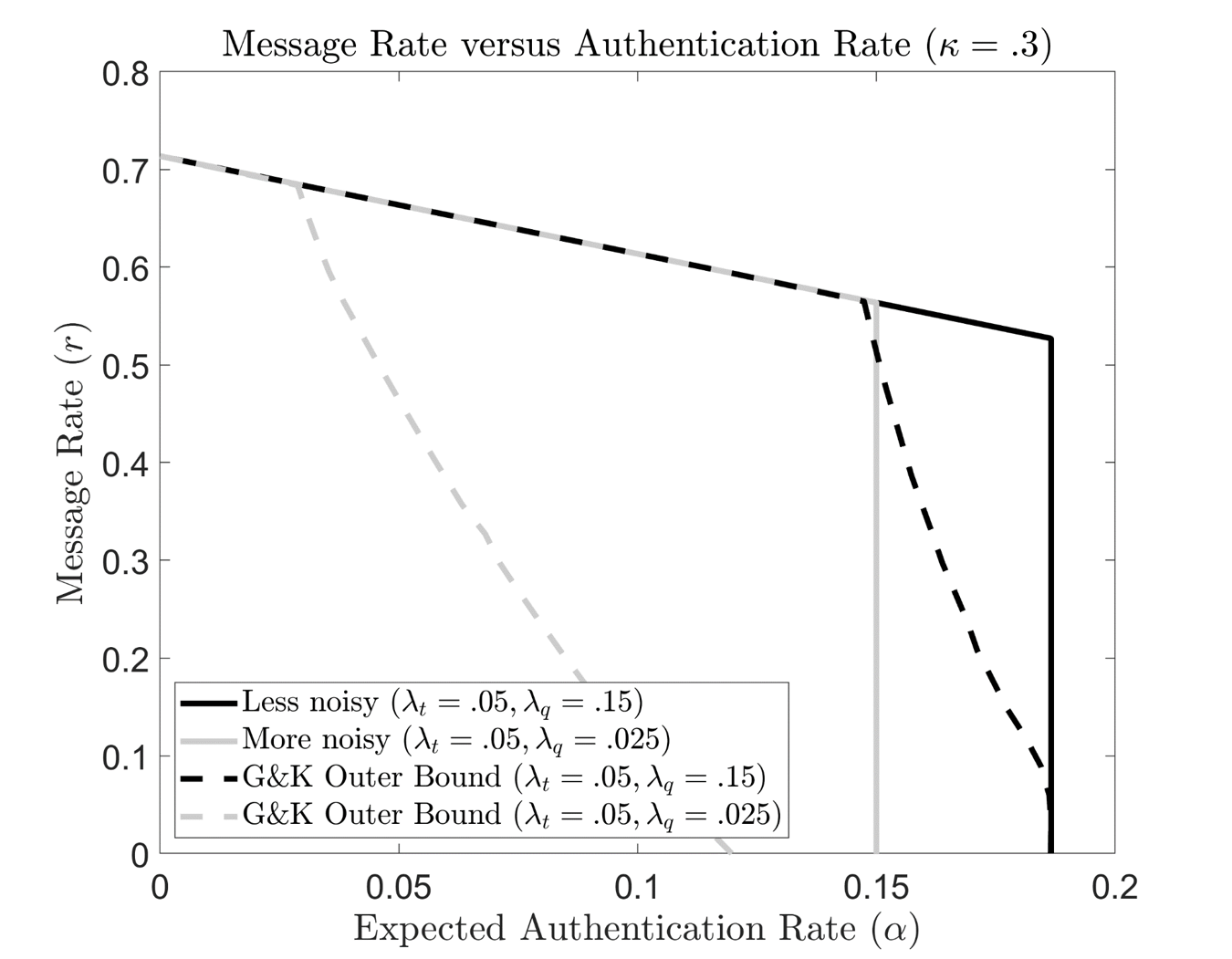}
	\caption{Message rate ($r$) vs. authentication rate ($\alpha$) for both less noisy and more noisy main channel for the inner bound of the average authentication capacity region (Eq. \eqref{eq:reg_lai:s2}) and Gungor and Koksal's outer bound (Eq. \eqref{eq:reg_gungor}).}
	\label{fig:RvsAlpha_wGungor}
\end{figure}

\begin{figure}
	\centering
	\includegraphics[scale=.5]{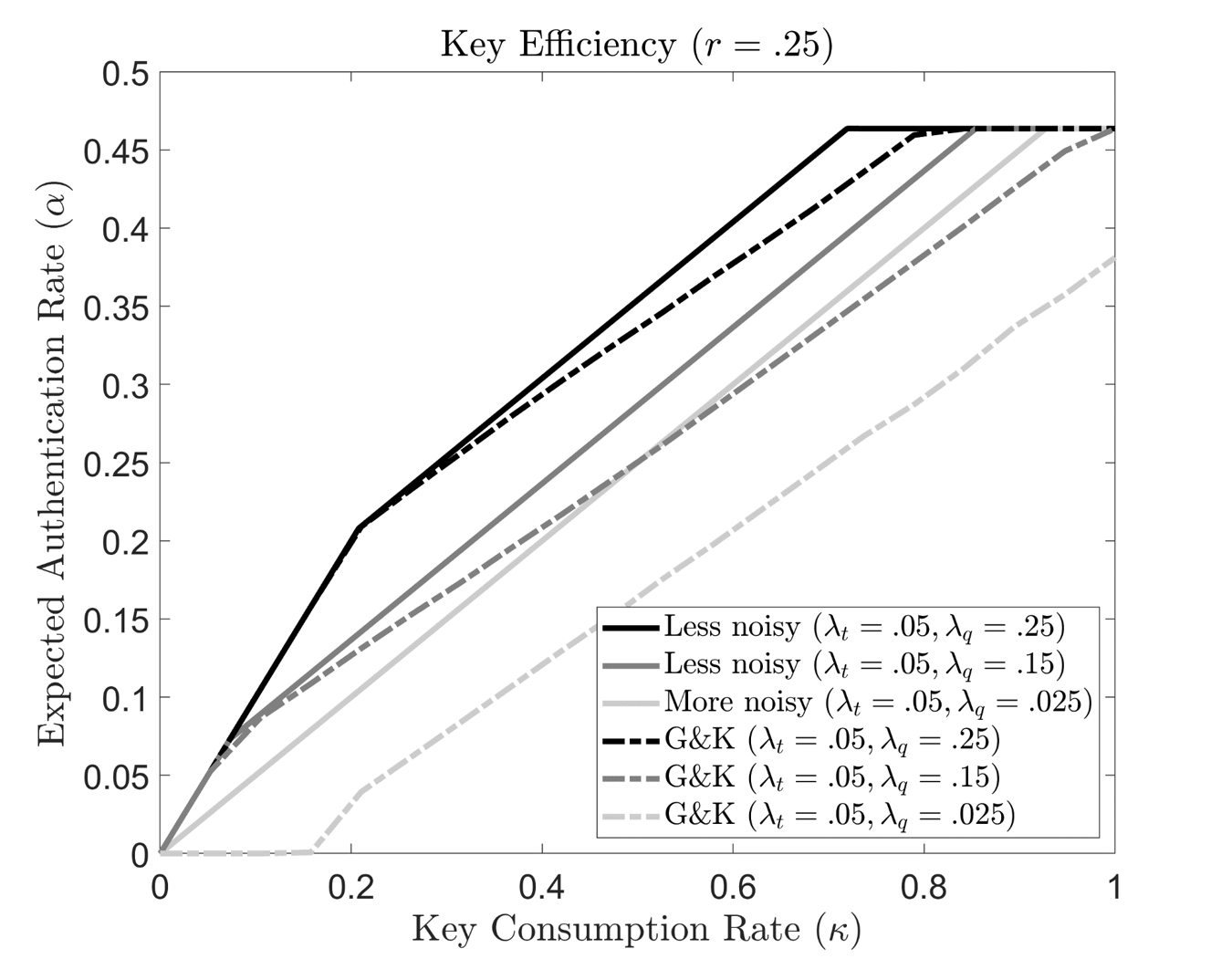}
	\caption{The amount of authentication rate gained per increase in key consumption rate is better when secrecy capacity is nonzero. Curves obtained from inner bound of the average authentication capacity region (Eq. \eqref{eq:reg_lai:s2}) and Gungor and Koksal's outer bound (Eq. \eqref{eq:reg_gungor}).}
	\label{fig:keyEfficiency_wGungor}
\end{figure}

\appendices

\section{Code construction primer}

The proofs of error bounds will rely heavily on the method of types. 
Specifically, we will require the following well known results (proofs of which can be found in~\cite[Chapter~2]{CK} or~\cite[Chapter~11]{CT}) for each $(\mbf{y},\mbf{x}) \in \mbcf{Y}\times \mbcf{X}$:
\begin{align}
- \frac{1}{n} \logt \mbf{t}(\mbf{y}|\mbf{x}) &=  \mathbb{H}(p_{\mbf{y}|\mbf{x}} |p_{\mbf{x}} ) + \mathbb{D}(p_{\mbf{y}|\mbf{x}} || t |p_{\mbf{x}} ) , \label{eq:app:lai:type:1}\\
- \frac{1}{n} \logt\abs{\mbcf{T}_{\mu}(\mbf{x})} & = \mathbb{H}(\mu|p_{\mbf{x}}) + O(n^{-1}\logt n) ,\label{eq:app:lai:type:2}\\
- \frac{1}{n} \logt \mbf{t}(\mbcf{T}_{\mu}(\mbf{x})|\mbf{x}) & =  \mathbb{D}( \mu || t |p_{\mbf{x}} ) + O(n^{-1}\logt n),\label{eq:app:lai:type:3}\\
\abs{\mcf{P}_n(\mcf{Y},\mcf{X})}  & = n^{O(1)}. \label{eq:app:lai:type:4}
\end{align}  
All orders are determined solely by the cardinalities of support sets. 
Equation~\eqref{eq:app:lai:type:4} will primarily be used to restrict summations to a particular type without requiring upper or lower bounds. 
In particular,
$$ \sum_{\mbf{x} \in \mbcf{X}} f(x)  = n^{O(1)}\max_{\nu \in \mcf{P}_n(\mcf{X})} \sum_{x \in \mbcf{T}_{\nu}} f(\mbf{x})   $$
for any function $f: \mbcf{X}  \rightarrow \mathbb{R}^+$ since
$$ \max_{\nu \in \mcf{P}_n(\mcf{X})} \sum_{x \in \mbcf{T}_{\nu}} f(\mbf{x}) \leq \sum_{\mbf{x} \in \mbcf{X}} f(x) \leq \sum_{\nu \in \mcf{P}_n(\mcf{X})} \max_{\nu \in \mcf{P}_n(X)} \sum_{x \in \mbcf{T}_{\nu}} f(\mbf{x})  . $$
In addition to these well known properties, we will need to develop a few lemmas, which extend the above concepts to more general cases.

The first of these lemmas will allow for us to chain together type class inclusions. 
This will help streamline analysis since, in general\footnote{Indeed, consider the case where $\mbf{w} = (0,0,1,1)$, $\mbf{u} = (1,0,1,0)$ and $\mbf{z} = (0,0,1,1)$. Clearly $p_{\mbf{z}|\mbf{w}} \neq p_{\mbf{z}|\mbf{u}} p_{\mbf{u}|\mbf{w}}$.}, $\mbf{z} \notin \mcf{T}_{\nu\sigma}(\mbf{w})$ when $\mbf{z} \in \mcf{T}_{\nu}(\mbf{u})$ and $\mbf{u} \in \mcf{T}_{\sigma}(\mbf{w})$. 

\begin{lemma}\label{lem:unique}
Let $\nu \in \mcf{P}(\mcf{Z}|\mcf{U})$, $\sigma \in \mcf{P}(\mcf{U} \gg \mcf{W})$ and $\tau \in \mcf{P}(\mcf{W})$. If $\mbf{z} \in \mbcf{T}_{\nu}(\mbf{u})$ and $\mbf{u} \in \mbcf{T}_{\sigma}(\mbf{w})$, then $\mbf{z} \in \mbcf{T}_{\nu\sigma}(\mbf{w})$. 
\end{lemma}
\begin{IEEEproof}
First, note that 
\[
\nu(c|b) = p_{\mbf{z}|\mbf{u}}(c|b) = \sum_{a} p_{\mbf{z}|\mbf{u},\mbf{w}}(c|b,a) p_{\mbf{w}|\mbf{u}}(a|b) = p_{\mbf{z}|\mbf{u},\mbf{w}}(c|b,a_b),
\]
where $a_b$ is the value such that $p_{\mbf{u}|\mbf{w}}(b|a_b)= \sigma(b|a_b)$ is non-zero. Hence,
\begin{align*}
p_{\mbf{z}|\mbf{w}}(c|a) &= \sum_{b}p_{\mbf{z}|\mbf{u},\mbf{w}}(c|b,a) p_{\mbf{u}|\mbf{w}}(b|a) \\
&= \sum_{b: a_b = b }p_{\mbf{z}|\mbf{u},\mbf{w}}(c|b,a) \sigma (b|a) + \sum_{b: a_b \neq b }p_{\mbf{z}|\mbf{u},\mbf{w}}(c|b,a) \sigma(b|a) \\
&= \sum_{b: a_b = b } \nu(c|b)\sigma(b|a) \\
&= \nu\sigma(c|a).
\end{align*}
\end{IEEEproof}

\begin{cor}\label{cor:unique}
Furthermore
$$\set{\mbf{\tilde u} : \mbf{z} \in \mbcf{T}_{\mu}(\mbf{\tilde u}),~\mbf{\tilde u} \in \mbcf{T}_{\sigma}(\mbf{w})} = \mbcf{T}_{\bar \mu}(\mbf{z},\mbf{w}) ,$$
where $\bar \mu \in \mcf{P}_n(\mcf{Z}|\mcf{U},\mcf{W})$ and more specifically $\bar \mu \times  \mu\sigma = \mu\times  \sigma$.
\end{cor}
\begin{IEEEproof}
We will complete the corollary in two steps, first showing $\mbcf{U}^* \defn  \set{\mbf{\tilde u} : \mbf{z} \in \mbcf{T}_{\mu}(\mbf{\tilde u}),~\mbf{\tilde u} \in \mbcf{T}_{\sigma}(\mbf{w})} \subseteq  \mbcf{T}_{\bar \mu}(\mbf{z},\mbf{w})$, and then $\mbcf{T}_{\bar \mu}(\mbf{z},\mbf{w}) \subseteq \mbcf{U}^* $.

To show $\mbcf{U}^* \subseteq  \mbcf{T}_{\bar \mu}(\mbf{z},\mbf{w})$, we need that $p_{\mbf{\tilde u}|\mbf{z},\mbf{w}} = \bar \mu$ for all $\mbf{\tilde u} \in \mbcf{U}^*$.
Towards this goal, for any $\mbf{\tilde u} \in \mbcf{U}^*$, consider the following two expansions of $p_{\mbf{z},\mbf{\tilde u}|\mbf{w}}$. 
First,
$$p_{\mbf{z},\mbf{\tilde u}|\mbf{w}} =  p_{\mbf{\tilde u}|\mbf{z},\mbf{w}} \times p_{\mbf{z}|\mbf{w}} =  p_{\mbf{\tilde u}|\mbf{z},\mbf{w}} \times \mu\sigma $$
since $\mbf{z} \in  \mbcf{T}_{\mu}(\mbf{u})$ and $\mbf{u} \in \mbcf{T}_{\sigma}(\mbf{w})$ implies $\mbf{z} \in \mbcf{T}_{\mu\sigma}(\mbf{w})$ by Lemma~\ref{lem:unique}.
Second,
$$p_{\mbf{z},\mbf{\tilde u}|\mbf{w}} =  p_{\mbf{z}|\mbf{\tilde u},\mbf{w}} \times p_{\mbf{\tilde u}|\mbf{w}} = p_{\mbf{z}|\mbf{\tilde u}} \times p_{\mbf{\tilde u}|\mbf{w}} = \mu  \times \sigma $$
since $\mbf{\tilde u}$ determines $\mbf{w}.$
Solving for $p_{\mbf{\tilde u}|\mbf{z},\mbf{w}}$ shows that $\bar \mu \times  \mu\sigma = \mu\times  \sigma$.

Next, to show $\mbcf{T}_{\bar \mu}(\mbf{z},\mbf{w}) \subseteq \mbcf{U}^*$, we need to show that $\mbf{\tilde u} \in \mbcf{T}_{\bar \mu}(\mbf{z},\mbf{w})$ implies that $p_{\mbf{\tilde u}|\mbf{w}} = \sigma$ and $p_{\mbf{z}|\mbf{\tilde u}} = \mu$. 
To this end, for all $\mbf{\tilde u} \in \mbcf{T}_{\bar \mu}(\mbf{z},\mbf{w})$, it must follow that 
$$ p_{\mbf{\tilde u}|\mbf{w}} = p_{\mbf{\tilde u}|\mbf{z},\mbf{w}}p_{\mbf{z}|\mbf{w}} = \sum_{z} \bar \mu(u|z,w) \mu\sigma(z|w)  = \sum_{z}  \mu(z|u) \sigma(u|w) = \sigma .$$
Note that this also proves that $\tilde u \in \mbcf{T}_{\bar \mu}(\mbf{z},\mbf{w})$ for at most one $\mbf{w}$, hence
$$p_{\mbf{z}|\mbf{\tilde u}} \times \sigma=  p_{\mbf{z}|\mbf{\tilde u},\mbf{w}} \times \sigma = p_{\mbf{z}|\mbf{\tilde u},\mbf{w}} \times p_{\mbf{\tilde u}|\mbf{w}} = p_{\mbf{\tilde u},\mbf{z}|\mbf{w}} = p_{\mbf{\tilde u}|\mbf{z},\mbf{w}} \times p_{\mbf{z}|\mbf{w}} =  \bar \mu \times \mu\sigma = \mu \times \sigma .$$

\end{IEEEproof}

The next lemma is needed because we will be using stochastic encoders which select an element at random from a type class, instead of selecting symbols independently for each of the $n$ channel inputs.
This will allow us to only penalize when the channel is bad since non-typical outputs from the stochastic encoder are eliminated. 
\begin{lemma} \label{lem:btx}
Given finite sets $\mcf{U}$, $\mcf{Y}$ and $\mcf{X}$: 
\begin{align}
&-\frac{1}{n} \logt \left(  \sum_{\mbf{x} \in \mbcf{T}_{q}(\mbf{u})}  |\mbcf{T}_{q}(\mbf{u})|^{-1} \mbf{t}(\mbf{y}|\mbf{x})\right)  = \mathbb{H}(p_{\mbf{y}|\mbf{u}}|p_{\mbf{u}}) +  \mathbb{F}_n(p_{\mbf{y}|\mbf{u}}||t, q|p_{\mbf{u}}) +  O(n^{-1} \logt n),
\end{align}
for each $\mbf{y} \in \mbcf{Y}$, $\mbf{u} \in \mbcf{U}$, and $q \in \mcf{P}_n(\mcf{X}|\mcf{U}; p_{\mbf{u}} )$.
\end{lemma}
\begin{IEEEproof}

First observe the following:
\begin{align}
&\sum_{\mbf{x} \in \mbcf{T}_{q}(\mbf{u})} |\mbcf{T}_{q}(\mbf{u})|^{-1} \mbf{t}(\mbf{y}|\mbf{x}) \notag \\
&= \sum_{\substack{p_{Y,U,X} \in \mcf{P}_n(\mcf{Y},\mcf{U},\mcf{X}): \\ \subalign{  p_{Y|U}  &= p_{\mbf{y}|\mbf{u}} ,\\ p_{X|U} &= q , \\ p_{U} &= p_{\mbf{u}} }}  }  \hspace{4pt} \sum_{\mbf{x} \in \mbcf{T}_{X|U,Y}(\mbf{u},\mbf{y})}  |\mbcf{T}_{q}(\mbf{u})|^{-1} \mbf{t}(\mbf{y}|\mbf{x}) \label{eq:btx:1}\\
&= \sum_{\substack{p_{Y,U,X} \in \mcf{P}_n(\mcf{Y},\mcf{U},\mcf{X}): \\ \subalign{  p_{Y|U}  &= p_{\mbf{y}|\mbf{u}} ,\\ p_{X|U} &= q , \\ p_{U} &= p_{\mbf{u}} }}  }  \hspace{4pt} \sum_{\mbf{x} \in \mbcf{T}_{X|U,Y}(\mbf{u},\mbf{y})}  2^{-n \left[ \mathbb{H}(q |p_{\mbf{u}}) + \mathbb{H}(p_{Y|X}|p_{X}) + \mathbb{D}(p_{Y|X}||t|p_{X}) \right]+ O(\logt n)} \label{eq:btx:2}\\
&=\sum_{\substack{p_{Y,U,X} \in \mcf{P}_n(\mcf{Y},\mcf{U},\mcf{X}): \\ \subalign{  p_{Y|U}  &= p_{\mbf{y}|\mbf{u}} ,\\ p_{X|U} &= q , \\ p_{U} &= p_{\mbf{u}} }}  }  2^{-n \left[- \mathbb{H}(p_{X|U,Y}|p_{U,Y})+ \mathbb{H}(p_{X|U} |p_{U }) + \mathbb{H}(p_{Y|X}|p_{X}) + \mathbb{D}(p_{Y|X}||t|p_{X}) \right]+ O(\logt n)} \label{eq:btx:3} \\
&= \max_{\substack{p_{Y,U,X} \in \mcf{P}_n(\mcf{Y},\mcf{U},\mcf{X}): \\ \subalign{  p_{Y|U}  &= p_{\mbf{y}|\mbf{u}} ,\\ p_{X|U} &= q , \\ p_{U} &= p_{\mbf{u}} }}  }  2^{-n \left[- \mathbb{H}(p_{X|U,Y}|p_{U,Y})+ \mathbb{H}(p_{X|U} |p_{U}) + \mathbb{H}(p_{Y|X}|p_{X}) + \mathbb{D}(p_{Y|X}||t|p_{X}) \right]+ O(\logt n)} ;\label{eq:btx:4}
\end{align}
where~\eqref{eq:btx:1} is because $\mbcf{T}_{q}(\mbf{u}) = \cup_{\subalign{p_{Y,U,X} &\in \mcf{P}_n(\mcf{Y},\mcf{U},\mcf{X}): \\  p_{Y|U}  &= p_{\mbf{y}|\mbf{u}},\\   p_{X|U} &= q, \\ p_{U} &= p_{\mbf{u}}  }  } \mbcf{T}_{p_{X|U,Y}}(\mbf{u},\mbf{y})$ and type classes have no intersection;~\eqref{eq:btx:2} is because of Equations~\eqref{eq:app:lai:type:1} and~\eqref{eq:app:lai:type:2};~\eqref{eq:btx:3} is by Equation~\eqref{eq:app:lai:type:2} and because the summands do not depend on the value of $\mbf{x}$; and~\eqref{eq:btx:4} is by Equation~\eqref{eq:btx:4}.

The lemma result now follows since 
\begin{align}
&- \mathbb{H}(p_{X|U,Y}|p_{U,Y})+ \mathbb{H}(p_{X|U} |p_{U}) + \mathbb{H}(p_{Y|X}|p_{X}) \notag\\
&\hspace{10pt}= \mathbb{H}(p_{Y|U}|p_{U}) + \mathbb{I}(p_{Y|U,X},p_{U|X}|p_{X}),
\end{align}
and
\begin{align}
&\min_{\substack{p_{Y,U,X} \in \mcf{P}_n(\mcf{Y},\mcf{U},\mcf{X}): \\ \subalign{ p_{Y|U}  &= p_{\mbf{y}|\mbf{u}} ,\\ p_{X|U} &= q , \\ p_{U} &= p_{\mbf{u}} }}  }  \mathbb{I}(p_{Y|U,X},p_{U|X}|p_{X}) + \mathbb{D}(p_{Y|X}||t|p_{X})\notag \\
&= \min_{\substack{p_{Y,U,X} \in \mcf{P}_n(\mcf{Y},\mcf{U},\mcf{X}): \\ \subalign{ p_{Y|U}  &= p_{\mbf{y}|\mbf{u}} ,\\ p_{X|U} &= q , \\ p_{U} &= p_{\mbf{u}} }}  } \sum_{\subalign{y&\in \mcf{Y},\\ x &\in \mcf{X}, \\u &\in \mcf{U}}} p_{Y,X,U}(y,x,u) \logt \frac{ p_{Y,X|U}(y,x|u)}{p_{X|U}(x|u)t(y|x)} \\
&= \min_{\substack{\zeta \in \mcf{P}_n(\mcf{Y}|\mcf{X},\mcf{U};q \times p_{\mbf{u}}): \\ \zeta q = p_{\mbf{y}|\mbf{u}} }} \sum_{\subalign{y&\in \mcf{Y},\\ x &\in \mcf{X}, \\u &\in \mcf{U}}} \zeta(y|x,u) q(x|u) p_{\mbf{u}}(u)  \logt \frac{\zeta(y|x,u)}{t(y|x)} \\
&=\mathbb{F}_{n}(p_{\mbf{y}|\mbf{u}}||t,q|p_{\mbf{u}}).
\end{align}
\end{IEEEproof}

Type class arguments are especially useful for code construction since elements over a type class are generally equiprobable, thus allowing for analysis by basic counting arguments.  
Here is no different, we will want to know the probability of choosing at random a value of the type class that is typical with a particular observation. 
The following lemma is in this spirit, and is, in fact, a minor result from Csisz{\'a}r and K{\"o}rner~\cite[Lemma~10.1]{CK}.
\begin{lemma}\textbf{\cite[Minor~result~from~Lemma~10.1]{CK}}\label{lem:obvious}
Let $\sigma \in \mcf{P}_{n}(\mcf{U}\gg \mcf{W})$ and $\tau \in \mcf{P}_{n}(\mcf{W})$. 
If $\mbf{U}$ is uniformly distributed over $\mbcf{T}_{\sigma}(\mbf{w})$ then
\[
-\frac{1}{n} \logt \Pr \left( \mbf{y} \in \mbcf{T}_{\mu}(\mbf{U}) 
\right)  =  \mathbb{I}(\mu,\sigma|\tau) + O(\logt n)
\]
for all $\mbf{y} \in \mbcf{T}_{\mu\sigma}(\mbf{w})$ and $\mbf{w} \in \mbcf{T}_{\tau}$.
\end{lemma}
\begin{IEEEproof}
If $\mbf{U}$ is chosen uniformly over $\mbcf{T}_{\sigma}(\mbf{w})$, then 
\begin{equation}\label{eq:lem:obvious:1}
 \Pr \left( \mbf{y} \in \mbcf{T}_{\mu}(\mbf{U}) \right) = \sum_{\mbf{u}} p_{\mbf{U}}(\mbf{u}) \idc{\mbf{y} \in \mbcf{T}_{\mu}(\mbf{u}) } = \frac{\abs{\set{\mbf{u} : \mbf{y} \in \mbcf{T}_{\mu}(\mbf{u}),~\mbf{u} \in \mbcf{T}_{\sigma}(\mbf{w})}} }{\abs{\mbcf{T}_{\sigma}(\mbf{w})}}
\end{equation}
for all $\mbf{y}\in \mbcf{T}_{\mu\sigma}(\mbf{w})$.
But, $\set{\mbf{u} : \mbf{y} \in \mbcf{T}_{\mu}(\mbf{u}),~\mbf{u} \in \mbcf{T}_{\sigma}(\mbf{w})} = \mbcf{T}_{\bar \mu}(\mbf{y},\mbf{w}),$
where $\bar \mu \in \mcf{P}_n(\mcf{Y}|\mcf{U},\mcf{W})$ is such that $\bar \mu \times  \mu\sigma = \mu\times  \sigma$, by Corollary~\ref{cor:unique}.
Hence, 
\begin{equation}
-\frac{1}{n} \logt \Pr \left( \mbf{y} \in \mbcf{T}_{\mu}(\mbf{U}) \right) =    \mathbb{H}(\sigma|\tau) - \mathbb{H}(\bar \mu | \mu\sigma \times \tau) + O(n^{-1}\logt n)
\end{equation}
by~\eqref{eq:app:lai:type:2}.
The final result now follows since
\begin{align*}
&\mathbb{H}(\sigma|\tau) - \mathbb{H}(\bar \mu | \mu\sigma \times \tau)  \\
&= -  \sum_{\subalign{c&\in \mcf{Y},\\b&\in \mcf{U},\\a&\in \mcf{W}}}  \mu(c|b) \sigma(b|a) \tau(a) \logt \sigma(b|a) +  \sum_{\subalign{c&\in \mcf{Y},\\b&\in \mcf{U},\\a&\in \mcf{W}}} \bar \mu(b|a,c) \mu\sigma(c|a) \tau(a) \logt \bar \mu(b|a,c)  \\
&= -  \sum_{\subalign{c&\in \mcf{Y},\\b&\in \mcf{U},\\a&\in \mcf{W}}}  \mu(c|b) \sigma(b|a) \tau(a) \logt \sigma(b|a) +  \sum_{\subalign{c&\in \mcf{Y},\\b&\in \mcf{U},\\a&\in \mcf{W}}}  \mu(c|b) \sigma(b|a)  \tau(a) \logt \frac{ \mu(c|b) \sigma(b|a) }{\mu\sigma(c|a) }  \\
&= \sum_{\subalign{c&\in \mcf{Y},\\b&\in \mcf{U},\\a&\in \mcf{W}}} \mu(c|b) \sigma(b|a) \tau(a) \logt \frac{\mu(c|b)}{ \mu\sigma(c|a) } \\
&= \mathbb{I}(\mu,\sigma|\tau).
\end{align*}
\end{IEEEproof}

While Lemma~\ref{lem:obvious} determines the probability of randomly selecting a single value of $\mbf{u}$ which will be typical with $\mbf{y}$, it will also be necessary to consider the probabilities of selecting multiple values of $\mbf{u}$ where at least one is typical for each in a sequence of observations. 
Such a result will be necessary to look at the probabilities of codewords after multiple observations by an adversary.
\begin{lemma}\label{lem:circum}
Fix $\tau \in \mcf{P}_n(\mcf{W})$ and $\sigma \in \mcf{P}_n(\mcf{U} \gg \mcf{W})$, and let $\mcf{M}_i$ be a finite set such that $|\mcf{M}_i| \geq 8j\ln j$ for each $i\in \{1,\dots, j\}$, and suppose $\mbf{w}_m \in \mbcf{T}_{\tau}$ for each $m \in \cup_{i=1}^j \mcf{M}_i.$
Given independent RVs $U_{m}$ uniformly distributed over $\mbcf{T}_{\sigma}(\mbf{w}_m)$ for each $m \in \cup_{i=1}^j \mcf{M}_i$, then
\begin{align}
& \logt \Pr \left(\bigcap_{i=1}^j \left\{ \mbf{z}_i \in \bigcup_{m \in \mcf{M}_i}\mbcf{T}_{\nu_i}(\mbf{U}_{m}) \right\} \right) \notag\\
&\hspace{10pt} \geq  \sum_{i=1}^j \left| \logt (|\mcf{M}_i|) -  n\mathbb{I}(\nu_i,\sigma|\tau) \right|^- + O( j \logt nj )
\end{align}
for all $\bigtimes_{i=1}^j \nu_i \in \bigtimes_{i=1}^{j} \mcf{P}_{n}(\mcf{Z}|\mcf{U};\sigma\tau),$ and $\bigtimes_{i=1}^j \mbf{z}_i \in \bigtimes_{i=1}^j \mbcf{Z}$ such that $\mbf{z}_i \in \bigcap_{m \in \mcf{M}_i} \mbcf{T}_{\nu_i\sigma}(\mbf{w}_{m})$ for each $i \in \{1,\dots,j\}.$
\end{lemma}
\begin{IEEEproof}

The major obstacle in proving the lemma is that events $\mbf{z}_i \in \bigcup_{ m \in \mcf{M}_i}\mbcf{T}_{\nu_i}(\mbf{U}_m )$ and $\mbf{z}_j \in \bigcup_{ m \in \mcf{M}_j}\mbcf{T}_{\nu_j}(\mbf{U}_m )$ are correlated\footnote{Indeed, for instance if $\mbf{z}_i = \mbf{z}_j$, $\nu_i = \nu_j,$ and $\mcf{M}_i = \mcf{M}_j$, then clearly $\idc{ \mbf{z}_i \in \bigcup_{m \in \mcf{M}_{i}} \mbcf{T}_{\nu_i}(\mbf{U}_{m})} = 1$ if and only if $\idc{ \mbf{z}_j \in \bigcup_{m \in \mcf{M}_{j}} \mbcf{T}_{\nu_j}(\mbf{U}_{m})} = 1$.}.
If there exists sets (which shall be constructed later) $\mcf{M}^*_i\subseteq \mcf{M}_i$, for each $i \in \{1,\dots,j\}$, such that $|\mcf{M}^*_i| \geq (2j)^{-1} |\mcf{M}_i|$ and  $\mcf{M}^*_i \cap \mcf{M}^*_j = \emptyset$ for $i \neq j$, then the lemma can be obtained as follows:
\begin{align}
\Pr \left(\bigcap_{i=1}^j \left\{ \mbf{z}_i \in \bigcup_{ m \in \mcf{M}_i} \mbcf{T}_{\nu_i}(\mbf{U}_{m}) \right\} \right) &\geq \Pr \left(\bigcap_{i=1}^j \left\{ \mbf{z}_i \in \bigcup_{ m \in \mcf{M}^*_i} \mbcf{T}_{\nu_i}(\mbf{U}_{m}) \right\} \right)  \label{eq:circum:1}\\
&= \prod_{i=1}^j \left[ 1 - \prod_{m \in \mcf{M}_i^*}  \Pr \left(  \mbf{z}_i \notin \mbcf{T}_{\nu_i}(\mbf{U}_{m}) \right) \right]  \label{eq:circum:2} \\
&= \prod_{i=1}^j \left[ 1 - \left( 1 - 2^{-n \mathbb{I}(\nu_i,\sigma|\tau) +O(\logt n) }\right)^{|\mcf{M}_i^*|}\right]  \label{eq:circum:3} \\
&\geq \prod_{i=1}^j \left[ 1 - \exp \left( -|\mcf{M}_i^*|  2^{-n  \mathbb{I}(\nu_i,\sigma|\tau) +O(\logt n) } \right)\right]  \label{eq:circum:4} \\
&\geq 2^{ \sum_{i=1}^j \left| \logt ( |\mcf{M}_i^*|)  -  n\mathbb{I}(\nu_i,\sigma|\tau) \right|^-  +O(j \logt n) }   \label{eq:circum:5}\\
&\geq 2^{ \sum_{i=1}^j \left| \logt ( |\mcf{M}_i|)  -  n\mathbb{I}(\nu_i,\sigma|\tau) \right|^-  +O(j \logt nj ) }   \label{eq:circum:6}
\end{align}
where~\eqref{eq:circum:2} is because $\mcf{M}_i^* \cap \mcf{M}_j^* = \emptyset$ for $i\neq j$ and hence the events are independent;~\eqref{eq:circum:3} is by Lemma~\ref{lem:obvious};~\eqref{eq:circum:4} is because\footnote{See~\cite[Lemma~10.5.3]{CT}.} $(1-a)^b \leq e^{-ab}$ for $a \in [0,1]$ and $b > 0$;~\eqref{eq:circum:5} is because $1-e^{-x} \geq 2^{|\logt x|^- -1}$; and~\eqref{eq:circum:6} is because $|\mcf{M}_i^*| \geq (2j)^{-1} |\mcf{M}_i|.$

The existence of the aforementioned sets $\mcf{M}_i^*$, for each $i \in \{1,\dots, j\}$, is guaranteed by the following random coding argument.
For each $m \in \bigcup_{i=1}^j \mcf{M}_i$, let $\mcf{I}(m) \defn \{i \in \{1,\dots,j\}: m \in \mcf{M}_i\}$ and let $I_m$ be independent RVs uniform over $\mcf{I}(m)$. 
Now let
\begin{align}
\mcf{M}_i^* \defn \{ m \in \mcf{M}_i : I_{m} = i \}
\end{align}
for each $i \in \{1,\dots, j\}$, and observe that 
\begin{align}
\Pr \left( \bigcap_{i=1}^j \left\{ \sum_{m \in \mcf{M}_i} \idc{I_m = i} \geq (2j)^{-1} |\mcf{M}_i| \right\} \right) &\geq 1 - \sum_{i=1}^j \Pr \left(  \sum_{m \in \mcf{M}_i} \idc{I_m = i} < (2j)^{-1} |\mcf{M}_i|  \right) \label{eq:circum:1b}\\
&\geq 1 - \sum_{i=1}^j \exp\left(- \left(  \frac{1}{2}\ln \frac{e}{2} \right) j |\mcf{M}_i| \right) \label{eq:circum:2b}\\
&> 1 - \min_{i\in \{1,\dots, j\}} \exp \left( -\frac{|\mcf{M}_i|}{8j} + \ln j\right) \geq 0 \label{eq:circum:3b}
\end{align}
where~\eqref{eq:circum:1b} is the union bound;~\eqref{eq:circum:2b} is by the (upcoming) Lemma~\ref{lem:ck} and because $\idc{ I_m = i}$ is Bernoulli$(b)$ with $b\geq \frac{1}{j}$; and~\eqref{eq:circum:3b} is because $|\mcf{M}_i| \geq 8j \ln j$ for all $i \in \{1,\dots, j\}.$
But
\begin{align}
\Pr \left( \bigcap_{i=1}^j \left\{ \sum_{m \in \mcf{M}_i} \idc{I_m = i} \geq (2j)^{-1} |\mcf{M}_i| \right\} \right) > 0
\end{align}
implies there must exist at least one selection of set of outcomes for $I_{m}$ which leads to orthogonal $\mcf{M}_i^*$ such that $|\mcf{M}_i^*| \geq (2j)^{-1} |\mcf{M}_i|$ for all $i \in \{1,\dots, j\}.$

\end{IEEEproof}

The next lemma appears in~\cite[Lemma~17.9]{CK}.

\begin{lemma}\label{lem:ck} \textbf{(\cite[Lemma~17.9]{CK})} The probability that in $k$ independent trials an event of probability $q$ occurs less/more than $\alpha q k$ times, according as $\alpha \lessgtr 1$, is bounded above by $e^{-c(\alpha)qk}$ where $c(\alpha) = \alpha \ln \alpha - \alpha + 1$.
\end{lemma}
In determining the number of sequences of $\mbf{u}$ chosen such that $\mbf{z} \in \mbcf{T}_{\nu}(\mbf{u})$, each independent selection of a sequence can be viewed as a trial. Lemma \ref{lem:ck} can be applied effectively, regardless of if $qk \lessgtr 1$. Although, the method by which it should be applied differs. To streamline the analysis, the following corollary of Lemma~\ref{lem:ck} will be used instead.

\begin{cor}\label{cor:ck}
For all real numbers $a\geq\frac{1}{n}$ and $b\geq 0$
\begin{equation}
\Pr \left( \bigtimes_{i=1}^{\lfloor 2^{na} \rfloor} V_i \in \mcf{\hat V}_{(ne)^2 } \right) \geq 1 - 2e^{-n^2/4}
\end{equation}
where
\[
\mcf{\hat V}_{\gamma} \defn \set{ \bigtimes_{i=1}^{\lfloor 2^{na} \rfloor}   v_i :  \left \lfloor\frac{1}{\gamma} 2^{-n\left| a - b\right|^+    } \right \rfloor < \sum_{i=1}^{\lfloor 2^{na} \rfloor} v_i \leq  \gamma 2^{-n\left| a - b\right|^+ } }
\]
and $\bigtimes_{i=1}^{\lfloor 2^{na} \rfloor} V_i$ are independent Bernoulli $(2^{-nb})$ random variables.
Furthermore, if $n(a - b)  > 1+ 2 \logt n$ then 
\begin{equation}
\Pr \left(\bigtimes_{i=1}^{\lfloor 2^{na} \rfloor} V_i \in \mcf{\hat V}_{2 e } \right) \geq 1 - 2e^{-n^2/4}.
\end{equation}
\end{cor}
\begin{IEEEproof}
Throughout this proof we will use that 
$$2^{na-1} \leq 2^{na} -1 \leq \lfloor 2^{na} \rfloor \leq 2^{na},$$
which follows because $a \geq \frac{1}{n}.$

If $n(a - b)  \geq 1+ 2 \logt n$, then
\begin{align}
\Pr \left( \sum_{i=1}^{\lfloor 2^{na} \rfloor } V_i < \left \lfloor (2e)^{-1} 2^{n \left| a -b \right|^+  } \right \rfloor \right) &\leq \Pr \left( \sum_{i=1}^{\lfloor 2^{na} \rfloor} V_i < e^{-1} \lfloor 2^{na} \rfloor 2^{-nb} \right) \notag \\
&< e^{-(1 - 2e^{-1})2^{n (a - b) -1 } } < e^{-n^2/4}\label{cor:2}
\end{align}
and
\begin{equation}
\Pr \left( \sum_{i=1}^{\lfloor 2^{na} \rfloor } V_i > (2e) 2^{n \left| a -b \right|^+ } \right) \leq \Pr \left( \sum_{i=1}^{\lfloor 2^{na} \rfloor } V_i > e \lfloor 2^{na}\rfloor 2^{-nb} \right) < e^{-2^{n (a - b) -1 }} < e^{-n^2}
\end{equation}
by Lemma~\ref{lem:ck}.

On the other hand, if $n(a-b)  < 1 + 2 \logt n$, then
\begin{align}
\Pr \left( \sum_{i=1}^{\lfloor 2^{na}\rfloor} V_i < \left \lfloor (en)^{-2} 2^{n \left| a -b \right|^+ } \right \rfloor \right) &= \Pr \left( \sum_{i=1}^{\lfloor 2^{na} \rfloor} V_i < 0 \right) 
= 0 \label{cor:3}
\end{align}
and
\begin{align}
\Pr \left( \sum_{i=1}^{\lfloor 2^{na} \rfloor} V_i > (ne)^2  2^{n \left| a -b \right|^+ } \right) &\leq \Pr \left( \sum_{i} V_i > (ne)^2  \right) \notag \\
&= \Pr \left( \sum_{i} V_i > \left( \frac{(ne)^2 }{\lfloor 2^{na}\rfloor 2^{-nb} } \right)   \lfloor 2^{na}\rfloor 2^{-nb} \right)   \notag \\
&< e^{-n^2/4} \label{cor:4}
\end{align}
by Lemma~\ref{lem:ck}, where in particular
\begin{align*}
c\left( \frac{(ne)^2 }{d} \right) d &= (ne)^2 \ln \frac{(ne)^2}{d}   - (ne)^2  + d\\
&> (ne)^2  \ln \frac{e^2}{2}   - (ne)^2 \\
&= (ne)^2 ( 1 - \ln(2)) > n^2/4
\end{align*}
for all real numbers $d< 2n^2$.
\end{IEEEproof}

We conclude with a technical proof, which will help to simplify presentation.
\begin{lemma}\label{lem:tbcut}
Let $\mcf{A}_1 , \dots , \mcf{A}_{k}$ be subsets of some finite set $\mcf{X}$.
\[
\frac{k \min_{j \in \{1,\dots,k\}} \abs{\mcf{A}_j} }{\max_{x \in \cup_{i} \mcf{A}_i} \abs{ \{ i : x \in \mcf{A}_i \} }  } \leq \abs{\cup_{i} \mcf{A}_i } < \frac{k \max_{j \in \{1,\dots,k\}} \abs{\mcf{A}_j} }{\min_{x \in \cup_{i} \mcf{A}_i} \abs{ \{i : x \in \mcf{A}_i \} }  }.
\]
\end{lemma}
\begin{IEEEproof}

Note that
\begin{equation}
\abs{\cup_{i} \mcf{A}_i} = \sum_{x \in \mcf{X} } \idc{x\in \cup_{i} \mcf{A}_i}.
\end{equation}
From this equation, it must also follow that
\begin{equation}
\abs{\cup_{i=1}^k \mcf{A}_i} = \sum_{x\in \mcf{X}} \sum_{i=1}^k \frac{\idc{x \in \mcf{A}_i}}{\abs{\set{i : x \in \mcf{A}_i}}}
\end{equation}
since
\[
\idc{x\in \cup_{i=1}^k \mcf{A}_i} = \sum_{i=1}^k \frac{\idc{x \in \mcf{A}_i}}{\abs{\set{i : x \in \mcf{A}_i}}}
\]
for every $x \in \cup_{i=1}^k \mcf{A}_i$. Swapping the order of the summations and replacing the denominator and numerator with the min or max proves the lemma. 
\end{IEEEproof}

\section{Proof of Theorem~\ref{thm:lai_strong}}\label{sec:p_lai_strong}

Appendices~\ref{app:cc1:cc} through \ref{app:cc1:t1e} are dedicated to proving the following theorem:
\begin{theorem}\label{thm:p_lai_strong}
Given finite sets $\mcf{U}$ and $\mcf{W}$, distributions $\rho \in \mcf{P}_n(\mcf{X}|\mcf{U};\sigma\tau)$, $\sigma \in \mcf{P}_{n} (\mcf{U} \gg \mcf{W})$, and $\tau \in \mcf{P}_n(\mcf{W})$ there exists a $$(r,\alpha ,\kappa, \epsilon,j,n)-\mathrm{AA~code~for~DM-AIC}(t,q),$$ 
where
\begin{align*}
-n^{-1} \logt \epsilon &\geq \min_{\mu \in \mcf{P}_n(\mcf{Y}|\mcf{U};\sigma\tau)} \mathbb{F}_n(\mu||t,\rho|\sigma\tau) - \left| j \kappa + \mathbb{S}_{\hat r,\tilde r}(\mu|\sigma,\tau) \right|^-  + O(n^{-1}\logt n) , \\
\alpha  &\geq  \min_{\nu^j \in \bigtimes_{i=1}^j \mcf{P}_{n}(\mcf{Z}|\mcf{U};\sigma\tau)}  \left| j \kappa + \sum_{i=1}^j \left| \mathbb{S}_{\hat r,\tilde r}(\nu_i|\sigma,\tau) \right|^- \right|^+  + \sum_{i=1}^j \mathbb{F}_n(\nu_i||q,\rho|\sigma\tau)   + O(j n^{-1}\logt n),
\end{align*}
for all $r \leq \tilde r + \hat r$, $\kappa \geq \tilde \kappa$, and large enough $n$.
\end{theorem}
Theorem~\ref{thm:p_lai_strong} will be proved by using a random code construction presented in Appendix~\ref{app:cc1:cc}. 
For each round of communication, the same code shall be used.
Next, the message error probability (i.e., the bound on $\epsilon$) is then derived in Appendix~\ref{app:cc1:mea} and the authentication rate (i.e., the bound on $\alpha$) in~\ref{app:cc1:t1e}.
The error probability does not change with the number of rounds, since the code is fixed over all rounds.
On the other hand, the authentication rate should decrease as the number of rounds increases due to the fact that the adversary continues to gain information. 
Because of this, we only need to consider the final round ($j$) of communication when analyzing the authentication rate.

Afterwards, using Theorem~\ref{thm:p_lai_strong}, we prove Theorem~\ref{thm:lai_strong} in Appendix~\ref{app:lai_strong_proof}. 
A proof for Theorem~\ref{thm:lai_strong} is still necessary in light of Theorem~\ref{thm:p_lai_strong} since Theorem~\ref{thm:p_lai_strong} does not concern itself with asymptotic behaviour, whereas the capacity region is defined as the limit point of a sequence of codes.

\begin{IEEEproof}
\subsection{Code construction}\label{app:cc1} \label{app:cc1:cc}

The code construction involves rate splitting, where the message set $\mcf{M} \defn \set{1, \dots, 2^{nr}}$ is split into the product of sets $\mcf{\tilde M} \defn \set{1, \dots, 2^{n\tilde r}}$ and $\mcf{\hat M} \defn \set{1, \dots, 2^{n\hat r}}$, such that $\tilde r + \hat r  = r$.
The set of keys is denoted $\mcf{K} = \set{1, \dots, 2^{nj \kappa}}$, where the same key is being used for all $j$ rounds of communication.
Furthermore, the same code will be used in every round (i.e., $f_1=f_2=\dots= f_j$ and $\varphi_1 = \varphi_2 = \dots =\varphi_j$) and thus we shall simply refer to this code as $f,\varphi$.



\noindent  \textbf{Random codebook generation:} Independently for each $\hat m \in \mcf{\hat M}$, select $\mbf{w}_{f}(\hat m)$ uniformly from $\mbcf{T}_{\tau}$. Then, independently for each $(\hat m,\tilde m, k)  \in \mcf{\hat M}\times \mcf{\tilde M} \times \mcf{K}$, select the value of $\mbf{u}_{f}(\hat m, \tilde m, k)$ uniformly from $\mbcf{T}_{\sigma}(\mbf{w}_{f}(\hat m))$.

\noindent  \textbf{Encoder:} For $(\hat M, \tilde M, K) = (\hat m, \tilde m, k)$ the encoder chooses $\mbf{X}$ uniformly from $\mbcf{T}_{\rho}(\mbf{u}(\hat m,\tilde m, k));$
i.e.,
\begin{equation}\label{eq:encoder}
f(\mbf{x} | \hat m,\tilde m,k) \defn \begin{cases} 
|\mbcf{T}_{\rho}(\mbf{u}_f(\hat m, \tilde m, k))|^{-1} &\text{if } \mbf{x} \in \mbcf{T}_{\rho}(\mbf{u}_f(\hat m, \tilde m, k)) \\
0 &\text{else}
\end{cases}.
\end{equation}

\noindent  \textbf{Decoder:} For received sequence $\mbf{Y} = \mbf{y}$ and key $K=k$, the decoder $\varphi$ chooses the message estimates according to the distribution
\begin{equation}\label{eq:decoder1}
\varphi(\hat m', \tilde m'|\mbf{y},k) \defn \begin{cases}
1 & \text{ if } \sum_{\mbf{x} \in \mbcf{T}_{\rho}(\mbf{u}_f(\hat m',\tilde m',k))} \mbf{t}( \mbf{y}|\mbf{x})   >  \sum_{\mbf{\tilde x} \in \mbcf{T}_{\rho}(\mbf{u}_f(i,j,l))} \mbf{t}  ( \mbf{y}|\mbf{\tilde x}) \\
& ~~ \forall (i,j,l)  \in \mcf{\hat M}\times \mcf{\tilde M} \times \mcf{K} - \set{(\hat m',\tilde m', k)}\\
0 & \text{ otherwise}
\end{cases},
\end{equation}
for all $(\hat m',\tilde m') \in \mcf{\hat M} \times \mcf{\tilde M}$, and instead declares intrusion according to
\begin{equation}\label{eq:decoder2}
\varphi(\mbf{!}|\mbf{y},k) \defn \begin{cases}
1 & \text{ if } \sum_{(\hat m,\tilde m) \in \mcf{\hat M} \times \mcf{\tilde M}} \varphi_f(\hat m',\tilde m'|\mbf{y},k) = 0\\
0 & \text{ otherwise}
\end{cases}.
\end{equation}
That is, for a given $\mbf{y}$, Bob declares $(\hat m',\tilde m')$ to be the message sent if $(\hat m',\tilde m',k)$ is the unique maximizer of $ \sum_{\mbf{x} \in \mbcf{T}_{\rho}(\mbf{u}_f(\hat m',\tilde m',k))} \mbf{t}( \mbf{y}|\mbf{x}) $, and declares $\mbf{!}$ if the maximizer is not unique.

\subsection{Message error analysis}\label{app:cc1:mea}
Let $F,\Phi$ be the random variables that represent the randomly chosen encoder and decoder.
By extension, $\mbf{w}_{F}(\hat m)$ and $\mbf{u}_{F}(\hat m,\tilde m,k)$ denote the RVs representing the randomly chosen $\mbf{w}_{f}(\hat m)$ and $\mbf{u}_{f}(\hat m,\tilde m, k)$.
To prove the error bound in Theorem~\ref{thm:p_lai_strong}, it is sufficient to show that
\begin{equation} \label{eq:proof_lai:fin2}
\mathbb{E}[\varepsilon_{F,\Phi}] \leq \max_{\mu \in \mcf{P}_n(\mcf{Y}|\mcf{U};\sigma\tau)} 2^{- n\left( \mathbb{F}_n(\mu||t,\rho|\sigma\tau) - \left|\mathbb{S}_{\hat r, \tilde r}(\mu|\sigma,\tau) + j\kappa \right|^- \right) + O(\logt n)} .
\end{equation}
Indeed,
\begin{equation} \label{eq:proof_lai:fin2alt}
\Pr \left( \varepsilon_{F,\Phi} \geq  n \mathbb{E}[\varepsilon_{F,\Phi}]   \right) \leq n^{-1} 
\end{equation}
directly follows from Equation~\eqref{eq:proof_lai:fin2}  by Markov's inequality, hence proving that as $n$ increases, the probability of a code constructed according to Appendix~\ref{app:cc1} satisfying 
$$ -n^{-1} \logt  \varepsilon_{f,\varphi} \geq  \mathbb{F}_n(\mu||t,\rho|\sigma\tau) - \left|\mathbb{S}_{\hat r, \tilde r}(\mu|\sigma,\tau) + j\kappa \right|^-  + O(n^{-1} \logt n) $$ converges to one.  

For the first step in proving Equation~\eqref{eq:proof_lai:fin2}, it will be helpful to express $\varepsilon_{f,\varphi}$ in terms of indicator functions determining when a code is in error. 
To do this, for each $\mu \in \mcf{P}_n(\mcf{Y}|\mcf{U};\sigma\tau)$, $\mbf{u} \in \mbcf{T}_{\sigma\tau}$ and $\mbf{y} \in \mbcf{T}_{\mu}(\mbf{u})$ let 
\begin{align}
\mcf{D}(\mu) &\defn \set{\mu'\in \mcf{P}_n(\mcf{Y}|\mcf{U};\sigma\tau)  :  \begin{array}{rl} \mathbb{H}(\mu|\sigma\tau) + \mathbb{F}_n(\mu||t,\rho|\sigma\tau)& \\
&\hspace{-70pt}\geq  \mathbb{H}(\mu'|\sigma\tau) + \mathbb{F}_n(\mu'||t,\rho|\sigma\tau) + O(n^{-1}\logt n)  \end{array}   } , \notag \\
 \mcf{E}(\mbf{u},\mbf{y}) &\defn \set{\mbf{\hat u} \in \mbcf{T}_{\sigma\tau} : \mbf{y} \in \bigcup_{\mu' \in \mcf{D}(p_{\mbf{y}|\mbf{u}})} \mbcf{T}_{\mu'}(\mbf{\hat u})   } \notag
\end{align}
where the order term is inherited from Lemma~\ref{lem:btx} so that $\mbf{\hat u} \in \mcf{E}(\mbf{u},\mbf{y})$ implies
$$\sum_{ \mbf{x} \in \mbcf{T}_{\rho}(\mbf{u}) } \mbf{t}(\mbf{y}|\mbf{x}) \leq \sum_{ \mbf{x} \in \mbcf{T}_{\rho}(\mbf{\hat u}) } \mbf{t}(\mbf{y}|\mbf{x}).$$
With these definitions, the function
\begin{align}
\lambda_{f,\varphi}(\hat m,\tilde m, k, \mbf{u},\mbf{y}) \defn   1\left\{ \mbf{u}_{f}(\hat m, \tilde m, k) = \mbf{u}  \middle\} 1 \middle\{ \bigcup_{(i_1,i_2,i_3) \in \mcf{\hat M}\times \mcf{\tilde M}\times \mcf{K} - (\hat m,\tilde m,k) }\mbf{u}_{f}(i_1,i_2,i_3) \in \mcf{E}(\mbf{u},\mbf{y}) \right\}  \notag 
\end{align}
is only non-zero if both $\mbf{u}_{f}(\hat m, \tilde m,k) = \mbf{u}$ and there exists a value of $(i_1,i_2,i_3) \neq (\hat m,\tilde m,k) $ such that $\mbf{u}_{f}(i_1,i_2,i_3) \in \mcf{E}(\mbf{u},\mbf{y}).$ 
The probability of error can now be written as
\begin{equation} \label{eq:proof_lai:error2}
\varepsilon_{f,\varphi} =  \sum_{\subalign{\hat m&\in \mcf{\hat M},\\\tilde m &\in \mcf{\tilde M},\\ k &\in \mcf{K},\\ \mbf{y} &\in \mbcf{Y}, \\ \mbf{u} &\in \mbcf{T}_{\sigma\tau} }}    \lambda_{f,\varphi}(\hat m, \tilde m, k,\mbf{u},\mbf{y}) 2^{-n( \hat r + \tilde r + j\kappa)} \sum_{\mbf{x} \in \mbcf{T}_{\rho}(\mbf{u})} |\mbcf{T}_{\rho}(\mbf{u})|^{-1} \mbf{t} (\mbf{y}|\mbf{x}) .
\end{equation} 
Or, as
\begin{align} 
\varepsilon_{f,\varphi} &=   \sum_{\substack{\mu \in \mcf{P}_n(\mcf{Y}|\mcf{U}; \sigma\tau) \\ \subalign{\hat m&\in \mcf{\hat M},\\\tilde m &\in \mcf{\tilde M},\\ k &\in \mcf{K}, \\ \mbf{u} &\in \mbcf{T}_{\sigma\tau}}} } \sum_{\mbf{y} \in \mbcf{T}_{\mu}(\mbf{u})}  \lambda_{f,\varphi}(\hat m, \tilde m, k,\mbf{u},\mbf{y}) 2^{-n( \hat r  + \tilde r+ j \kappa + \mathbb{ F}_n(\mu||t,\rho|\sigma\tau) + \mathbb{H}(\mu|\sigma\tau) ) + O(\logt n)} \label{eq:proof_lai:error3}
\end{align}
by splitting the summation of $\mbf{y}$ depending on the empirical distribution $p_{\mbf{y}|\mbf{u}}$ and using Lemma~\ref{lem:btx}.

Next, assume that
\begin{align}
&\mathbb{E} \left[ \lambda_{F,\Phi}(\hat m, \tilde m, k,\mbf{u},\mbf{y}) \right]
\leq   2^{-n \mathbb{H}(\sigma\tau) + O(\logt n)} 2^{n \left| \tilde r +j\kappa  + \max\left(\hat r -\mathbb{I}(\mu^\dagger,\sigma\tau), -\mathbb{I}(\mu^*,\sigma\tau)  \right)  \right|^- }   \label{eq:proof_lai:mea'},
\end{align}
where 
\begin{align}
\mu^\dagger &= \argmax_{\gamma \in \mcf{D}(\mu):\gamma\sigma\tau = 
\mu\sigma\tau}\mathbb{H}(\gamma|\sigma\tau) \\
\text{and } \mu^* &= \argmax_{\gamma \in \mcf{D}(\mu): \gamma\sigma = \mu \sigma }\mathbb{H}(\gamma|\sigma\tau),
\end{align}
for each $\mu \in \mcf{P}_{n}(\mcf{Y}|\mcf{U};\sigma\tau),~\hat m \in \mcf{\hat M},~\tilde m\in \mcf{\tilde M},~ k\in \mcf{K},~\mbf{u} \in \mbcf{T}_{\sigma\tau}$, and $\mbf{y} \in \mbcf{T}_{\mu}(\mbf{u}).$
Later, Equation~\eqref{eq:proof_lai:mea'} will be justified. For now, though, under the assumption that Equation~\eqref{eq:proof_lai:mea'} is true,
\begin{align} 
\mathbb{E} [\varepsilon_{F,\Phi}] 
&=   \sum_{\substack{\mu \in \mcf{P}_n(\mcf{Y}|\mcf{U}; \sigma\tau) \\ \subalign{\hat m&\in \mcf{\hat M},\\\tilde m &\in \mcf{\tilde M},\\ k &\in \mcf{K}, \\ \mbf{u} &\in \mbcf{T}_{\sigma\tau}}}}  \sum_{\mbf{y} \in \mbcf{T}_{\mu}(\mbf{u})}  2^{-n( \hat r  + \tilde r+ j\kappa + \mathbb{ F}_n(\mu||t,\rho|\sigma\tau) + \mathbb{H}(\mu|\sigma\tau) ) + O(\logt n)} \mathbb{E} \left[ \lambda_{F,\Phi}(\hat m, \tilde m, k,\mbf{u},\mbf{y}) \right]  \label{eq:proof_lai:avg} \\
&\leq \max_{  \mu \in \mcf{P}_n(\mcf{Y}|\mcf{U}; \sigma\tau)}  2^{-n\mathbb{ F}_n(\mu||t,\rho|\sigma\tau) + O(\logt n)} 2^{n \left| \tilde r +j\kappa  + \max\left(\hat r- \mathbb{I}(\mu^\dagger,\sigma\tau) , -\mathbb{I}(\mu^*,\sigma|\tau)  \right)  \right|^- } \label{eq:proof_lai:avg2}
\end{align}
follows from~\eqref{eq:proof_lai:error3} by using the linearity of the expected value and carrying out the summations using Equation~\eqref{eq:app:lai:type:2} and~\eqref{eq:app:lai:type:4} when applicable. 
To obtain Equation~\eqref{eq:proof_lai:fin2}, observe
for all $\mu \in \mcf{P}_n(\mcf{Y}|\mcf{U},\mcf{X};\sigma\tau)$ and $\mu^\ddagger \in \{ \mu^* , \mu^\dagger\}$ that
$$-\mathbb{F}_n(\mu^\ddagger||t,\rho|\sigma\tau) + O(n^{-1} \logt n) \geq -\mathbb{F}_n(\mu||t,\rho|\sigma\tau) - \mathbb{H}(\mu|\sigma\tau) + \mathbb{H}(\mu^\ddagger|\sigma\tau) \geq -\mathbb{F}_n(\mu||t,\rho|\sigma\tau) $$
since $\mu^\ddagger \in \mcf{D}(\mu)$ and $\mathbb{H}(\mu^\ddagger|\sigma\tau) \geq \mathbb{H}(\mu|\sigma\tau).$
Hence,
\begin{align} 
\mathbb{E} [\varepsilon_{F,\Phi}] 
&\leq \max_{  \mu \in \mcf{P}_n(\mcf{Y}|\mcf{U}; \sigma\tau)} \begin{cases}
2^{n\left( -\mathbb{F}_n(\mu^\dagger||t,\rho|\sigma\tau) + \left| \hat r+ \tilde r +j\kappa  - \mathbb{I}(\mu^\dagger,\sigma\tau)   \right|^-\right) + O(\logt n)}  &\text{ if } \hat r - \mathbb{I}(\mu^\dagger,\sigma\tau) > -\mathbb{I}(\mu^*,\sigma|\tau) \\
2^{n\left( - \mathbb{F}_n(\mu^*||t,\rho|\sigma\tau) + \left| \tilde r +j\kappa  - \mathbb{I}(\mu^*,\sigma|\tau)   \right|^- \right) + O(\logt n)}  &\text{ else }  
\end{cases}
\label{eq:proof_lai:avg3}\\
&\leq  \max_{  \mu \in \mcf{P}_n(\mcf{Y}|\mcf{U}; \sigma\tau)} 2^{n\left( -\mathbb{F}_n(\mu||t,\rho|\sigma\tau) + \left| j\kappa + \mathbb{S}_{\hat r,\tilde r}(\mu|\sigma,\tau)  \right|^-\right) + O(\logt n)},
\end{align}
proving Equation~\eqref{eq:proof_lai:fin2}.

Now, we return to proving Equation~\eqref{eq:proof_lai:mea'}, which can be done by upper bounding $\lambda_{F,\Phi}(\hat m,\tilde m, k,\mbf{u},\mbf{y})$ in two ways.
The first upper bound can be constructed directly as follows
\begin{align}
\mathbb{E}\left[ \lambda_{F,\Phi}(\hat m,\tilde m, k,\mbf{u},\mbf{y}) \right] &\leq \mathbb{E}\left[ 1\left\{ \mbf{u}_{F}(\hat m, \tilde m, k) = \mbf{u}  \right\} \right] \label{eq:proof_lai:ab:1}\\
&= \Pr \left( \mbf{u}_{F}(\hat m, \tilde m, k) = \mbf{u}  \right)  \label{eq:proof_lai:ab:2}\\
&= 2^{-n \mathbb{H}(\sigma \tau) + O(\logt n)} \label{eq:proof_lai:ab:3};
\end{align}
where~\eqref{eq:proof_lai:ab:1} is because $\lambda_{f,\varphi} (\hat m,\tilde m, k,\mbf{u},\mbf{y}) \leq 1\{\mbf{u}_{f}(\hat m, \tilde m, k) = \mbf{u} \}$ for all codes;~\eqref{eq:proof_lai:ab:2} is because the expectation of an indicator is the probability of that indicator; and~\eqref{eq:proof_lai:ab:3} is by Equation~\eqref{eq:app:lai:type:3} since $\mbf{u}_{F}$ is uniformly distributed over $\mbcf{T}_{\sigma\tau}$.
For the second upper bound, observe that
\begin{align}
\mathbb{E} \left[ \lambda_{F,\Phi}(\hat m, \tilde m, k,\mbf{u},\mbf{y}) \right] &\leq  \sum_{(i_1,i_2,i_3) \in \mcf{\hat M}\times \mcf{\tilde M} \times \mcf{K} -  \{ (\hat m,\tilde m,k) \} }  \mathbb{E} \left[  \idc{ \mbf{u}_{F}(\hat m, \tilde m, k) = \mbf{u}  } \idc{\mbf{u}_{F}(i_1,i_2,i_3) \in \mcf{E}(\mbf{u},\mbf{y}) }  \right]\\
&=   \sum_{(i_1,i_2,i_3) \in \mcf{\hat M}\times \mcf{\tilde M} \times \mcf{K} -  \{ (\hat m,\tilde m,k) \} }   \Pr \left( \mbf{u}_{F}(\hat m, \tilde m, k) = \mbf{u} , \mbf{u}_{F}(i_1,i_2,i_3) \in \mcf{E}(\mbf{u},\mbf{y}) \right) ,\label{eq:cc1:mea:1}
\end{align}
since the expectation of an event indicator is the probability of the event.
The value of $\Pr ( \mbf{u}_{F}(\hat m, \tilde m, k) = \mbf{u} , \mbf{u}_{F}(i_1,i_2,i_3) \in \mcf{E}(\mbf{u},\mbf{y}) )$ depends on if $\hat m = i$.  
If $\hat m \neq i$, then $\mbf{u}_{F}(\hat m,\tilde m, k)$ and $\mbf{u}_{F}(i_1,i_2,i_3)$ are independently and uniformly chosen over $\mbcf{T}_{\sigma\tau}$, and thus
\begin{align}
&\hspace{-10pt}\Pr ( \mbf{u}_{F}(\hat m, \tilde m, k) = \mbf{u} , \mbf{u}_{F}(i,j,l) \in \mcf{E}(\mbf{u},\mbf{y}) ) \notag\\
&=\abs{\mcf{T}_{\sigma\tau}}^{-1} \sum_{\subalign{\mu' &\in \mcf{D}(p_{\mbf{y}|\mbf{u}}): \\ \mu'\sigma\tau &= p_{\mbf{y}|\mbf{u}}\sigma\tau} } \Pr( \mbf{y} \in \mcf{T}_{\mu'}(\mbf{u}_F) ) \label{eq:cc1:mea:1.1:-2}  \\
&= 2^{-n \mathbb{H}(\sigma\tau) + O(\logt n) } \sum_{\subalign{\mu' &\in \mcf{D}(p_{\mbf{y}|\mbf{u}}): \\ \mu'\sigma\tau &= p_{\mbf{y}|\mbf{u}}\sigma\tau}} 2^{-n \mathbb{I}(\mu',\sigma\tau)} \label{eq:cc1:mea:1.1:-1} \\
&=  2^{-n(\mathbb{H}(\sigma\tau)+  \mathbb{I}(\mu^\dagger,\sigma\tau))  + O(\logt n) } \label{eq:cc1:mea:1.1};
\end{align}
where~\eqref{eq:cc1:mea:1.1:-2} is because $$\mu'\sigma\tau  = p_{\mbf{y}|\mbf{u}'} p_{\mbf{u}'} = p_{\mbf{y}} = p_{\mbf{y}|\mbf{u}}p_{\mbf{u}} = p_{\mbf{y}|\mbf{u}} \sigma\tau $$
for each $\mbf{u}'\in \mbcf{T}_{\sigma\tau}$ and $\mbf{y} \in \mbcf{T}_{\mu'}(\mbf{u}')$;~\eqref{eq:cc1:mea:1.1:-1} is by Equation~\eqref{eq:app:lai:type:2} and Lemma~\ref{lem:obvious}; and~\eqref{eq:cc1:mea:1.1} is by Equation~\eqref{eq:app:lai:type:4} and because 
$$\min_{\subalign{\mu' &\in \mcf{D}(p_{\mbf{y}|\mbf{u}}): \\ \mu'\sigma\tau &= p_{\mbf{y}|\mbf{u}}\sigma\tau}} \mathbb{I}(\mu',\sigma\tau) = \mathbb{H}(\mu\sigma\tau) - \max_{\subalign{\mu' &\in \mcf{D}(p_{\mbf{y}|\mbf{u}}): \\ \mu'\sigma\tau &= p_{\mbf{y}|\mbf{u}}\sigma\tau}} \mathbb{H}(\mu'|\sigma\tau) = \mathbb{I}(\mu^\dagger,\sigma\tau).$$
Otherwise, if $\hat m = i_1$, then $\mbf{u}_{F}(\hat m, \tilde m, k)$ and $\mbf{u}_{F}(\hat m,i_2,i_3)$ are independently chosen from $\mbcf{T}_{\sigma}(\mbf{w})$ for some $\mbf{w} \in \mbcf{T}_{\tau}$, and furthermore, this value of $\mbf{w}$ can be determined by the value of $\mbf{u}_{f}(\hat m, \tilde m, k)$ since $\sigma \in \mcf{P}(\mcf{U} \gg \mcf{W})$. 
Thus, when $\hat m = i$, a similar analysis to before shows that 
\begin{align}
\Pr ( \mbf{u}_{F}(\hat m, \tilde m, k) = \mbf{u} , \mbf{u}_{F}(i_1,i_2,i_3) \in \mcf{E}(\mbf{u},\mbf{y}) ) 
&=  2^{-n ( \mathbb{H}(\sigma\tau) + \mathbb{I}(\mu^*,\sigma|\tau) ) + O(\logt n) } \label{eq:cc1:mea:1.2}.
\end{align}
Plugging~\eqref{eq:cc1:mea:1.1} and~\eqref{eq:cc1:mea:1.2} into~\eqref{eq:cc1:mea:1} and carrying out the summations yields 
\begin{align}
\mathbb{E} \left[ \lambda_{F,\Phi}(\hat m, \tilde m, k,\mbf{u},\mbf{y}) \right] &\leq \sum_{\substack{i_1 \in \mcf{\hat M}-\{\hat m\} ,\\ (i_2,i_3)\in \mcf{\tilde M}\times \mcf{K}}} 2^{-n(\mathbb{H}(\sigma\tau)+  \mathbb{I}(\mu^\dagger,\sigma\tau))  + O(\logt n) } \notag \\
&\hspace{5pt} + \sum_{ (i_2,i_3)\in \mcf{\tilde M}\times \mcf{K} - \{(\tilde m, k)\} } 2^{-n(\mathbb{H}(\sigma\tau)+  \mathbb{I}(\mu^*,\sigma|\tau))  + O(\logt n) }\\
&= 2^{-n \mathbb{H}(\sigma\tau) + n\left( \tilde r + j\kappa + \max \left( \hat r - \mathbb{I}(\mu^\dagger,\sigma\tau) , - \mathbb{I}(\mu^*,\sigma|\tau)\right) \right) + O(\logt n) }. \label{eq:proof_lai:ab:4}
\end{align}
Combining Equations~\eqref{eq:proof_lai:ab:3} and~\eqref{eq:proof_lai:ab:4} validates Equation~\eqref{eq:proof_lai:mea'}.

\subsection{Authentication rate analysis}\label{app:cc1:t1e}

The proof of the authentication rate of Theorem~\ref{thm:p_lai_strong} follows first by proving that
\begin{equation}\label{eq:avg:t1e:ub}
-n^{-1} \logt \omega_{f^j,\varphi^j,j} \geq  \min_{\nu^j \in \bigtimes_{i=1}^j \mcf{P}_{n}(\mcf{Z}|\mcf{U};\sigma\tau)}  \left| j \kappa + \sum_{i=1}^j \left| \mathbb{S}_{\hat r,\tilde r}(\nu_i|\sigma,\tau)\right|^- \right|^+  + \sum_{i=1}^j \mathbb{F}_n(\nu_i||q,\rho|\sigma\tau)   + O(j n^{-1}\logt n)
\end{equation} 
for all codes (chosen according to Appendix~\ref{app:cc1:cc}) $(f,\varphi)$ such that $f \in \mcf{F}^* \cap \mcf{F}^\dagger$, where $\mcf{F}^*$ is the set of all $f$ such that
\begin{align}
n^{-1}\logt \abs{\mcf{\bar M}_f(\mbf{z},\nu,k)} &=  \left|\mathbb{S}_{\hat r,\tilde r}(\nu|\sigma,\tau) \right|^+ + O(n^{-1}\logt n), \label{eq:avg:suff:1} \\
\mcf{\bar M}_{f}(\mbf{z},\nu,k) &\defn \{ (\hat m,\tilde m) \in \mcf{\hat M} \times \mcf{\tilde M}: \mbf{z} \in \mbcf{T}_{\nu}(\mbf{u}_{f}(\hat m,\tilde m,k)) \} \notag 
\end{align}
for all $\nu \in \mcf{P}_n(\mcf{Z}|\mcf{U};\sigma\tau)$ and $\mbf{z} \in \mbcf{Z}_{f}(\nu,k)   \defn  \bigcup_{\subalign{\hat m&\in \mcf{\hat M},\\ \tilde m &\in \mcf{\tilde M}}} \mbcf{T}_{\nu}(\mbf{u}_{f}(\hat m, \tilde m, k))$, and $\mcf{F}^\dagger$ is the set of all $f$ such that
\begin{align}
n^{-1}\logt \abs{\mcf{ K}_f(\mbf{z}^j, \nu^j) } &\geq \left| j\kappa + \sum_{i=1}^j | \mathbb{S}_{\hat r,\tilde r}(\nu_i|\sigma,\tau)|^- \right|^+ + O(n^{-1}\logt n), \label{eq:avg:suff:2}\\
\mcf{K}_f( \mbf{z}^j,  \nu^j) &\defn \left\{ k \in \mcf{K} : \cap_{i=1}^j \{ \mbf{z}_i \in \mbcf{Z}_{f}(\nu_i,k)]\} \right\}, 
\end{align}
for all $\nu^j \in \bigtimes_{i=1}^j \mcf{P}_{n} (\mcf{Z}|\mcf{U};\sigma\tau)$ and $\mbf{z}^j \in \bigtimes_{i=1}^j \mbcf{Z}_{f}(\nu_i,k) $. 
Having shown $f \in \mcf{F}^*\cap \mcf{F}^\dagger$ imply~\eqref{eq:avg:t1e:ub}, for the second part of the proof, it is shown that 
\begin{equation} \label{eq:avg:t1e:lbp}
\Pr \left( F \in \mcf{F}^* \cap \mcf{F}^\dagger \right) \geq 1-e^{-n^2/4 + O(n)}.
\end{equation}
Hence for large enough $n$, the code construction of Appendix~\ref{app:cc1:cc} produces a code in $\mcf{F}^*\cap \mcf{F}^\dagger$, and therefore~\eqref{eq:avg:t1e:ub} holds.

These two goals have been subdivided into their own subsections for simplicity.

\subsubsection{Authentication rate for $f \in \mcf{F}^\dagger \cap \mcf{F}^*$}

For this section, it will be assumed that the code satisfies~\eqref{eq:avg:suff:1} and~\eqref{eq:avg:suff:2}.
The bound presented in Theorem~\ref{thm:p_lai_strong} can be obtained through direct manipulation, but first, two observations will prove helpful.

The first observation is that for any deterministic decoder (i.e., the range of $\phi$ is $\{0,1\}$), the type I error can be simplified to 
\begin{equation}\label{eq:avg:t1e_redux}
\omega_{f,\varphi,j} 
= \sum_{\mbf{z}^j \in \mbcf{Z}^j} \max_{k \in \mcf{K}} p(\mbf{z}^j,k).
\end{equation}
Indeed, clearly setting $\psi(\mbf{y}|\mbf{z}^j) = 1\{ g(\mbf{z}^j) = \mbf{y}\}$, where $g: \mbcf{Z}^j \rightarrow \mbcf{Y}$ is any deterministic function for which 
\[
\sum_{k } p(\mbf{z}^j,k) \varphi(\mcf{M}|g(\mbf{z}^j),k)  = \max_{\mbf{y} \in \mbcf{Y}}\sum_{k \in \mcf{K} } p(\mbf{z}^j,k) \varphi(\mcf{M}|\mbf{y},k) 
\]
for all $\mbf{z}^j$, maximizes the type I error.
Thus, Equation~\eqref{eq:avg:t1e_redux} since $\varphi(\mcf{M}|\mbf{y},k) = 1$ for at most one $k$ in the code construction.

The second observation is that 
\begin{align}
&-\frac{1}{n} \logt \sum_{\mbf{z}^j \in \mbcf{Z}^j }   \max_{k \in \mcf{K}} \prod_{i=1}^j 1\{ \mbf{z}_i \in \mbcf{Z}_{f}(\nu_i,k) \} \notag \\
&\hspace{10pt} \geq  \left| j\kappa + \sum_{i=1}^j | \mathbb{S}_{\hat r,\tilde r}(\nu_i|\sigma,\tau)|^- \right|^+ 
+ n\sum_{i=1}^j \left[\hat r + \tilde r + \kappa + \mathbb{H}(\nu_i|\sigma\tau) - |\mathbb{S}_{\hat r,\tilde r}(\nu_i|\sigma,\tau)|^+  \right] + O(j\logt n) ,    \label{eq:avg:t1e:pr:so}
\end{align}
for each $\nu^j \in \bigtimes_{i=1}^j \mcf{P}_{n}(\mcf{Z}|\mcf{U};\sigma\tau)$.
This observation follows since
\begin{align}
 \sum_{\mbf{z}^j \in \mbcf{Z}^j }   \max_{k \in \mcf{K}} \prod_{i=1}^j 1\{ \mbf{z}_i \in \mbcf{Z}_{f}(\nu_i,k) \} & = \left| \bigcup_{k \in \mcf{K}} \bigtimes_{i=1}^j \mbcf{Z}_{f}(\nu_i,k) \right|  \label{eq:avg:t1e:pr:so:1}\\
 &\leq \frac{2^{n j \kappa} \max_{k\in \mcf{K} } \prod_{i=1}^j \left|\mbcf{Z}_{f}(\nu_i,k) \right|  }{\min_{\mbf{z}^j  \in \bigcup_{k\in\mcf{K}} \mbcf{Z}^j_{f}(\nu_i,k) } |\mcf{ K}_f(\mbf{z}^j , \nu^j ) |  } \label{eq:avg:t1e:pr:so:2}\\
&\leq 2^{nj\kappa - n \left| j\kappa + \sum_{i=1}^j | \mathbb{S}_{\hat r,\tilde r}(\nu_i|\sigma,\tau)|^- \right|^+ + O(\logt n)}   \max_{k\in \mcf{K} } \prod_{i=1}^j \left|\mbcf{Z}_{f}(\nu_i,k) \right| \label{eq:avg:t1e:pr:so:3} \\
&\leq 2^{n\left[ j\kappa -  \left| j\kappa + \sum_{i=1}^j | \mathbb{S}_{\hat r,\tilde r}(\nu_i|\sigma,\tau)|^- \right|^+ \right] + O(\logt n)}  \notag \\
&\hspace{20pt} \cdot \max_{k\in \mcf{K} } \prod_{i=1}^j  \frac{2^{n\left[\hat r + \tilde r + \mathbb{H}(\nu_i|\sigma\tau)\right] + O(\logt n)} }{\min_{\mbf{z} \in \mbcf{Z}_{f}(\nu_i,k)} |\mcf{\bar M}_f(\mbf{z},\nu_i,k) |}  \label{eq:avg:t1e:pr:so:4} \\
&\leq 2^{-n  \left| j\kappa + \sum_{i=1}^j | \mathbb{S}_{\hat r,\tilde r}(\nu_i|\sigma,\tau)|^- \right|^+  + O(\logt n)}  \notag \\
&\hspace{20pt} \cdot 2^{n\sum_{i=1}^j \left[\hat r + \tilde r + \kappa + \mathbb{H}(\nu_i|\sigma\tau) - |\mathbb{S}_{\hat r,\tilde r}(\nu_i|\sigma,\tau)|^+  \right] + O(j\logt n)}   \label{eq:avg:t1e:pr:so:5}
\end{align}
where~\eqref{eq:avg:t1e:pr:so:1} is because $ \max_{k \in \mcf{K}} \prod_{i=1}^j 1\{ \mbf{z}_i \in \mbcf{Z}_{f}(\nu_i,k) \} = 1$ for all $\mbf{z}^j \in \bigtimes_{i=1}^j \mbcf{Z}_{f}(\nu_i,k)$ and is $0$ otherwise;~\eqref{eq:avg:t1e:pr:so:2} is by Lemma~\ref{lem:tbcut};~\eqref{eq:avg:t1e:pr:so:3} is because $f \in \mcf{F}^\dagger$ ;~\eqref{eq:avg:t1e:pr:so:4} is by Lemma~\ref{lem:tbcut} and Equation~\eqref{eq:app:lai:type:2};
finally~\eqref{eq:avg:t1e:pr:so:5} is because $f \in \mcf{F}^*$.

With these observations, noting that Appendix~\ref{app:cc1:cc} produces deterministic codes, the following direct calculation yields Equation~\eqref{eq:avg:t1e:ub}:
\begin{align}
&\omega_{f^j,\varphi^j} \notag\\
&= \sum_{\mbf{z}^j \in \mbcf{Z}^j} \hspace{-3pt} 2^{-nj\kappa} \max_{k \in \mcf{K}} \prod_{i=1}^j p( \mbf{z}_i|k)  \label{eq:avg:t1e:pr:d:1} \\
&= \sum_{\mbf{z}^j \in \mbcf{Z}^j  } \hspace{-3pt}  2^{-nj\kappa} \max_{k \in \mcf{K}} \prod_{i=1}^j \left( \sum_{\subalign{\hat m &\in \mcf{\hat M},\\ \tilde m &\in \mcf{\tilde M}}} \sum_{\mbf{x}\in \mbcf{T}_{\rho}(\mbf{u}_{f}(\hat m,\tilde m, k)) } 2^{-n(\hat r + \tilde r)}  |\mbcf{T}_{\rho}(\mbf{u}_{f}(\hat m,\tilde m, k) )|^{-1} \mbf{q}(\mbf{z}_i|\mbf{x}) \right) \label{eq:avg:t1e:pr:d:2}\\
&= \sum_{\mbf{z}^j \in \mbcf{Z}^j  } \hspace{-3pt} 2^{-nj(\hat r + \tilde r+ \kappa)} \max_{k \in \mcf{K}} \prod_{i=1}^j \left( \sum_{\nu \in \mcf{P}_{n}(\mcf{Z}|\mcf{U};\sigma\tau) }  \hspace{-15pt} |\mcf{\bar M}(\mbf{z}_i,\nu,k)| 2^{-n[\mathbb{H}(\nu|\sigma\tau) + \mathbb{F}_n(\nu||q,\rho|\sigma\tau) ] + O(\logt n)}   \right)  \label{eq:avg:t1e:pr:d:3}  \\
&= \sum_{\mbf{z}^j \in \mbcf{Z}^j  } \hspace{-3pt}  2^{-nj(\hat r + \tilde r+ \kappa)} \max_{k \in \mcf{K}} \prod_{i=1}^j \left( \sum_{\nu \in \mcf{P}_{n}(\mcf{Z}|\mcf{U};\sigma\tau) } \hspace{-15pt}1\{ \mbf{z}_i \in \mbcf{Z}_{f}(\nu,k) \}  2^{-n\left[\mathbb{H}(\nu|\sigma\tau) + \mathbb{F}_n(\nu||q,\rho|\sigma\tau) - |\mathbb{S}_{\hat r,\tilde r}(\nu_i|\sigma,\tau)|^+ \right] + O(\logt n)}   \right) \label{eq:avg:t1e:pr:d:4} \\
&=    \sum_{\nu^j \in \bigtimes_{i=1}^j \mcf{P}_{n}(\mcf{Z}|\mcf{U};\sigma\tau)}  2^{-n\sum_{i=1}^j \left[\hat r + \tilde r + \kappa + \mathbb{H}(\nu_i|\sigma\tau) + \mathbb{F}_n(\nu_i||q,\rho|\sigma\tau) - |\mathbb{S}_{\hat r,\tilde r}(\nu_i|\sigma,\tau)|^+ \right] + O(j \logt n)} \notag \\
&\hspace{25pt} \cdot \sum_{\mbf{z}^j\in \mbcf{Z}^j }   \max_{k \in \mcf{K}} \prod_{i=1}^j 1\{ \mbf{z}_i \in \mbcf{Z}_{f}(\nu,k) \} \label{eq:avg:t1e:pr:d:5} \\
&\leq  \max_{\nu^j \in \bigtimes_{i=1}^j \mcf{P}_{n}(\mcf{Z}|\mcf{U};\sigma\tau)}  2^{-n \left[  \left| j \kappa + \sum_{i=1}^j \left| \mathbb{S}_{\hat r,\tilde r}(\nu_i|\sigma,\tau)\right|^- \right|^+  + \sum_{i=1}^j \mathbb{F}_n(\nu_i||q,\rho|\sigma\tau) \right] + O(j\logt n)} \label{eq:avg:t1e:pr:d:6};
\end{align}
where~\eqref{eq:avg:t1e:pr:d:1} is because $\mbf{Z}_1,\dots,\mbf{Z}_j$ are independent given $k \in \mcf{K}$, and because the key is uniformly distributed over $\{ 1,\dots,2^{nj\kappa}\};$~\eqref{eq:avg:t1e:pr:d:2} is by the law of total probability and definition of $\mbf{Z}_1,\dots,\mbf{Z}_j;$~\eqref{eq:avg:t1e:pr:d:3} is by further subdividing the summation based upon the empirical distribution of $\mbf{z}_i|\mbf{u}_{f}(\hat m,\tilde m,k)$, using Lemma~\ref{lem:btx}, and then performing the summations over $\hat m$ and $\tilde m$;~\eqref{eq:avg:t1e:pr:d:4} is because $f \in \mcf{F}^*$;~\eqref{eq:avg:t1e:pr:d:5} is a result of exchanging the summations and product; finally,~\eqref{eq:avg:t1e:pr:d:6} is by~\eqref{eq:avg:t1e:pr:so} and because of Equation~\eqref{eq:app:lai:type:4}.

\subsubsection{Probability of randomly choosing a code in $\mcf{F}^*\cap \mcf{F}^\dagger$}

Once again, let RV $F$ denote the randomly chosen encoder from Appendix~\ref{app:cc1:cc}.
To verify Equation~\eqref{eq:avg:t1e:lbp}, it will be helpful to first show that 
\begin{align}
\Pr \left( F \notin \mcf{F}_1 \right) < 2\abs{\mcf{Z}}^n |\mcf{P}_n(\mcf{Z}|\mcf{U};\sigma\tau)| e^{-n^2/4} = e^{-n^2/4 + O(n)} \label{eq:avg:suff:1.1}
\end{align}
where $\mcf{F}_1$ is the set of all encoders $f$ such that
\begin{align} 
\frac{1}{n} \logt \abs{\set{\hat m : \mbf{z} \in \mbcf{T}_{\nu\sigma}(\mbf{w}_{f}(\hat m)) }}  &=  \left| \hat r - \mathbb{I}(\nu\sigma,\tau) + O(n^{-1} \logt n) \right|^+ + O(n^{-1}\logt n)  \notag \\
&=  \left| \hat r - \mathbb{I}(\nu\sigma,\tau)  \right|^+ + O(n^{-1}\logt n) 
\end{align}
for each $\mbf{z} \in \bigcup_{\hat m\in \mcf{\hat M}} \mbcf{T}_{\nu\sigma}(\mbf{w}_{f}(\hat m))$ and $\nu \in \mcf{P}_{n}(\mcf{Z}|\mcf{U};\sigma\tau)$.
To show Equation~\eqref{eq:avg:suff:1.1}, observe that $$\abs{\set{\hat m : \mbf{z} \in \mbcf{T}_{\nu\sigma}(\mbf{w}_{F}(\hat m)) }} = \sum_{\hat m \in \mcf{\hat M}} \idc{\mbf{z} \in \mbcf{T}_{\nu\sigma}(\mbf{w}_{F}(\hat m))} $$
for each $\mbf{z} \in \mbcf{Z}$ and $\nu \in \mcf{P}(\mcf{Z}|\mcf{U}).$
Furthermore, for each $\hat m \in \mcf{\hat M}$, the $\idc{\mbf{z} \in \mbcf{T}_{\nu\sigma}(\mbf{w}_{F}(\hat m))}$ are Bernoulli $(2^{-n\mathbb{I}(\nu\sigma,\tau) + O(\logt n)})$ RVs by Lemma~\ref{lem:obvious}, and independent by definition.  
Therefore, Equation~\eqref{eq:avg:suff:1.1} follows by Corollary~\ref{cor:ck}, the union bound, and because the restriction to $\mbf{z} \in \bigcup_{\hat m \in \mcf{\hat M}} \mbcf{T}_{\nu\sigma}(\mbf{w}_{F}(\hat m))$ guarantees that $\{\hat m : \mbf{z} \in \mbcf{T}_{\nu\sigma}(\mbf{w}_{f}(\hat m))\}\neq \emptyset$.

Similar proofs will be used to show 
\begin{align} 
\Pr \left( F \notin \mcf{F}^* \right) \leq e^{-n^2/4 + O(n)}, \label{eq:avg:suff:sat:p:end}
\end{align}
and
\begin{align}
\Pr \left( F \notin \mcf{F}^\dagger\right) \leq e^{-n^2/4 + O(n)}; \label{eq:cc1:fdagger:0}
\end{align}
Equation~\eqref{eq:avg:t1e:lbp} directly follows from Equations~\eqref{eq:avg:suff:sat:p:end},~\eqref{eq:cc1:fdagger:0}, and the union bound.
In order to prove both Equation~\eqref{eq:avg:suff:sat:p} and~\eqref{eq:cc1:fdagger:0}, it will be helpful to fix the values of $\mbf{w}_{F}$, while leaving $\mbf{u}_{F}$ random.
It will also be helpful to ensure that the code will be contained in $\mcf{F}_1$ for the fixed values of $\mbf{w}_{F}$.
To this end, for each $\mbf{w}^{|\mcf{\hat M}|} \defn \bigtimes_{i=1}^{|\mcf{\hat M}|} \mbf{w}_i \in \bigtimes_{i=1}^{|\mcf{\hat M}|} \mbcf{T}_{\tau}$, let $F_{\mbf{w}^{|\mcf{\hat M}|}} = F|\{F \in \mcf{F}(\mbf{w}^{|\mcf{\hat M}|})\}$, where $\mcf{F}(\mbf{w}^{|\mcf{\hat M}|})$ is the set of encoders $f$ such that $ \mbf{w}_{f}(i) = \mbf{w}_{i}$ for each $i \in \mcf{\hat M}$.
It is important to observe that $\mcf{F}(\mbf{w}^{|\mcf{\hat M}|})$ is either a subset of, or mutually exclusive with, $\mcf{F}_1$ for each $\mbf{w}^{|\mcf{\hat M}|}$.
The set of $\mbf{w}^{|\mcf{\hat M}|}$ for which $\mcf{F}(\mbf{w}^{|\mcf{\hat M}|}) \subseteq \mcf{F}_1$ will be helpful to distinguish, thus let $\mbcf{W}_1 \defn \{ \mbf{w}^{|\mcf{\hat M}|} : \mcf{F}(\mbf{w}^{|\mcf{\hat M}|}) \subseteq \mcf{F}_1\}.$

Now, to prove Equation~\eqref{eq:avg:suff:sat:p:end}, first observe that 
\begin{align}
\Pr \left( F \notin \mcf{F}^* \right)&=  \sum_{\mbf{w}^{|\mcf{\hat M}|} \in \bigtimes_{i=1}^{|\mcf{\hat M}|} \mbcf{T}_{\tau}(\mbf{w}) } \Pr \left( F_{\mbf{w}^{|\mcf{\hat M}|}} \notin \mcf{F}^* \right) \Pr \left( F \in \mcf{F}(\mbf{w}^{|\mcf{\hat M}|})\right)  \notag     \\
& \leq  \Pr \left( F \notin \mcf{F}_1\right) + \sum_{\mbf{w}^{|\mcf{\hat M}|} \in \mbcf{W}_1} \Pr \left( F_{\mbf{w}^{|\mcf{\hat M}|}} \notin \mcf{F}^* \right) \Pr \left( F \in \mcf{F}(\mbf{w}^{|\mcf{\hat M}|})\right)  \notag     \\  
& \leq  \Pr \left( F \notin \mcf{F}_1\right) + \max_{\mbf{w}^{|\mcf{\hat M}|} \in \mbcf{W}_1} \Pr \left( F_{\mbf{w}^{|\mcf{\hat M}|}} \notin \mcf{F}^* \right) 
\end{align}
because the events $F\in \mcf{F}(\mbf{w}^{|\mcf{\hat M}|})$ are mutually exclusive for each $\mbf{w}^{|\mcf{\hat M}|}$.
At the same time
\begin{align}
\abs{\mcf{\bar M}_{F_{\mbf{w}^{|\mcf{\hat M}|}}}(\mbf{z},\nu,k)}  &=  \sum_{\subalign{\hat m &\in \mcf{\hat M} \\ \tilde m &\in \mcf{\tilde M}} }   \idc{\mbf{z} \in \mbcf{T}_{\nu}(\mbf{u}_{F_{\mbf{w}^{|\mcf{\hat M}|}}}(\hat m,\tilde m, k))}  ,
\label{eq:avg:stuff:1}
\end{align}
for each $\mbf{w}^{|\mcf{\hat M}|}$, $\mbf{z} \in \mbcf{Z}$, and $\nu \in \mcf{P}_n(\mcf{Z}|\mcf{U};\sigma\tau)$.
Note $\idc{\mbf{z} \in \mbcf{T}_{\nu}(\mbf{u}_{F_{\mbf{w}^{|\mcf{\hat M}|}}}(\hat m,\tilde m, k))}$ is a Bernoulli$(2^{-n \mathbb{I}(\nu,\sigma|\tau) + O(\logt n)})$ RV for all $\hat m$ such that $\mbf{z} \in \mbcf{T}_{\nu\sigma}(\mbf{w}_{\hat m})$ by Lemma~\ref{lem:obvious}, while $\idc{\mbf{z} \in \mbcf{T}_{\nu}(\mbf{u}_{F_{\mbf{w}^{|\mcf{\hat M}|}}}(\hat m,\tilde m, k))} = 0 $ for all $\hat m$ such that $\mbf{z} \notin \mbcf{T}_{\nu\sigma}(\mbf{w}_{\hat m})$ by Lemma~\ref{lem:unique}. 
What's more, each value of $\idc{\mbf{z} \in \mbcf{T}_{\nu}(\mbf{u}_{F_{\mbf{w}^{|\mcf{\hat M}|}}}(\hat m,\tilde m, k))}$ in the sum is independent by the code construction.
Therefore,
\begin{align}  
 \max_{\mbf{w}^{|\mcf{\hat M}|} \in \mbcf{W}_1} \Pr \left( F_{\mbf{w}^{|\mcf{\hat M}|}} \notin \mcf{F}^* \right)  \leq |\mcf{Z}|^n |\mcf{K}| |\mcf{P}_{n}(\mcf{Z}|\mcf{U};\sigma\tau)|e^{-n^2/4 + O(n)} = e^{-n^2/4 + O(n)}  \label{eq:avg:suff:sat:p}
\end{align}
follows by Corollary~\ref{cor:ck} and the union bound, since $\abs{\mcf{\bar M}_{F_{\mbf{w}^{|\mcf{\hat M}|}}}(\mbf{z},\nu,k)}$ is a sum of $|\set{\hat m : \mbf{z} \in \mbcf{T}_{\nu\sigma}(\mbf{w}_{\hat m})}| 2^{n \tilde r} = 2^{n[ \tilde r +|\hat r - \mathbb{I}(\nu\sigma,\tau)|^+ ] + O(\logt n)} $ independent Bernoulli$(2^{-n \mathbb{I}(\nu,\sigma|\tau) + O(\logt n)})$ RVs for each $\mbf{z} \in \bigcup_{\hat m} \mbcf{T}_{\nu\sigma}(\mbf{w}_{\hat m})$.

Similarly to prove Equation~\eqref{eq:cc1:fdagger:0}, as before
\begin{align}
 \Pr \left( F \notin \mcf{F}^\dagger \right)
& \leq   \Pr \left( F \notin \mcf{F}_1 \right)  + \max_{\mbf{w}^{|\mcf{\hat M}|} \in  \mbcf{W}_1 }  \Pr \left( F_{\mbf{w}^{|\mcf{\hat M}|}} \notin \mcf{F}^\dagger \right)    \label{eq:cc1:fdagger:1}.
\end{align}
For each $\mbf{z}^j \in \bigtimes_{i=1}^j \bigcup_{\hat m \in \mcf{\hat M}} \mbcf{T}_{\nu_i\sigma}(\mbf{w}_{\hat m})$ and $\nu^j \in \bigtimes_{i=1}^j \mcf{P}_n(\mcf{Z}|\mcf{U};\sigma\tau)$, the term $|\mcf{K}_{F_{\mbf{w}^{|\mcf{\hat M}|}}}(\mbf{z}^j, \nu^j)| $ is a sum of indicator RVs, specifically
\begin{align}
|\mcf{K}_{F_{\mbf{w}^{|\mcf{\hat M}|}}}( \mbf{z}^j, \nu^j)| = \sum_{k \in \mcf{K}} \idc{ k \in \mcf{K}_{F_{\mbf{w}^{|\mcf{\hat M}|}}}(\mbf{z}^j,  \nu^j)}. \label{eq:cc1:needsummands}
\end{align}
Each $\idc{ k \in \mcf{K}_{F_{\mbf{w}^{|\mcf{\hat M}|}}}( \mbf{z}^j,  \nu^j)}$ is Bernoulli$(b)$, where
\begin{align}
- \frac{1}{n} \logt b \leq -\frac{j}{n} \logt j +  \sum_{i=1}^j | \tilde r + |\hat r - \mathbb{I}(\nu_i \sigma,\tau)|^+ - \mathbb{I}(\nu_i,\sigma|\tau) |^- + O(n^{-1}\logt n),
\end{align}
by Lemma~\ref{lem:circum} because 
\begin{align}
\idc{ k \in \mcf{K}_{F_{\mbf{w}^{|\mcf{\hat M}|}}}( \mbf{z}^j,  \nu^j)} &= \idc{ \cap_{i=1}^j \{ \mbf{z}_i \in \bigcup_{\subalign{\hat m &\in \mcf{\hat M}, \\ \tilde m & \in \mcf{\tilde M}}} \mbcf{T}_{\nu_i}(\mbf{u}_{F_{\mbf{w}^{|\mcf{\hat M}|}}}(\hat m,\tilde m , k))\} }     \notag  \\
&= \idc{ \cap_{i=1}^j \{ \mbf{z}_i \in \bigcup_{\subalign{\hat m &\in \mcf{\hat M}: \mbf{z}_i \in \mbcf{T}_{\nu_i\sigma}(\mbf{w}_{\hat m})  , \\ \tilde m & \in \mcf{\tilde M}}} \mbcf{T}_{\nu_i}(\mbf{u}_{F_{\mbf{w}^{|\mcf{\hat M}|}}}(\hat m,\tilde m , k))\} }  \notag,
\end{align}
$\mbf{w}^{|\mcf{\hat M}|} \in  \mbcf{W}_1$, and because $2^{n\tilde r} \geq 8j \ln j$ for $\tilde r > 0$ and large enough $n$.
Furthermore, $\idc{ k \in \mcf{K}_{F_{\mbf{w}^{|\mcf{\hat M}|}}}( \mbf{z}^j,  \nu^j)}$ are independent for different $k$ since
 \begin{align}
\idc{ k \in \mcf{K}_{F_{\mbf{w}^{|\mcf{\hat M}|}}}( \mbf{z}^j,  \nu^j)} &= \idc{ \cap_{i=1}^j \{ \mbf{z}_i \in \bigcup_{\subalign{\hat m &\in \mcf{\hat M}, \\ \tilde m & \in \mcf{\tilde M}}} \mbcf{T}_{\nu_i}(\mbf{u}_{F_{\mbf{w}^{|\mcf{\hat M}|}}}(\hat m,\tilde m , k))\} } \\
&= \prod_{i=1}^j \left[ 1 - \prod_{\subalign{\hat m &\in \mcf{\hat M}, \\ \tilde m & \in \mcf{\tilde M}}} 1 \left\{  \mbf{z}_i \notin \mbcf{T}_{\nu_i} (\mbf{u}_{F_{\mbf{w}^{|\mcf{\hat M}|}}}(\hat m,\tilde m,k)) \right\} \right],
\end{align}
and $\mbf{u}_{F_{\mbf{w}^{|\mcf{\hat M}|}}}(\hat m,\tilde m, k)$ is independently chosen from $\mbcf{T}_{\sigma} (\mbf{w}_{\hat m})$ for each value of $k\in \mcf{K}$.
Now, 
\begin{align}
\max_{\mbf{w}^{|\mcf{\hat M}|} \in  \mbcf{W}_1 }  \Pr \left( F_{\mbf{w}^{|\mcf{\hat M}|}} \notin \mcf{F}^\dagger \right) \leq |\mcf{Z}|^{nj}  |\mcf{P}_{n}(\mcf{Z}|\mcf{U};\sigma\tau)|^j e^{-n^2/4 } =e^{-n^2/4 + O(n) } \label{eq:cc1:fdagger:2}
\end{align}
follows by Corollary~\ref{cor:ck} and the union bound; hence Equation~\eqref{eq:cc1:fdagger:0} is due to Equations~\eqref{eq:avg:suff:1.1},~\eqref{eq:cc1:fdagger:1}, and~\eqref{eq:cc1:fdagger:2}.

\end{IEEEproof}

\subsection{Proof of Theorem~\ref{thm:lai_strong}} \label{app:lai_strong_proof}

\begin{IEEEproof} 

First fix finite sets $\mcf{U}$ and $\mcf{W}$, and distributions $\rho \in \mcf{P}(\mcf{X}|\mcf{U})$, $\sigma \in \mcf{P}(\mcf{U}\gg\mcf{W})$, and $\tau \in \mcf{P}(\mcf{W})$ such that $\mathbb{L}(t,q|\rho,\sigma,\tau) > 0.$
The proof follows by first showing that if $(r,\alpha,\kappa,\hat r, \tilde r , \tilde \kappa) \in \mcf{R}_{\gamma_1,\gamma_2}(\rho,\sigma,\tau)$, where
\begin{align}
&\mcf{R}_{\gamma_1,\gamma_2}(\rho,\sigma,\tau) \notag \\
&\defn \set{\begin{array}{ll} 
 \hspace{-7pt}(r,\alpha,\kappa,\hat r, \tilde r, \tilde  \kappa)  :& \\
r - \tilde r - \hat r &\leq 0\\
\tilde \kappa - \kappa  &\leq 0 \\
\tilde r + j \tilde \kappa &= \mathbb{I}(t\rho,\sigma|\tau) - \gamma_1 \\
\tilde r + \hat r + j \tilde \kappa &= \mathbb{I}(t\rho,\sigma\tau) - \gamma_2 \\
\alpha - j\tilde \kappa  &\leq 0 \\
j\tilde \kappa  &\leq  \mathbb{L}_{\gamma_1,\gamma_2} (t,q|\rho,\sigma,\tau)  
\end{array}}
\end{align}
and
\[
\mathbb{L}_{\gamma_1,\gamma_2}(t,q|\rho,\sigma,\tau)\defn \min_{\nu \in \mcf{P}(\mcf{Z}|\mcf{U}) } \mathbb{F}(\nu||q,\rho|\sigma\tau) + \mathbb{S}_{-\gamma_1,-\gamma_2+\gamma_1}(t\rho, \nu| \sigma,\tau)
\]
for some strictly positive $\gamma_1$ and $\gamma_2$, then $(r,\alpha, \kappa)$ is achievable.
From there, that all $(r,\alpha,\kappa) \in \mcf{R}^*_{\gamma_1,\gamma_2}(\rho,\sigma,\tau)$, where 
\begin{align}
&\mcf{R}^*_{\gamma_1,\gamma_2}(\rho,\sigma,\tau) \notag \\
&\defn \left\{ \begin{array}{ll} 
\hspace{-7pt}(r,\alpha,\kappa)  :& \\
\alpha - j\kappa &\leq 0\\
\alpha + r &\leq \mathbb{I}(t\rho,\sigma\tau) - \gamma_2 \\
\alpha &\leq \mathbb{I}(t\rho,\sigma|\tau)-\gamma_1 \\
\alpha &\leq \mathbb{L}_{\gamma_1,\gamma_2} (t,q|\rho,\sigma,\tau) 
\end{array}
\right\},
\end{align}
are achievable directly follows by applying Fourier-Motzkin elimination
(see~\cite[Appendix~D]{GNIT}) to $\mcf{R}_{\gamma_1,\gamma_2}(\rho,\sigma,\tau)$ in order to eliminate $\hat r$, $\tilde r$, and $\hat \kappa$.
Thus, as the final step, showing that $\bigcup_{\gamma_1,\gamma_2} \mcf{R}^*_{\gamma_1,\gamma_2}(\rho,\sigma,\tau) = \lim_{(\gamma_1,\gamma_2)\rightarrow (0,0)} \mcf{R}^*_{\gamma_1,\gamma_2}(\rho,\sigma,\tau)$ which yields the stated rate region.

For the first step, demonstrating that $(r,\alpha,\kappa,\hat r, \tilde r , \tilde \kappa) \in \mcf{R}_{\gamma_1,\gamma_2}(\rho,\sigma,\tau)$ implies $(r,\alpha,\kappa)$ is achievable can be done essentially via Theorem~\ref{thm:p_lai_strong}.
But notice, Theorem~\ref{thm:p_lai_strong} is concerned with constructing codes for a finite $n$, and clearly the empirical distributions of the code will vary with the value of $n$. 
Because of this, define the following approximating distributions and error terms
\begin{align*}
\tau_n &:= \argmin_{\tau' \in \mcf{P}_n(\mcf{W}) } \abs{\tau' - \tau}\\
\sigma_n &:= \argmin_{\sigma' \in \mcf{P}_n(\mcf{U}\gg\mcf{W}): \text{ if } \sigma(u|w) = 0 \text{ then } \sigma'(u|w) = 0} \abs{\sigma' - \sigma}\\
\rho_n &:= \argmin_{\rho' \in \mcf{P}_n(\mcf{X}|\mcf{U};\sigma_n\tau_n) } \abs{\rho' - \rho}\\
\Delta_{1,n} &:= \sup_{\mu \in \mcf{P}(\mcf{Y}|\mcf{U}) } \sup_{n'>n}  \abs{\mathbb{F}_n(\mu||t,\rho_n|\sigma_n\tau_n) - \mathbb{F}(\mu||t,\rho |\sigma\tau) }\\
\Delta_{2,n} &:= \sup_{\nu \in  \mcf{P}(\mcf{Z}|\mcf{U})} \sup_{n'>n}  \abs{\mathbb{F}_n(\nu|| q,\rho_n|\sigma_n\tau_n) - \mathbb{F}(\nu||q,\rho|\sigma\tau) }\\
\Delta_{3,n} &:= \sup_{\mu \in \mcf{P}(\mcf{Y}|\mcf{U}) \bigcup \mcf{P}(\mcf{Z}|\mcf{U})} \sup_{n'>n}  \abs{\mathbb{I}(\mu,\sigma_{n'}|\tau_{n'}) - \mathbb{I}(\mu,\sigma|\tau) } \\
\Delta_{4,n} &:= \sup_{\mu \in \mcf{P}(\mcf{Y}|\mcf{U}) \bigcup \mcf{P}(\mcf{Z}|\mcf{U})} \sup_{n'>n}  \abs{\mathbb{I}(\mu,\sigma_{n'}\tau_{n'}) - \mathbb{I}(\mu,\sigma\tau) } \\
\Delta_{n} &:= \max_{i \in \set{1,2,3,4}} \Delta_{i,n}.
\end{align*}
It is important to note that $\Delta_{n}$ goes to zero monotonically due to the continuity of entropy.
Next, suppose that $(r,\alpha,\kappa,\hat r, \tilde r , \tilde \kappa) \in \mcf{R}_{\gamma_1,\gamma_2}(\rho,\sigma,\tau)$, and observe that for all large enough $n$, there exists a $(r,\alpha_n,\kappa,\epsilon_n,n)$-AA code such that 
\begin{align}
-n\logt \epsilon_{n}  &\geq  \min_{\mu \in \mcf{P}_n(\mcf{Y}|\mcf{U};\sigma_n\tau_n)}  \mathbb{ F}_n(\mu||t,\rho_n|\sigma_n\tau_n) - \left| j\tilde \kappa + \mathbb{S}_{\hat r,\tilde r}(\mu|\sigma_n,\tau_n)\right|^-  + O(n^{-1}\logt n) ,  \label{eq:lais:me2} 
\end{align}
and 
\begin{align}
\alpha_n \geq \min_{\nu^j \in \bigtimes_{i=1}^j \mcf{P}_{n}(\mcf{Z}|\mcf{U};\sigma\tau) } \left| j \tilde \kappa + \sum_{i=1}^j |\mathbb{S}_{\hat r,\tilde r}(\nu_i|\sigma_n,\tau_n)|^{-} \right|^+ + \sum_{i=1}^j \mathbb{F}_n(\nu_i||q,\rho_n|\sigma_n\tau_n)  + O(j n^{-1}\logt n) , \label{eq:lais:t1e}
\end{align}
by Theorem~\ref{thm:p_lai_strong} since $r\leq \hat r + \tilde r$ and $\kappa \geq \tilde \kappa$ are requirements of $(r,\alpha,\kappa,\hat r,\tilde r, \tilde \kappa) \in \mcf{R}_{\gamma_1,\gamma_2}(\rho,\sigma,\tau)$.
Hence, proving that $\lim_{n \rightarrow \infty} \epsilon_n = 0$ and $\lim_{n\rightarrow \infty} \alpha_n \geq \alpha$ also proves $(r,\alpha,\kappa)$ is achievable.
Towards the goal of proving $\lim_{n \rightarrow \infty} \epsilon_n = 0$, using Equation~\eqref{eq:lais:me2} it follows that
\begin{align}
\epsilon_{n}  &\leq  \max_{\mu \in \mcf{P}(\mcf{Y}|\mcf{U};\sigma_n\tau_n)} 2^{- n\left( \mathbb{F}_n(\mu||t,\rho_n|\sigma_n\tau_n) - \left|\mathbb{S}_{-\gamma_1,-\gamma_2+\gamma_1}(t\rho,\mu|\sigma_n,\tau_n) \right|^- \right) + O(\logt n)} , \notag\\
&\leq \max_{\mu \in \mcf{P}(\mcf{Y}|\mcf{U};\sigma_n\tau_n)} 2^{- n\left( \mathbb{F}(\mu||t,\rho|\sigma\tau) -  \left|\mathbb{S}_{-\gamma_1,-\gamma_2+\gamma_1}(t\rho,\mu|\sigma,\tau) \right|^- \right) + 3n \Delta_n + O(\logt n)} \notag\\
&\leq \max_{\mu \in \mcf{P}(\mcf{Y}|\mcf{U})}  2^{- n\left( \mathbb{F}(\mu||t,\rho|\sigma\tau) -  \left|\mathbb{S}_{-\gamma_1,-\gamma_2+\gamma_1}(t\rho,\mu|\sigma,\tau) \right|^- \right) + 3n \Delta_n + O(\logt n)} \label{eq:lais:eub2}  
\end{align}
by the definition of $\Delta_n$.
If the maximum of Equation~\eqref{eq:lais:eub2} is achieved by $\mu = t\rho$, then Equation~\eqref{eq:lais:eub2} gives
\begin{equation}
\lim_{n\rightarrow \infty} \epsilon_{n} \leq \lim_{n\rightarrow \infty} 2^{-n \left[ \min(\gamma_1,\gamma_2) -  3\Delta_n - O(n^{-1}\logt n)\right] } \leq \lim_{n\rightarrow \infty} 2^{-\frac{n}{2} \min(\gamma_1,\gamma_2)  } = 0,
\end{equation}
since $\gamma_1$ and $\gamma_2$ are strictly positive constants, while $\lim_{n\rightarrow \infty} \Delta_n = 0$ and $\lim_{n\rightarrow \infty} n^{-1} \logt n =0$. 
Alternatively, if the maximum of Equation~\eqref{eq:lais:eub2} is not achieved when $\mu = t\rho$ then $\mathbb{F}(\mu||t,\rho|\sigma\tau) >0$, and accordingly Equation~\eqref{eq:lais:eub2} yields
\begin{align}
\lim_{n\rightarrow \infty} \epsilon_{n} \leq \lim_{n\rightarrow \infty} 2^{-n \left[ \mathbb{F}(\mu||t,\rho|\sigma\tau)  - 3 \Delta_n - O( n^{-1}\logt n) \right] } \leq \lim_{n\rightarrow \infty}  2^{-\frac{n}{2}  \mathbb{F}(\mu||t,\rho|\sigma\tau)   }  = 0.
\end{align}
In both cases, $\lim_{n\rightarrow \infty} \epsilon_n = 0.$
The inequality $\lim_{n \rightarrow \infty} \alpha_n \geq \alpha$ from Equation~\eqref{eq:lais:t1e} is derived as follows
\begin{align}
&\alpha_{n} + O(j n^{-1}\logt n)  + 3j \Delta_n  \notag\\
&\geq \min_{\subalign{\nu_1 &\in \mcf{P}_{n}(\mcf{Z}|\mcf{U};\sigma\tau) ,\vspace{-5pt}\\ &\vdots \\ \nu_j &\in \mcf{P}_{n}(\mcf{Z}|\mcf{U};\sigma\tau) }} \left| j \tilde \kappa + \sum_{i=1}^j |\mathbb{S}_{\hat r,\tilde r}(\nu_i|\sigma_n,\tau_n)|^{-} \right|^+ + \sum_{i=1}^j \mathbb{F}_n(\nu_i||q,\rho_n|\sigma_n\tau_n)  + 3j \Delta_n ,\notag\\
&\geq \min_{\subalign{\nu_1 &\in \mcf{P}(\mcf{Z}|\mcf{U}) ,\vspace{-5pt}\\ &\vdots \\ \nu_j &\in \mcf{P}(\mcf{Z}|\mcf{U}) }} \left| j \tilde \kappa + \sum_{i=1}^j |\mathbb{S}_{\hat r,\tilde r}(\nu_i|\sigma,\tau)|^{-} \right|^+ + \sum_{i=1}^j \mathbb{F}(\nu_i||q,\rho|\sigma\tau)  \label{eq:lais:t1eub1} \\ 
&=\min_{\subalign{\nu_1 &\in \mcf{P}(\mcf{Z}|\mcf{U}) ,\vspace{-5pt}\\ &\vdots \\ \nu_j &\in \mcf{P}(\mcf{Z}|\mcf{U}) }} \left| j \tilde \kappa - \sum_{i=1}^j |j \tilde \kappa - \mathbb{S}_{-\gamma_1,-\gamma_2+\gamma_1}(t\rho,\nu_i|\sigma,\tau)|^{+} \right|^+ +\sum_{i=1}^j \mathbb{F}(\nu_i||q,\rho|\sigma\tau)  \label{eq:lais:t1eub2} \\
&\geq \min_{\subalign{\nu_1 &\in \mcf{P}(\mcf{Z}|\mcf{U}) ,\vspace{-5pt}\\ &\vdots \\ \nu_j &\in \mcf{P}(\mcf{Z}|\mcf{U}) }} \left| j \tilde \kappa - \sum_{i=1}^j \mathbb{F}(\nu_i||q,\rho|\sigma\tau)  \right|^+ + \sum_{i=1}^j \mathbb{F}(\nu_i||q,\rho|\sigma\tau)   \label{eq:lais:t1eub3} \\
&\geq \min_{\subalign{\nu_1 &\in \mcf{P}(\mcf{Z}|\mcf{U}) ,\vspace{-5pt}\\ &\vdots \\ \nu_j &\in \mcf{P}(\mcf{Z}|\mcf{U}) }}  \max \left( j \tilde \kappa ,\sum_{i=1}^j \mathbb{F}(\nu_i||q,\rho|\sigma\tau) \right)  \geq j \tilde \kappa \geq \alpha    \label{eq:lais:t1eub4};
\end{align}
where~\eqref{eq:lais:t1eub1} is by the definition of $\Delta_n;$~\eqref{eq:lais:t1eub2} is because 
\begin{align*}
\mathbb{S}_{\hat r,\tilde r}(\nu_i|\sigma,\tau) &= \tilde r - \mathbb{I}(\nu_i,\sigma|\tau) + |\hat r - \mathbb{I}(\nu_i \sigma ,\tau)|^+ \\
&= \mathbb{I}(t\rho,\sigma|\tau) - \gamma_1 - j\tilde \kappa - \mathbb{I}(\nu_i,\sigma|\tau) + |\mathbb{I}(t \rho\sigma ,\tau) - \gamma_2 + \gamma_1  - \mathbb{I}(\nu_i \sigma ,\tau)|^+ \\
&= -(j\tilde \kappa - \mathbb{S}_{-\gamma_1 ,-\gamma_2 + \gamma_1}(t\rho, \nu_i|\sigma,\tau)  ); 
\end{align*}
\eqref{eq:lais:t1eub3} is because $j\tilde k \leq \mathbb{F}(\nu||q,\rho|\sigma\tau) +   \mathbb{S}_{-\gamma_1,-\gamma_2+\gamma_1}(t\rho,\nu|\sigma,\tau) $ and $ \mathbb{F}(\nu||q,\rho|\sigma\tau) \geq 0$ for all $\nu \in \mcf{P}(\mcf{Z}|\mcf{U})$;
finally~\eqref{eq:lais:t1eub4} is because  if $a\geq b$ then $|a-b|^+ +b = a$ while if $a< b$ then $|a-b|^++b = b$.
Hence, $\lim_{n \rightarrow \infty} \alpha_n \geq \alpha$ from Equation~\eqref{eq:lais:t1eub4} since $\lim_{n\rightarrow\infty} n^{-1}\logt n \rightarrow 0$ and $\lim_{n\rightarrow \infty} \Delta_{n} = 0.$

Having shown that $(r,\alpha,\kappa) \in \mcf{R}^*_{\gamma_1,\gamma_2}(\rho,\sigma,\tau)$ are achievable, what remains is to show that $$\bigcup_{\gamma_1,\gamma_2} \mcf{R}^*_{\gamma_1,\gamma_2}(\rho,\sigma,\tau) = \lim_{(\gamma_1,\gamma_2) \rightarrow(0,0)} \mcf{R}^*_{\gamma_1,\gamma_2}(\rho,\sigma,\tau).$$
This can be proven by showing
\begin{equation}\label{eq:lais:first_reduction}
\mcf{R}^*_{\gamma_1,\gamma_2}(\rho,\sigma ,\tau) \subseteq \mcf{R}^*_{\gamma_1^*,\gamma_2^*}(\rho,\sigma,\tau)
\end{equation}
for all $0 < \gamma_1^* \leq \gamma_1$ and $0 < \gamma_2^*\leq \gamma_2$.  
To this end, suppose that $(r,\alpha,\kappa) \in \mcf{R}^*_{\gamma_1,\gamma_2}(\rho,\sigma,\tau)$ and immediately observe that
\begin{align}
\alpha - j\kappa &\leq 0, \label{eq:lais:toes1}\\
\alpha + r &\leq \mathbb{I}(t\rho,\sigma\tau) - \gamma_2 \leq \mathbb{I}(t\rho,\sigma\tau) - \gamma_2^*, \\
 \alpha &\leq \mathbb{I}(t\rho,\sigma|\tau) - \gamma_1 \leq \mathbb{I}(t\rho,\sigma\tau) - \gamma_1^*. \label{eq:lais:toes3}
\end{align}
Additionally,  
\begin{align}
\alpha \leq  \mathbb{L}_{\gamma_1^*,\gamma_2^*}(t,q|\rho,\sigma,\tau) \label{eq:lais:toes4}
\end{align}
since
\begin{align*}
a - \gamma_1 + | b + \gamma_1 - \gamma_2|^+ = a + \max(b-\gamma_2,-\gamma_1) \leq a + \max(b-\gamma_2^*,-\gamma_1^*) = a - \gamma_1^* + | b + \gamma_1 - \gamma_2^*|^+
\end{align*}
for all real numbers $a$ and $b$. 
Combining Equations~\eqref{eq:lais:toes1}--\eqref{eq:lais:toes4} proves that $(r,\alpha,\kappa) \in \mcf{R}^*_{\gamma_1^*,\gamma_2^*}(\rho,\sigma,\tau).$

\end{IEEEproof}

\section{Proof of Theorem~\ref{thm:-t2}}\label{sec:-t}\label{app:cc2}

Similar to Appendix~\ref{sec:p_lai_strong}, Appendices~\ref{app:cc2:cc} through \ref{app:cc2:t1e} are dedicated to proving the following theorem:
\begin{theorem}\label{thm:cc2:n}
If there exists an $(r,\alpha, \kappa,\epsilon,j,n)$-AA code for DM-ASC$(t,q)$, and $j \sqrt{\epsilon} < 1,$ then for large enough $n$ and positive real number $\beta \leq r$ there also exists a $$(r-\beta,\alpha + \beta - 2n^{-1} \logt ne , \kappa + [1 + j^{-1}]  \beta ,\sqrt{\epsilon},j,n)-\mathrm{AA}~\text{code for DM-AIC}(t,q).$$ 
\end{theorem}

\begin{IEEEproof}

To prove Theorem~\ref{thm:cc2:n}, we use the random coding construction in Appendix~\ref{app:cc2:cc}. 
Appendix~\ref{app:cc2:cc}'s code construction, with probability greater than $1- j\sqrt{\epsilon}$, produces a code $(\tilde f^j,\tilde \varphi^j)$ where $\varepsilon_{\tilde f^j,\tilde \varphi^j} \leq \sqrt{\epsilon}$ as shown in Appendix~\ref{app:cc2:mea}.
Furthermore, with probability greater than $1- je^{-n^2/4+j(n+1)r}$, Appendix~\ref{app:cc2:cc}'s code construction produces a code such that $ \omega_{\tilde f^j,\tilde \varphi^j} < 2^{-n (\alpha+\beta)  + 2\logt ne }$, as shown in~\ref{app:cc2:t1e}.
For large enough $n$, these results guarantee the existence of a code satisfying the theorem statement since $j\sqrt{\epsilon}<1$ and $\lim_{n\rightarrow \infty} je^{-n^2/4+j(n+1)r} = 0.$

\subsection{Code Construction}\label{app:cc2:cc}

For a given positive real number $\beta \leq r$, we shall use the following construction to transform codes to send $j$-messages chosen uniformly from $\mcf{M}\defn \{ 1,\dots, 2^{nr}\}$ with a secret key drawn uniformly from $\mcf{K}_1 \defn \{ 1, \dots, 2^{nj\kappa}\}$, into codes to send $j$-messages chosen uniformly from $\mcf{\tilde M} \defn \{ 1,\dots, 2^{n(r-\beta)} \}$ with a secret key drawn uniformly from $\mcf{K}_1 \times \mcf{K}_2,$ where $\mcf{K}_2 \defn \{ 1,\dots, 2^{n (j+1)\beta} \}.$
The starting codes will be denoted $(f_i,\varphi_i) \in \mcf{P}(\mbcf{X}|\mcf{M}, \mcf{K}_1) \times \mcf{P}(\mcf{M} \cup \set{\mbf{!}} | \mbcf{Y}, \mcf{K}_1)$, and the resulting codes after the transformation will be denoted $(\tilde f_i,\tilde \varphi_i) \in \mcf{P}(\mbcf{X}|\mcf{\tilde M}, \mcf{K}_1,\mcf{K}_2) \times \mcf{P}(\mcf{\tilde M} \cup \set{\mbf{!}} | \mbcf{Y}, \mcf{K}_1,\mcf{K}_2)$ for each round $i \in \{1,\dots, j\}$. 
New codes are necessary for each round in order to prevent attacks where the adversary sends a previous rounds' message.



\noindent  \textbf{Random codebook generation:} 
Independently for each $k_2 \in \mcf{K}_2$ and $i \in \{1,\dots,j\}$, select a mapping $g_{k_2,i}: \mcf{\tilde M} \rightarrow \mcf{M}$ uniformly from the set of all injective mappings from $\mcf{\tilde M}$ to $\mcf{M}$.

\noindent  \textbf{Encoders:}
\[
\tilde f_i(\mbf{x} | m',k_1,k_2) \defn f_i(\mbf{x} | g_{k_2,i}(m'),k_1)
\]
for each $(\mbf{x},m',k_1,k_2) \in \mbcf{X} \times \mcf{\tilde M} \times \mcf{K}_1 \times \mcf{K}_2$ and $i \in \{1,\dots,j\}$.

\noindent  \textbf{Decoders:}
\[
\tilde \varphi_i(m'|\mbf{y},k_1,k_2) = \begin{cases}
\varphi_i(g_{k_2,i}(m')|\mbf{y},k_1) &\text{ if } m' \neq \mbf{!}  \\
\varphi_i(\mbf{!} | \mbf{y},k) + \varphi_i(\mcf{M} - g_{k_2,i}(\mcf{\tilde M})|\mbf{y},k_1) &\text{ otherwise} 
\end{cases},
\]
for all $(\mbf{y},k_1, k_2) \in \mbcf{Y}\times \mcf{K}_1 \times \mcf{K}_2$,  $m' \in \mcf{\tilde M} \cup \mbf{!}$ and $i \in \{ 1,\dots,j\}$. 

\subsection{Message error analysis}\label{app:cc2:mea}
The average probability of message error over all possible $(\tilde f_i,\tilde \varphi_i)$ is equal to the probability of message error for $(f_i,\varphi_i)$. Indeed, this is a direct consequence of
\begin{align}
\varepsilon_{\tilde f_i,\tilde \varphi_i}(m',k_1,k_2) &= 1- \sum_{\subalign{\mbf{y}&\in \mbcf{Y}, \\ \mbf{x}&\in \mbcf{X} }} \tilde \varphi_i(m'|\mbf{y},k_1,k_2)  \mbf{t}(\mbf{y}|\mbf{x}) \tilde f_i(\mbf{x}|m',k_1,k_2) \notag \\
&= 1- \sum_{\subalign{\mbf{y}&\in \mbcf{Y}, \\ \mbf{x}&\in \mbcf{X} }}  \varphi_i(g_{k_2,i}(m')|\mbf{y},k_1)  \mbf{t}(\mbf{y}|\mbf{x}) f_i(\mbf{x}|g_{k_2,i}(m'),k_1) \notag \\
&= \varepsilon_{f_i,\varphi_i}(g_{k_2,i}(m'),k_1),
\end{align}
and the fact that the mapping $g_{k_2,i}$ is chosen uniformly from the set of of all injective mappings.
Therefore,
\begin{equation}
\mathbb{E}[ \varepsilon_{\tilde F_i,\tilde \Phi_i} ]  = \sum_{m' \in \mcf{\tilde M}, k_1 \in \mcf{K}_1 } 2^{-n(r - \beta + \kappa)} \left( \sum_{m \in \mcf{M}} 2^{-n r } \varepsilon_{f_i,\varphi_i}(m,k_1)\right) = \varepsilon_{f_i,\varphi_i} 
\end{equation}
since $g_{k_2,i}$ is chosen uniformly from the set of all injective mappings $\mcf{\tilde M} \rightarrow \mcf{M}$. 

Now, $$\Pr \left( \varepsilon_{\tilde F_i,\tilde \Phi_i} \geq \sqrt{\epsilon} \right) \leq \frac{\varepsilon_{f_i,\phi_i}}{\sqrt{\epsilon}} \leq \sqrt{\epsilon}$$
directly follows from Markov's inequality for each $i \in \{1,\dots,j\}$.
Then, 
$$\Pr \left( \cup_{i=1}^j \{ \varepsilon_{\tilde F_i,\tilde \Phi_i} \geq \sqrt{\epsilon} \}  \right) \leq j \sqrt{\epsilon}$$
comes directly from the union bound.

\subsection{Authentication rate analysis}\label{app:cc2:t1e}

Here we shall show that given the original code has average authentication rate $\alpha$, the new average authentication rate will be  
\begin{equation}\label{eq:thm:-t:tt1e}
\tilde \alpha  \geq \alpha + \beta - 2n^{-1} \logt ne
\end{equation}
as long as $(\tilde f^j,\tilde \varphi^j) \in \mcf{G}^*$, where $\mcf{G}^*$ is the set of $(\tilde f^j,\tilde \varphi^j)$ for which 
\begin{equation}\label{eq:thm:-t:suff2}
\abs{\set{k_2 : \{ m' \in g_{k_2,\ell}(\mcf{\tilde M}) - \{ m_\ell\} \} \cap_{i=1}^j \{ m_i \in g_{k_2,i}(\mcf{\tilde M}) \}  }} \leq (ne)^2
\end{equation}
for all $m' \times m^j \in \mcf{M} \times \mcf{M}^j.$
Furthermore, we shall show that 
\begin{equation}\label{eq:thm:-t:p(code)}
\Pr \left( (\tilde F^j,\tilde \Phi^j) \notin \mcf{G}^* \right) \leq  j e^{-\frac{n^2}{4}+ n(j+1)r}.
\end{equation}
Hence, as $n\rightarrow \infty$ the probability of a code satisfying~\eqref{eq:thm:-t:tt1e} goes to one.

First, we shall demonstrate that Equation~\eqref{eq:thm:-t:suff2} is sufficient to guarantee~\eqref{eq:thm:-t:tt1e}.
Afterwards, we shall show that~\eqref{eq:thm:-t:p(code)} is a lower bound on the probability of a randomly chosen codes constructed according to Appendix~\ref{app:cc2:cc} satisfying~\eqref{eq:thm:-t:suff2}.

\subsubsection{Authentication rate given $(\tilde f^j,\tilde \varphi^j) \in \mcf{G}^*$}

To show 
\begin{align} 
\tilde \alpha   &\geq \alpha +\beta - 2n^{-1} \logt ne  \label{eq:app:cc2:t1e:1:wts}  
\end{align}
for all $(\tilde f^j,\tilde \varphi^j) \in \mcf{G}^*(f^j,\varphi^j)$ and $i\in \{1,\dots,j\}$, it will be helpful to first recognize that
\begin{align}
2^{-n\alpha} 
&=\max_{\subalign{i &\in \{1,\dots,j\},\\ \psi &\in \mcf{P}(\mbcf{Y}|\mbcf{Z}^j)} } \sum_{\subalign{ \mbf{x}^j &\in \mbcf{X}^j, \\  \mbf{y} &\in  \mbcf{Y} , \\ \mbf{z}^j &\in \mbcf{Z}^j,\\   m^j &\in \mcf{ M}^j,\\ k_1 &\in \mcf{K}_1}}  \hspace{-5pt}\psi(\mbf{y}|\mbf{z}^j)  2^{-nj(r+\kappa )} \varphi_i(\mcf{ M}-\{m_i\}| \mbf{y},k_1) \prod_{\ell=1}^j \left[\mbf{q}(\mbf{z}_\ell|\mbf{x}_\ell)  f_\ell(\mbf{x}_\ell|m_\ell,k_1)\right] \label{eq:app:cc2:t1e:1:-1a} 
\end{align}
and
\begin{align}
2^{-n\tilde \alpha} &= \max_{\subalign{i &\in \{1,\dots,j\},\\ \psi &\in \mcf{P}(\mbcf{Y}|\mbcf{Z}^j)} } \sum_{\subalign{ \mbf{x}^j &\in \mbcf{X}^j, \\  \mbf{y} &\in  \mbcf{Y} , \\ \mbf{z}^j &\in \mbcf{Z}^j,\\   \tilde m^j &\in \mcf{\tilde M}^j,\\ k_1 &\in \mcf{K}_1, \\ k_2 & \in \mcf{K}_2 }}  \hspace{-5pt} \psi(\mbf{y}|\mbf{z}^j)  2^{-nj(r+\kappa ) - n \beta } \tilde \varphi_i(\mcf{\tilde M}-\{\tilde m_i\}| \mbf{y},k_1,k_2) \prod_{\ell=1}^j \left[\mbf{q}(\mbf{z}_\ell|\mbf{x}_\ell)  \tilde f_\ell(\mbf{x}_\ell|\tilde m_\ell,k_1,k_2)\right] \label{eq:app:cc2:t1e:1:-1b} 
\end{align}
both follow by applying the law of total probability to the definition of the average authentication rate.
The goal of the proof is to express an upper bound on the RHS of~\eqref{eq:app:cc2:t1e:1:-1b} in terms of the RHS of~\eqref{eq:app:cc2:t1e:1:-1a}.
Towards this goal, the RHS of~\eqref{eq:app:cc2:t1e:1:-1b} can be expressed in terms of $(f_i,\varphi_i)$, specifically as 
\begin{align}
&\max_{\subalign{i &\in \{1,\dots,j\},\\ \psi &\in \mcf{P}(\mbcf{Y}|\mbcf{Z}^j)} }  \sum_{\subalign{ \mbf{x}^j &\in \mbcf{X}^j, \\  \mbf{y} &\in  \mbcf{Y} , \\ \mbf{z}^j &\in \mbcf{Z}^j,\\  \tilde m^j &\in \mcf{\tilde M}^j,\\ k_1 &\in \mcf{K}_1,\\k_2 &\in \mcf{K}_2}}  \hspace{-5pt}\psi(\mbf{y}|\mbf{z}^j)  2^{-nj(r+\kappa ) - n\beta} \varphi_i(g_{k_2,i}(\mcf{\tilde M})-\{g_{k_2,i}(\tilde m_i)\}| \mbf{y},k_1) \prod_{\ell=1}^j \left[\mbf{q}(\mbf{z}_\ell|\mbf{x}_\ell)  f_\ell(\mbf{x}_\ell|g_{k_2,\ell}(\tilde m_\ell),k_1)\right] \label{eq:app:cc2:t1e:1:-1} 
\end{align}
by using the definition of $(\tilde f_i,\tilde \varphi_i)$, and recognizing that $g_{k_2,i}(\mcf{\tilde M} - \{\tilde m_i\}) = g_{k_2,i}(\mcf{\tilde M}) - \{g_{k_2,i}(\tilde m)\}$ since $g_{k_2,i}$ is an injective function for all $k_2 \in \mcf{K}_2$ and $i \in \{1,\dots, j\}$. 
Notice now that all terms in~\eqref{eq:app:cc2:t1e:1:-1} that contain a $\tilde m^j \in \mcf{\tilde M}^j$ are only dependent on the value after the mapping of $g_{k_2,i}$ is applied. 
In fact,~\eqref{eq:app:cc2:t1e:1:-1} is equal to
\begin{align}
&\max_{\subalign{i &\in \{1,\dots,j\},\\ \psi &\in \mcf{P}(\mbcf{Y}|\mbcf{Z}^j)} }  \sum_{\subalign{ \mbf{x}^j &\in \mbcf{X}^j, \\  \mbf{y} &\in  \mbcf{Y} , \\ \mbf{z}^j &\in \mbcf{Z}^j,\\  k_1 &\in \mcf{K}_1,\\ m^\ell &\in \mcf{M}^\ell ,\\ m' &\in \mcf{M}}}  \hspace{-5pt}\psi(\mbf{y}|\mbf{z}^j )  2^{-nj(r+\kappa ) - n\beta} \varphi_i( m'| \mbf{y},k_1) \left( \prod_{\ell=1}^j \left[\mbf{q}(\mbf{z}_\ell|\mbf{x}_\ell)  f_\ell(\mbf{x}_\ell|m_\ell,k_1)  \right] \right)   \notag \\
&\hspace{60pt} \cdot \sum_{k_2 \in \mcf{K}_2} \idc{ m' \in g_{k_2,i} (\mcf{\tilde M}) -  \{m_i\} } \prod_{\ell=1}^j \idc{m_\ell \in  g_{k_2,\ell}(\mcf{\tilde M})} 
\label{eq:app:cc2:t1e:1:1}
\end{align}
since $g_{k_2,i}$ is an injective function for all $k_2 \in \mcf{K}_2$ and $i \in \{1,\dots, j\}$.
But, $(\tilde f^j, \tilde \varphi^j) \in \mcf{G}^*(f^j,\varphi^j)$ requires that the summation over $\mcf{K}_2$ in~\eqref{eq:app:cc2:t1e:1:1} has upper bound $(ne)^2$ when $m' \neq m_i$ and $0$ when $m'=m_i$, and hence
\begin{align}
2^{-n\alpha} &\leq \max_{\subalign{i &\in \{1,\dots,j\},\\ \psi &\in \mcf{P}(\mbcf{Y}|\mbcf{Z}^j)} } (ne)^2 \sum_{\subalign{ \mbf{x}^j &\in \mbcf{X}^j, \\  \mbf{y} &\in  \mbcf{Y} , \\ \mbf{z}^j &\in \mbcf{Z}^j,\\   k_1 &\in \mcf{K}_1,\\ m^\ell &\in \mcf{M}^\ell ,\\ m' &\in \mcf{M}}}  \hspace{-5pt}\psi(\mbf{y}|\mbf{z}^j )  2^{-nj(r+\kappa ) - n\beta} \varphi_i( m'| \mbf{y},k_1) \left( \prod_{\ell=1}^j \left[\mbf{q}(\mbf{z}_\ell|\mbf{x}_\ell)  f_\ell(\mbf{x}_\ell|m_\ell,k_1)  \right] \right) \idc{m'\neq m} \notag \\
&=\max_{\subalign{i &\in \{1,\dots,j\},\\ \psi &\in \mcf{P}(\mbcf{Y}|\mbcf{Z}^j)} }(ne)^2 2^{-n\beta} \sum_{\subalign{ \mbf{x}^j &\in \mbcf{X}^j, \\  \mbf{y} &\in  \mbcf{Y} , \\ \mbf{z}^j &\in \mbcf{Z}^j,\\ k_1 &\in \mcf{K}_1,\\ m^\ell &\in \mcf{M}^\ell }}  \hspace{-5pt}\psi(\mbf{y}|\mbf{z}^j )  2^{-nj(r+\kappa )} \varphi_i( \mcf{M}-\{m_i\}| \mbf{y},k_1) \left( \prod_{\ell=1}^j \left[\mbf{q}(\mbf{z}_\ell|\mbf{x}_\ell)  f_\ell(\mbf{x}_\ell|m_\ell,k_1)  \right] \right) .
\label{eq:app:cc2:t1e:1:2}
\end{align}
The combination of Equations~\eqref{eq:app:cc2:t1e:1:-1a} and~\eqref{eq:app:cc2:t1e:1:2} proves~\eqref{eq:app:cc2:t1e:1:wts}.

\subsubsection{Probability of randomly choosing a code in $\mcf{G}^*$}

Equation~\eqref{eq:thm:-t:p(code)} follows from Corollary~\ref{cor:ck} and the union bound after demonstrating that (which we shall return to later)
\begin{align}
\Pr \left( \{ m' \in G_{k_2,\ell}(\mcf{\tilde M}) - \{ m_\ell\} \} \cap_{i=1}^j \{ m_i \in G_{k_2,i}(\mcf{\tilde M}) \}  \right) \leq 2^{-n(j+1)\beta},\label{eq:app:cc2:t1e:2:nts}
\end{align}
where $G_{k_2,i}$ is the RV denoting randomly chosen injective mapping in the code construction\footnote{It should be noted that the set of mappings $g_{k_2,i}$ in conjunction with $(f^j,\varphi^j)$ determines the value of $(\tilde f^j,\tilde \varphi^j)$.},
for a give $m' \times m^j \in \mcf{M} \times \mcf{M}^j$, $\ell \in \{1,\dots,j\}$, and $k_2 \in \mcf{K}_2$.
Indeed, it is required that 
\begin{align}
\sum_{k_2 \in \mcf{K}_2}  \idc{ m' \in g_{k_2,\ell}(\mcf{\tilde M})- \{ m_{\ell}\} } \prod_{i=1}^j \idc{ m_i \in g_{k_2,i}(\mcf{\tilde M})}  \leq (ne)^2 \label{eq:app:cc2:t1e:2:req}
\end{align}
for all $m' \times m^j \in \mcf{M} \times \mcf{M}^j$ and $\ell \in \{1,\dots,j\}$ in order for $(\tilde f^j,\tilde \varphi^j) \in \mcf{G}^*$.
But, the mappings $g_{k_2,i} : \mcf{\tilde M} \rightarrow \mcf{M}$ are independently chosen for each $k_2 \in \mcf{K}_2$ and $i \in \{1,\dots,j\}$.
As a result, the probability that a set of mappings is chosen that satisfy~\eqref{eq:app:cc2:t1e:2:req} is equivalent to the probability that
\begin{align}
\sum_{k_2 \in \mcf{K}_2}  \idc{ m' \in G_{k_2,\ell}(\mcf{\tilde M})- \{ m_{\ell}\} } \prod_{i=1}^j \idc{ m_i \in G_{k_2,i}(\mcf{\tilde M})}  \leq (ne)^2 \label{eq:app:cc2:t1e:2:req2}.
\end{align}
The probability of~\eqref{eq:app:cc2:t1e:2:req2} is greater than $1-e^{-n^2/4}$ by Corollary~\ref{cor:ck} since the LHS of~\eqref{eq:app:cc2:t1e:2:req2} is a sum of $|\mcf{K}_2| = 2^{n(j+1)\beta}$ independent Bernoulli RVs whose parameter is bounded above in Equation~\eqref{eq:app:cc2:t1e:2:nts}.  
Hence, Equation~\eqref{eq:thm:-t:p(code)} by applying the union bound to consider all $m' \times m^j \in \mcf{M} \times \mcf{M}^j$ and $\ell \in \{1,\dots,j\}$ simultaneously.

Returning now to prove Equation~\eqref{eq:app:cc2:t1e:2:nts}. 
First note that 
\begin{align}
&\Pr \left(  m' \in G_{k_2,\ell}(\mcf{\tilde M}) - \{ m_\ell\} \} \cap_{i=1}^j \{ m_i \in G_{k_2,i}(\mcf{\tilde M}) \}  \right) \notag \\
&= \Pr \left( \{ m',m_{\ell} \} \subset G_{k_2,\ell}(\mcf{\tilde M}) \} \right) \prod_{i=1 (\neq \ell)}^j \Pr \left(  m_i \in G_{k_2,i}(\mcf{\tilde M})  \right) ,\label{eq:app:cc2:t1e:2:nts1}
\end{align}
since $G_{k_2,i}$ is independent for each $i \in \{1,\dots,\ell\}.$
The probability that $m \in G_{k_2,i}(\mcf{\tilde M})$ (alternatively ($\{m,m'\} \subset G_{k_2,i}(\mcf{\tilde M})$)  is equal to the ratio of the number of subsets $\mcf{A} \subset \mcf{M}$ such that $m \in \mcf{A}$ (resp. $\{m,m'\} \in \mcf{A}$) and $|\mcf{A}| = |\mcf{\tilde M}|$ to the number of subsets $\mcf{B} \subset \mcf{M}$ such that $|\mcf{B}|= |\mcf{\tilde M}|,$ since $G_{k_2,i}$ is uniform over the set of all injective mappings from $\mcf{\tilde M}$ to $\mcf{M}$.
Hence,
\begin{align}
\Pr \left( m\in  G_{k_2,i}(\mcf{\tilde M}) \right) &=\left.\left( \begin{matrix} 2^{nr}-1 \\ 2^{n(r-\beta)}-1 \end{matrix} \right)\middle/ \left( \begin{matrix} 2^{nr} \\ 2^{n(r-\beta)} \end{matrix} \right) \right. 
=2^{-n\beta} \label{eq:app:cc2:t1e:nts2}
\end{align}
and similarly
\begin{align}
\Pr \left( \{m,m'\} \subset  G_{k_2,i}(\mcf{\tilde M}) \right) &=\left.\left( \begin{matrix} 2^{nr}-2 \\ 2^{n(r-\beta)}-2 \end{matrix} \right)\middle/ \left( \begin{matrix} 2^{nr} \\ 2^{n(r-\beta)} \end{matrix} \right) \right. 
\notag \\ &
=2^{-n\beta} \frac{2^{n(r-\beta)}-1}{2^{nr} - 1} 
= 2^{-2n \beta} \frac{1 - 2^{n(\beta - r) }}{1 - 2^{-nr}} \leq 2^{-2n\beta}. \label{eq:app:cc2:t1e:nts3}
\end{align}
Combining Equations~\eqref{eq:app:cc2:t1e:2:nts1}--\eqref{eq:app:cc2:t1e:nts3} yields Equation~\eqref{eq:app:cc2:t1e:2:nts}.

\end{IEEEproof}

\subsection{Proof of Theorem~\ref{thm:-t2}}

\begin{IEEEproof}

If $(r,\alpha,\kappa) \in \mcf{C}_{\text{AA}(j)}(t,q)$, then there exists a sequence of $(r_n,\alpha_n,\kappa_n,\epsilon_n,j,n)$-AA$(j)$ codes, $(f_n^j,\varphi_n^j)$, such that 
\begin{align}
\lim_{n\rightarrow \infty} |(r_n,\alpha_n,\kappa_n,\epsilon_n,n) - (r,\alpha,\kappa,0,n) |= 0.
\end{align}
Hence, there exists a sequence $\beta_n \leq r_n $  such that $\lim_{n\rightarrow \infty} \beta = \beta \leq r$, as well as a sequence of $(r_n-\beta_n , \alpha_n + \beta_n - 2n^{-1} \logt ne, \kappa + [1+j^{-1}] \beta_n, \sqrt{\epsilon_n},j, n)$-AA codes, $(\tilde f^j_n,\tilde \varphi^j_n)$, by Theorem~\ref{thm:cc2:n} since $j$ is fixed and thus $\lim_{n\rightarrow \infty} j\sqrt{\epsilon_n} = 0.$
Therefore, 
\begin{align} 
(r-\beta,\alpha+\beta, \kappa + [1+j^{-1}]\beta) \in \mcf{C}_{\text{AA}(j)}(t,q)
\end{align} 
since 
\begin{align}
\lim_{n\rightarrow \infty} |(r_n-\beta_n,\alpha_n+\beta_n - 2n^{-1} \logt ne,\kappa_n + [1+j^{-1}]\beta_n,\sqrt{\epsilon_n},n) - (r-\beta,\alpha+\beta,\kappa+[1+j^{-1}]\beta,0,n) |= 0.
\end{align}

\end{IEEEproof}
\section{Error in~\cite{gungor2016basic}}\label{app:gungordonescrewedup}

For this section, we switch to using the notation of~\cite{gungor2016basic}. We make no effort to reproduce or explain their notation here. The error occurs in the code construction (Appendix A), and revolves around how their decoder is defined. To understand this error, we start at~\cite[page~4535]{gungor2016basic}, and discuss their decoder. 

First introduced is set
\[
\mcf{D}_{k_1m|k_2} \defn \set{\mbf{y}: V_{k_1k_2m} \prec V_{\hat k_1 k_2 \hat m}~\forall~[\hat k_1,\hat m] \neq [k_1,m]}
\]
where $V_{k_1k_2m}  \in \mcf{P}(\mcf{Y}|\mcf{X})$, and $\prec$ is defined by if $ V_{k_1k_2m} \prec V_{\hat k_1 k_2 \hat m}$, then
\[
D(V_{k_1k_2m} ||W_{t} | P) + \left| I(P,V_{\hat k_1 k_2 \hat m}) - R_{K_1} - R_{M}  \right| < \xi.
\]
It is important to note that $V_{k_1k_2m} \prec V_{\hat k_1 k_2 \hat m}$ does not imply $V_{\hat k_1 k_2 \hat m}  \not \prec V_{k_1k_2m} $, and hence the regions $\mcf{D}_{k_1m|k_2}$ are not distinct for different $k_1,m$. The decoder is then defined by
\[
\phi(\mbf{y}, k_1,k_2) = \begin{cases}
m &\text{ if } \mbf{y} \in \mcf{D}_{k_1m|k_2} \\
0 &\text{ otherwise }
\end{cases},
\]
with the caveat that if there exists a $\hat k_1,\hat m$ such that $V_{k_1k_2m} \prec V_{\hat k_1 k_2 \hat m}$ and $V_{\hat k_1 k_2 \hat m}  \prec V_{k_1k_2m} $, then one of the messages is chosen arbitrarily\footnote{Although we will not use this fact, we feel compelled to point out that there potentially exist an exponential number of such $k_1,m$ for which $\mbf{y} \in \mcf{D}_{k_1m|k_2}$. }. This arbitrary choice is never defined; this is problematic since different choices will cause their analysis to fail in different sections. 

Let us, in good faith, assume that Equation~$(63)$ does in fact equal Equation~$(62)$, that is 
\begin{align*}
&\frac{1}{e^{nR}}\sum_{k_1,k_2,m} \mathbb{P}(\alpha_{na}|m,k_1,k_2,c) \\
&=\frac{1}{e^{nR}}\sum_{k_1,k_2,m} W_t^n ( \mcf{D}_{0|k_2} \cup \cup_{\hat k_1 \neq k_1, \hat m} \mcf{D}_{\hat k_1 \hat m|k_2} | \mbf{x}_{k_1k_2m}).
\end{align*}
This would imply that an erasure occurs if $\mbf{y} \in \mcf{D}_{\hat k_1 \hat m|k_2}$ for some $\hat k_1 \neq k_1$, hence the arbitrary choice mentioned before would always be in favor of $[k_1',m']$ for which $k_1' \neq k_1$. But now, the unjustified Equation~$(64)$ cannot follow from Equation~$(63)$ since $\mcf{D}_{k_1m|k_2}$ are not disjoint. Indeed, if $\cup_{\hat k_1 \neq k_1, \hat m} \mcf{D}_{\hat k_1 \hat m|k_2}  \cap \mcf{D}_{k_1m|k_2} \neq \emptyset$, then 
\[
\left( \mcf{D}_{0|k_2} \cup \cup_{\hat k_1 \neq k_1, \hat m} \mcf{D}_{\hat k_1 \hat m|k_2} \right) \not \subset \mcf{\bar D}_{k_1m|k_2}.
\]
This is actually a rather extreme error as it allows for reliable transmissions above the channel's capacity.

So, now we rectify this error and, in good faith, assume the rest of their paper is correct. Starting with Equation~$(63)$, which can be bounded as follows
\begin{align*}
&\frac{1}{e^{nR}}\sum_{k_1,k_2,m} W_t^n ( \mcf{D}_{0|k_2} \cup \cup_{\hat k_1 \neq k_1, \hat m} \mcf{D}_{\hat k_1 \hat m|k_2} | \mbf{x}_{k_1k_2m}) \\
&\leq \frac{1}{e^{nR}}\sum_{k_1,k_2,m} W_t^n ( \mcf{D}_{0|k_2} |\mbf{x}_{k_1k_2m} ) + W_t^n(\cup_{\hat k_1 \neq k_1, \hat m} \mcf{D}_{\hat k_1 \hat m|k_2} | \mbf{x}_{k_1k_2m}) \\
&\leq \frac{1}{e^{nR}}\sum_{k_1,k_2,m} W_t^n ( \mcf{\bar D}_{k_1m|k_2} |\mbf{x}_{k_1k_2m} ) + W_t^n(\cup_{\hat k_1 \neq k_1, \hat m} \mcf{D}_{\hat k_1 \hat m|k_2} | \mbf{x}_{k_1k_2m}).
\end{align*}
At this point, it should also be noted that
\[
\max \left( \frac{1}{e^{nR}} \sum_{k_1,k_2,m} W_t^n ( \mcf{\bar D}_{k_1m|k_2} |\mbf{x}_{k_1k_2m} ),\frac{1}{e^{nR}} \sum_{k_1,k_2,m}  W_t^n(\cup_{\hat k_1 \neq k_1, \hat m} \mcf{D}_{\hat k_1 \hat m|k_2} | \mbf{x}_{k_1k_2m}), \right)
\]
constitutes a \emph{lower bound} on the total probability error (erasure and undetected) under no attack. Hence, these two summations can be analyzed individually without impact to the total performance. A tight upper bound on the first summation is already derived in Equations~$(65)$--$(69)$ and is less than
\[
\max_{V,\hat V: V \not \prec \hat V} \exp-n\left[D(V||W_t|P) + \left| I(P,\hat V) - R_M - R_{K_1} \right|^+ \right].
\]
The second summation can be bounded as follows
\begin{align*}
&\frac{1}{e^{nR}}  \sum_{k_1,k_2,m}  W_t^n(\cup_{\hat k_1 \neq k_1, \hat m} \mcf{D}_{\hat k_1 \hat m|k_2} | \mbf{x}_{k_1k_2m}) \\
&=\frac{1}{e^{nR}} \sum_{k_1,k_2,m}  W_t^n(\cup_{V} \mcf{T}_{V}(\mbf{x}_{k_1k_2m}) \cap \cup_{\hat k_1 \neq k_1, \hat m} \cup_{\hat V: \hat V \prec V } \mcf{T}_{\hat V}(\mbf{x}_{\hat k_1,k_2,\hat m}) | \mbf{x}_{k_1k_2m}) \\
&=\frac{1}{e^{nR}} \sum_{k_1,k_2,m}  W_t^n(\cup_{V, \hat V : \hat V \prec V}   \mcf{T}_{V}(\mbf{x}_{k_1k_2m})  \cap  \cup_{\hat k_1 \neq k_1, \hat m} \mcf{T}_{\hat V}(\mbf{x}_{\hat k_1,k_2,\hat m}) | \mbf{x}_{k_1k_2m}) \\
&\dot \leq \frac{1}{e^{nR}}  \sum_{k_1,k_2,m} \sum_{V,\hat V: \hat V \prec V} e^{-n D(V||W_t|P)} \frac{\abs{  \mcf{T}_{V}(\mbf{x}_{k_1k_2m})  \cap  \cup_{\hat k_1 \neq k_1, \hat m} \mcf{T}_{\hat V}(\mbf{x}_{\hat k_1,k_2,\hat m})}}{\abs{\mcf{T}_{V}(\mbf{x}_{k_1k_2m})}} \\
&\dot = \frac{1}{e^{nR}}  \sum_{k_1,k_2,m} \max_{V,\hat V: \hat V \prec V} e^{-n D(V||W_t|P) } e^{-n \left| I(P,\hat V) - R_{K} - R_{M} \right|^+} \\
&= \max_{V,\hat V : \hat V \prec V} \exp -n\left[D(V||W_t|P) + \left| I(P,\hat V) - R_M - R_{K_1} \right|^+ \right],
\end{align*}
where the inequality is due to~\cite[Lemma~1]{gungor2016basic}, and everything else is by type class properties. Furthermore, it should be clear that this bound is indeed tight as well due to the tightness of~\cite[Lemma~1]{gungor2016basic}.

Hence, their error exponent can be properly expressed as 
\[
\max_{V,\hat V : V \not\prec \hat V \text{ or } \hat V \prec V } \exp-n\left[D(V||W_t|P) + \left| I(P,\hat V) - R_M - R_{K_1} \right|^+ \right].
\]
This is very important because it establishes that 
\[
R_{M} + R_{K_1} < \mathbb{I}(P,W_t)
\]
is required for their code to work. To see this, observe that the pair $(W_t,W_t)$ always satisfies $W_t \prec W_t$ or $W_t \not \prec W_t$. Hence, their lower bound is always greater than 
\begin{align*}
&\max_{V,\hat V : V \not\prec \hat V \text{ or } \hat V \prec V } \exp-n\left[D(V||W_t|P) + \left| I(P,\hat V) - R_M - R_{K_1} \right|^+ \right] \\
&\geq  \exp \left( -n \left( D(W_t||W_t|P) + \left|I(P, W_t) - R_{M} - R_{K_1} \right|^+ \right)\right) \\
&= \exp \left( -n \left|I(P, W_t) - R_{M} - R_{K_1} \right|^+ \right).
\end{align*}

\bibliographystyle{ieeetr}
\bibliography{this}

\end{document}